\documentclass[prb,aps,twocolumn,amsfonts,showpacs,eqsecnum,superscriptaddress,floatfix,multicol]{revtex4}
\usepackage{epsfig}
\usepackage{amsmath}
\usepackage{psfrag}
\usepackage{color}
\usepackage{graphicx}
\usepackage{subfigure}
\usepackage{bm}
\def\bea{\begin{eqnarray}}
\def\eea{\end{eqnarray}}
\def\be{\begin{equation}}
\def\ee{\end{equation}}

\begin{document}
\title{Dynamical Spin Response of Doped Two-Leg Hubbard-like Ladders}
\author{F.H.L. Essler}
\affiliation{The Rudolf Peierls Centre for Theoretical Physics\\
Oxford University, 1 Keble Road, Oxford OX1 3NP, UK\\}
\author{R.M. Konik}
\affiliation{Brookhaven National Laboratory,
Building 510A, Upton, NY 11973, USA}
\date{\today}

\begin{abstract}
We study the dynamical spin response of doped two-leg Hubbard-like 
ladders in the framework of a low-energy effective field theory 
description given by the SO(6) Gross Neveu model. 
Using the integrability of the SO(6) Gross-Neveu model, we derive the 
low energy dynamical magnetic susceptibility.  The susceptibility
is characterized by an incommensurate coherent mode
near $(\pi,\pi)$ and by broad two excitation scattering continua
at other $k$-points.
In our computation we are able to estimate the relative weights of these contributions.

All calculations are performed using form-factor expansions which yield exact low energy results
in the context of the SO(6) Gross-Neveu model.  To employ this expansion,
a number of hitherto undetermined form factors were computed.  To do so, we developed a
general approach for the computation of matrix elements of semi-local SO(6) Gross-Neveu 
operators.
While our computation takes place in the context of SO(6) Gross-Neveu, 
we also consider the effects of perturbations away from an SO(6) symmetric model, showing
that small perturbations at best quantitatively change the physics.
\end{abstract}
\maketitle
\section{Introduction}
Ladder materials \cite{dagotto} bridge the great divide between the well
understood one dimensional (doped) Mott insulators and their
perplexing two-dimensional analogs. The desire to understand their
fascinating physical properties \cite{experiment} as well as the
expectation that an understanding of ladder materials would culminate
in the theoretical understanding of doped, two dimensional Mott
insulators has led to intense, sustained theoretical interest in
the problem. Analytical
\cite{Fabr93,dima,schulz96,balents,lin,so8,schulz,varma,controzzi2,LAB,tsuchiizu,fradkin}
as well as numerical \cite{noack,noack95,endres,eric,weihong,orignac}
techniques have been used to deduce the zero temperature phase
diagrams of ladder models. The dynamical properties are less well
characterized. In fact, in contrast to the half-filled case\cite{halffilled},
the accurate determination of dynamical response functions for two-leg
ladder lattice models at finite doping remains a challenging open problem. 
One approach to calculating dynamical correlations at finite doping is
to utilize numerical methods such as quantum Monte-Carlo techniques or
exact diagonalization. The single-particle spectral function and the
dynamical spin and charge susceptibilities for both undoped and doped
Hubbard ladders were computed by quantum Monte-Carlo methods in
Ref. (\onlinecite{endres}) and compared to analytical random-phase
approximation calculations for a two-leg Hubbard ladder. In
Ref. (\onlinecite{orignac}) the dynamical structure factor was computed
by exact diagonalization of a $2\times 16$ t-J ladder model. 

A different approach is to utilize field theory methods to describe
the low-energy degrees of freedom. In Ref. (\onlinecite{lin}) the
structure of various dynamical response functions for the half-filled
case was determined in the framework of a low-energy description in
terms of the SO(8) Gross-Neveu model \cite{gross}. 
Detailed calculations utilizing the integrability of the SO(8)
Gross-Neveu model were carried out in Ref. (\onlinecite{so8}). In the
doped case it has been argued in Refs. (\onlinecite{Fabr93}) and (\onlinecite{schulz96})
that two-leg ladders can be described by the SO(6) Gross-Neveu model $\otimes$
U(1) Luttinger liquid at low energies.  The Luttinger liquid describes the gapless charge
degrees of freedom on the ladder while the SO(6) Gross-Neveu model governs the behavior
of the gapped degrees of freedom in the spin and orbital sectors.
A prominent feature in the magnetic response due to
bound state formation was analyzed in the framework of this theory and
compared to numerical computations for a two-leg t-J ladder in
Ref. (\onlinecite{orignac}). 

It is important to go beyond the existing calculations in order to
address questions raised by recent inelastic neutron scattering
experiments on doped cuprate two-leg ladders
\cite{bella}. Furthermore, the dynamical susceptibilities of doped
ladders can be used to develop a theory for systems of weakly coupled
ladders, which in turn can be used as toy models for the striped phase
in the cuprates \cite{stripes}.

In the present work we determine the dynamical structure factor of
doped Hubbard-like two-leg ladders at low energies in the framework
of the SO(6) Gross-Neveu description. We go beyond Ref. (\onlinecite{orignac}) 
in that (1) we consider all momentum space
regions with non-vanishing low-energy response;
and (2) we consider multiparticle contributions to the dynamical structure
factor.  This allows us to evaluate the contribution to the structure
factor of the bound
state identified in Ref. (\onlinecite{orignac}) relative to other contributions.

Given that we employ the SO(6) Gross-Neveu model as a low energy description of the doped
ladders it is worthwhile dwelling briefly on the nature of this description.  The SO(6) Gross-Neveu
model represents the attractive fixed point of an RG flow of a doped ladder with weak generic 
interactions\cite{schulz96}.  The flow has been established to two-loop order\cite{Fabr93}.  
We point out that the flow to the SO(6) Gross-Neveu does not assume that the bonding and antibonding bands
of the Hubbard ladder have identical velocities.
In the work of Ref. (\onlinecite{Fabr93}), it is shown that differing Fermi velocities will renormalize towards one 
another (an effect present
only at two loops).

As the
RG is perturbative in nature, the bare couplings must not be particularly strong.  How strong they
may be before the RG becomes unreliable is indicated in Ref. (\onlinecite{azaria}).  In this work,
the related problem of half-filled ladders was studied where the RG flow takes the ladders to a sector
where their low energy behavior is described by the SO(8) Gross-Neveu model.  There it was found
that $U/t$ ($U$ being representative of the strength of some generic short ranged interaction) may
be as large as $6$ before the RG flow is a poor indicator of the low energy physics.
At larger values of $U$, it was found the charge and spin/orbital sectors decoupled.  This
decoupling is already present in the U(1) Luttinger liquid $\otimes$ SO(6) Gross-Neveu model and
so the RG analysis for the doped ladders may well be descriptive for
$U/t > 6$.

Beyond a question of whether the bare couplings are sufficiently weak, it may be asked
whether the RG flow accurately represents the physics in a wider sense.  Where things
may go wrong is illustrated by an appeal to the U(1) Thirring model\cite{azaria1}.  Such a model has
a Lagrangian of the form
\begin{equation}\label{eIi}
{\cal{L}} = 
\bar\Psi_\alpha \gamma^\mu\partial_\mu\Psi_\alpha + {1\over 4}g_\parallel
(j_z)^2 + {1\over 4}g_\perp [(j_x)^2+(j_y)^2],
\end{equation}
where $j^\mu_a = \bar\Psi\gamma^\mu\sigma_a\Psi$ and $\psi$,$\bar{\psi}$ are
Dirac spinors.  The one-loop RG equations
for this model indicate that there is generic symmetry restoration to
an SU(2) symmetric model.  However the RG equations are misleading 
in the sector $\pi-|g_\perp|>-g_\parallel>|g_\perp|>0$.  In this
sector the model maps onto the sine-Gordon model\cite{shura} with interaction,
$\cos(\beta\Phi)$, where $\beta$ is given by
\begin{equation}\label{eIii}
\beta^2 = 8\pi - 8\mu ; ~~~~~ \mu = 
{\cos ^{-1} [\cos (g_\parallel ) / \cos (g_\perp)]}.
\end{equation}
$\beta$ completely describes the physics of the model.  And while the
two coupling constants, $g_\parallel$ and $g_\perp$, flow under the RG,
$\beta$ does not, being an RG invariant.  Thus the dynamically enhanced symmetry
seemingly promised by the RG is here illusory.

This case is however special.  $\beta$ as an RG invariant in the U(1)
Thirring model is reflective of the presence of an SU(2) based quantum group symmetry.  The RG flow
cannot restore the SU(2) symmetry in this case as the symmetry has merely
been deformed rather than broken\cite{anis}.  There is however no counterpart of this
exotic symmetry in the case of the SO(6) Gross-Neveu model and so this scenario of false
RG-induced symmetry restoration will not appear here.

While the flow to an SO(6) symmetric model seems reasonably robust, we take a more cautious
approach and understand the RG flow merely to move the model to a more symmetric state.
Thus we should think of the low energy sector of the Hubbard ladders as an SO(6) Gross-Neveu model
plus smallish perturbations.  But because the SO(6) Gross-Neveu model is gapped, perturbation
theory about it is well controlled: small perturbations lead only to small changes in the physics.
We show in Section III that corrections induced by perturbations to SO(6) Gross-Neveu can indeed be straightforwardly
handled and lead only to quantitative changes in the results.

To extract the dynamic magnetic susceptibility from the integrability of the SO(6) Gross-Neveu model, we employ 
a so called ``form-factor'' expansion of the spin-spin correlation function.
This expansion yields exact results for response functions below certain energy thresholds.  
At the origin of this feature of the form-factor expansion is the notion of a well
defined elementary excitation or particle in an integrable model.  Particles in 
an integrable model have infinite lifetimes -- integrable models are thus in a sense
superior versions of Fermi liquids.  The interactions between particles in an integrable model
exist only in restricted forms.  The scattering of N particles can always be reduced to a
series of two body scattering amplitudes.  This reduction occurs because of the presence
of an infinite series of conservation laws.  This simplified scattering yields that generic
eigenfunctions of the SO(6) Gross-Neveu model can always be thought of as some distinct collection
of the elementary excitations.  In an non-integrable model this would not be possible because of
particle production and decay.

The particle content of SO(6) Gross-Neveu is relatively simple.  The elementary
excitations are found in two
four dimensional representations, the kink ({\bf 4}) and the anti-kink (${\bf\bar4}$).
These excitations represent the elementary fermionic excitations of the ladder (as
discussed in greater detail in Section II).
An interesting feature of SO(6) Gross-Neveu is that two kinks or two anti-kinks can form
bound states.  These bound states transform as the six dimensional vector representation ({\bf 6})
of SO(6).  A portion of this six dimensional vector carries the quantum numbers of an SU(2) triplet.
It is this bound state that provides a coherent response near $(\pi,\pi)$ in the doped ladders.

With the spectrum of SO(6) Gross-Neveu in hand, we can readily write down
the form factor expansion of any two-point correlation function (imaginary time ordered 
at $T=0$) in the model:
\vbox{
\begin{eqnarray}\label{eIiii}
G^{\cal O}_T(x,\tau) &=& \langle 0|T\bigl (
{\cal O}(x,\tau){\cal O}^{\dagger}(0,0)
\bigr)|0\rangle \cr
&=& \hskip -.1in \sum_{n=0}^{\infty}
\sum_{s_n}
\langle 0|{\cal O}(x,0)|n;s_n\rangle e^{- \tau E_{s_n}}\times\cr
&& \hskip .35in \langle n;s_n|{\cal O}^{\dagger}(0,0)|0\rangle, ~~ (\tau>0), 
\end{eqnarray}}
where $E_{s_n}$ is the energy
of the eigenstate, $|n;s_{n}\rangle$, and
$s_n$ marks all possible quantum numbers carried by the state.
In the second line we have inserted a resolution of the identity
between the two fields of the correlator.  This resolution 
is composed of eigenstates (i.e. multi-particle states) of the SO(6) Gross-Neveu
Hamiltonian.  We have thus reduced the computation of any correlation function
to the calculation of a series of matrix elements representing transition
amplitudes between the ground state and multi-particle states.

The integrability of the SO(6) Gross-Neveu model allows the computation of these matrix
elements in principle.  Each matrix element satisfies a number of algebraic constraints
which can be solved.  However as the particle number in the eigenstate $|n:s_n\rangle$ increases,
solving these constraints becomes increasingly cumbersome.  Fortunately, if one is only interested
in the behavior of the low energy spectral function (i.e. the imaginary portion of the 
corresponding retarded correlator), 
it is only necessary to compute matrix elements
involving a small number of particles.
The spectral function is given by
\begin{widetext}
\begin{eqnarray}\label{eIiv}
&&-{1 \over \pi} {\rm Im} G^{\cal O}_{T}(x, -i\omega + \delta ) =
\sum_{n=0}^{\infty}\sum_{s_n} \bigg\{ 
\langle 0|{\cal O}(x,0)|n;s_n\rangle
\langle n;s_n|{\cal O}^{\dagger}(0,0)|0\rangle  \delta(\omega- E_{s_n})\cr
\cr
&& \hskip 2.1in - \epsilon \langle 0|{\cal O}^\dagger(0,0)|n;s_n\rangle
\langle n;s_n|{\cal O}(x,0)|0\rangle  \delta(\omega + E_{s_n})\bigg\},
\end{eqnarray}
\end{widetext}
\noindent where $\epsilon = \pm$ 
for fields, $\cal{O}$, that are bosonic/fermionic.  The delta functions present
in the above equation guarantee that only eigenstates with energy, $\omega$, contribute
to the spectral function.  As each particle has a fixed mass, higher particle
states will not contribute if the sum of their masses is greater than the energy,
$\omega$, of interest.

In this article, we compute only the one and lowest two particle form-factors of
appropriate fields.  This allows us to compute the spin response exactly to an energy of
$2\sqrt{2}$ times the spin gap in the region of k-space near $(\pi,\pi)$.  In
other regions of the Brillouin zone we are able to obtain exact results 
to energies twice the spin gap.  However in these latter regions, the spectral
intensity is considerable smaller than that found near $(\pi,\pi)$.

While our results are exact up to certain energies, it is likely
that in certain cases (in particular, our computation of the spin response
in the region of $(\pi,\pi)$) they remain extremely accurate beyond these thresholds.  As a rule,
form-factor expansions have been found to be extremely convergent with
only the first terms yielding a significant contribution \cite{oldff1,delone,deltwo,review}.
As one example, in Ref. (\onlinecite{review}) the spectral function for the spin-spin
correlation function in the Ising model was computed.  The two-particle eigenstates
produced a contribution 1000 times greater than the four-particle eigenstates and
a contribution $10^8$ times greater than the six particle contribution.  For practical
purposes then the two particle form-factor in this case gave ``exact'' results for
energies far in excess of the two particle threshold.

The outline of this paper is as follows.  In Section II, we review the 
connection between doped Hubbard-like ladders and the SO(6)
Gross-Neveu model \cite{Fabr93,schulz,lin}.  In particular we outline the
renormalization flow that takes generically interacting doped ladders to the SO(6) Gross-Neveu model.
We also delineate the connections between the excitations and operators in SO(6) Gross-Neveu
and their corresponding ladder lattice counterparts.

In Section III, we present our results for the dynamical magnetic susceptibility.  We consider
this quantity for a variety of dopings ranging from 1\% to 25\%.  We also compare
the results derived from our field theoretic treatment to an RPA analysis as well as to non-interacting
ladders.  We find there are significant differences.  In particular the RPA cannot reproduce accurately the
most prominent feature in the spin response, the coherent mode near $(\pi,\pi)$. 

In Section IV, we present our results on form-factors in the SO(6) Gross-Neveu.
The form-factors that are needed for the computation were not all known
prior to this work.  In part this came about because the spin response involve
operators in SO(6) Gross-Neveu which are not standard objects of field theoretic interest.
These operators are unusual in that they are semi-local while not being
SO(6) currents.
To derive the matrix of elements of these operators it was first necessary to 
establish the algebraic constraints satisfied by these form-factors.  While it was not the primary
intent of this work, we developed a general scheme by which the constraints for all semi-local
operators in SO(6) Gross-Neveu (and by straightforward extension, SO(2N) Gross-Neveu
with N odd) may be written down.  Portions of this developments are technical.
The basics of the derivation of all the form factors may be found in Section IV with certain
details related to the semi-local operators relegated to Appendix B.  For those interested
in only the end product, a summary of all the relevant form-factors may be found
in Section IV H.

\section{From Doped Ladders to the SO(6) Gross-Neveu Model}
\label{sec:GN}
Here we will briefly review the connection between doped Hubbard ladders
and the $SO(6)$ Gross-Neveu model \cite{Fabr93,schulz,lin}. We discuss
the reduction of doped ladders to their field theoretic equivalent
together with how to understand the excitation spectrum and fields of
the $SO(6)$ Gross-Neveu model in terms of the original ladder model.

\subsection{Relation of doped D-Mott Hubbard Ladders to the $SO(6)$
Gross-Neveu model}

\newcommand{\ccr}{c^\dagger_{Rj\alpha}}
\newcommand{\car}{c_{Rj\alpha}}
\newcommand{\ccl}{c^\dagger_{Lj\alpha}}
\newcommand{\calf}{c_{Lj\alpha}}
\def\a{{\bm{\alpha}}}
\def\nn{\nonumber\\} 
We consider a weakly interacting ladder with generic repulsive
interactions. It is well established \cite{Fabr93} 
that the system exhibits algebraically decaying d-wave pairing. We
begin with non-interacting electrons hopping on a ladder: 
\begin{eqnarray}\label{eIIi}
H_0 = -\sum_{x,\alpha }
&& \bigl( ta^\dagger_{1\alpha}(x+1)a_{1\alpha}(x)
+ ta^\dagger_{2\alpha}(x+1)a_{2\alpha}(x) \cr
&& ~~~~~ + t_{\perp}a^\dagger_{1\alpha}(x)a_{2\alpha}(x) + {\rm h.c.}
\bigr).
\end{eqnarray}
Here the $a_l$/$a_l^\dagger$ are the electron annihilation/creation
operators for the electrons on rung $l$ of the ladder, $x$ is a discrete
coordinate along the ladder, and $\alpha = \uparrow ,\downarrow$
describes electron spin.  $t$ and $t_\perp$ describe respectively
hopping between and
along the ladder's rung.

The first step in the map is to reexpress the $a$'s of $H_0$ in terms
of bonding/anti-bonding variables:
\begin{equation}\label{eIIii}
c_{j\alpha} = \frac{1}{\sqrt{2}}(a_{1\alpha} + (-1)^j a_{2\alpha}),
\end{equation}
(noting $j=1=ab$ and $j=2=b$).
With this transformation, the Hamiltonian can be diagonalized in
momentum space in terms of two bands.  
As we are interested in the low energy behavior of
the theory, the $c_{j\alpha}$'s are
linearized about the Fermi surface, $k_{Fj}$:
\begin{equation}\label{eIIiii}
c_{j\alpha} \sim \car e^{ik_{Fj}x} + \calf e^{-ik_{Fj}x},
\end{equation}
where $L,R$ corresponding to the right and left moving modes about
the Fermi surface.
With this $H_0$ becomes,
\begin{equation}\label{eIIiv}
H_0 = \int dx \sum_{j\alpha} v_{Fj}
\bigl [
 c^\dagger_{Rj\alpha} i\partial_x c_{Rj\alpha} - c^\dagger_{Lj\alpha} 
i\partial_x c_{Lj\alpha}
\bigr ].
\end{equation}
Away from half-filling, the $v_{Fj}$ will be generically different.
However for small to moderate dopings, the difference of the two Fermi
velocities will be small. We thus set $v_{Fab}=v_{Fb}$ in order to
simplify matters.
 
The next step is to consider adding interactions to the system.
We consider adding all possible allowed interactions to this Hamiltonian.  To organize this addition,
we introduce various $SU(2)$ scalar and vector currents\\[-8mm]~
\begin{eqnarray}\label{eIIv}
I_{ij}^P &=& c_{Pi\sigma}\epsilon_{\sigma\sigma'}c_{Pj\sigma'}\,; 
~~~ {\bf
I}_{ij}^P=\frac{1}{2}c_{Pi\tau}(\epsilon{\bm{\sigma}})_{\tau\tau'}
c_{Pj\tau'}\,;\cr
J^{P}_{ij} &=& c^\dagger_{Pi\sigma}c_{Pj\sigma}\,; 
~~~ {\mathbf J}^P_{ij}=\frac{1}{2}
c^\dagger_{Pi\tau} {\bm{\sigma}}_{\tau\tau'}c_{Pj\tau'}\,.
\end{eqnarray}
Here $P=R,L$, ${\bf\sigma}$ are the Pauli matrices and
$\epsilon_{\sigma\sigma'}$ is the $\epsilon$-tensor. The
crystal-momentum conserving interactions divide themselves into
forward and backward scattering terms: 
\begin{eqnarray}\label{eIIvi}
H_{B} &=& \sum_{i,j=1,2} b^{\rho}_{ij}J^R_{ij}J^L_{ij}-b^{\sigma}_{ij}{\bf J}^R_{ij}\cdot {\bf J}^L_{ij}\,;\cr\cr
H_{F} &=& \sum_{i\neq j=1,2}
f^{\rho}_{ij}J^R_{ii}J^L_{jj}-f^{\sigma}_{ij}{\bf J}^R_{ii}\cdot {\bf
J}^L_{jj}\,. 
\end{eqnarray}
Here $f$ and $b$ are the forward and backward scattering amplitudes. 
From hermiticity and parity, we have $b_{12}=b_{21}$ and
$f_{12}=f_{21}$.  Furthermore, in the same spirit of taking the Fermi
velocities of the two bands to be the same, $b_{11}=b_{22}$.  Thus we
obtain six independent amplitudes.  As we are working away from
half-filling in the chains, we do not consider Umklapp interactions.

To facilitate the analysis of these interactions, we invoke a change
to variables.  We begin by bosonizing the $c$'s:
\begin{equation}\label{eIIvii}
c_{Pj\alpha} = \kappa_{j\alpha} e^{i\phi_{Pj\alpha}},
~~~~ {\rm P = +,- = R,L}\,.
\end{equation}
Here $\kappa_{j\alpha}$ are Klein factors satisfying
\begin{equation}\label{eIIviii}
\{\kappa_{j\alpha},\kappa_{i\beta}\} = 2 \delta_{ij}\delta_{\alpha\beta}\,.
\end{equation}
In terms of these four Bose fields, four new Bose fields are defined
(effectively separating charge and spin):
\begin{eqnarray}\label{eIIix}
\phi_{P1} &=& \frac{1}{2}(\phi_{P1\uparrow} + \phi_{P1\downarrow} +
\phi_{P2\uparrow} + \phi_{P2\downarrow})\, ;\cr
\phi_{P2} &=& \frac{1}{2}(\phi_{P1\uparrow} - \phi_{P1\downarrow} +
\phi_{P2\uparrow} - \phi_{P2\downarrow})\, ;\cr
\phi_{P3} &=& \frac{1}{2}(\phi_{P1\uparrow} - \phi_{P1\downarrow} -
\phi_{P2\uparrow} + \phi_{P2\downarrow})\, ;\cr
\phi_{P4} &=& \frac{P}{2}(\phi_{P1\uparrow} + \phi_{P1\downarrow} -
\phi_{P2\uparrow} - \phi_{P2\downarrow}) .
\end{eqnarray}
Note that $\phi_{P4}$ has a relative sign between the right and left movers.
This sign effectively masks the underlying $SO(6)$ symmetry of the original 
Hamiltonian.  The first and second bosons describe charge and spin fluctuations 
respectively.   The latter two bosons, $\phi_3$ and $\phi_4$, are associated with
less transparent quantum numbers (as we discuss shortly).

From these chiral bosons, one can define pairs of conjugate bosons in
the standard fashion
\begin{eqnarray}\label{eIIx}
\varphi_{i} &=& \phi_{Ri}+\phi_{Li}\,;\cr
\theta_{i} &=& \phi_{Ri}-\phi_{Li}\,,
\end{eqnarray}
which obey the commutation relations,
\begin{equation}\label{eIIxi}
[\varphi (x),\theta (x')] = -i4\pi\Theta (x'-x)\,.
\end{equation}
where $\Theta (x-x')$ is the Heaviside step function.

In terms of these variables the free part of the Hamiltonian can be written
\begin{equation}\label{eIIxii}
H_0 = \frac{v_F}{8\pi} \sum_a \big\{ (\partial_x\theta_a)^2+(\partial_x\varphi_a)^2\big\}\,.
\end{equation}
The momentum conserving interactions can be written as
\begin{eqnarray}\label{eIIxiii}
H_B + H_F &=& \frac{1}{16\pi^2}\sum^4_{a=1} A_a\big\{ (\partial_x\theta_a)^2 - 
(\partial_x\varphi_a)^2\big\}\cr\cr
&& \hskip -.75 in - 2b^\sigma_{12}\cos(\theta_4)\cos(\theta_2) 
- (b^\sigma_{12}-4b^\rho_{12})\cos(\varphi_3)\cos(\theta_4)\cr\cr
&& \hskip -.75in + \cos(\theta_2)(2b^\sigma_{11}\cos(\theta_3) +2f^\sigma_{12}\cos(\varphi_3))\cr\cr
&& \hskip -.75in - \cos(\theta_4)(b^\sigma_{12}+4b^\rho_{12})\cos(\theta_3),
\end{eqnarray}
where the coefficients $A_a$ are equal to $A_{1/4}= \pm 2(b^\rho_{11} \pm f^\rho_{12})$ and 
$A_{2/3} = -(b^\sigma_{11}\pm f^\sigma_{12})/2$.

For generically repulsive scattering amplitudes, the various couplings $f$ and $b$ 
flow to fixed ratios under the RG \cite{Fabr93,schulz,lin}:
\begin{eqnarray}\nonumber
b^\rho_{12} &=& \frac{1}{4}b^\sigma_{12} = f^\rho_{12} =
-\frac{1}{4}b^\sigma_{11} = g > 0;\cr 
f^\sigma_{12} &=& b^\rho_{11} = 0.
\end{eqnarray}
\begin{eqnarray}\label{eIIxiv}
b^\rho_{12} &=& \frac{1}{4}b^\sigma_{12} = 2f^\rho_{12} =
-2 b^\rho_{11} = -\frac{1}{4}b^\sigma_{11} = g > 0;\cr 
f^\sigma_{12} &=&  = 0.
\end{eqnarray}
With these values, the interaction Hamiltonian dramatically simplifies to
\begin{eqnarray}\label{eIIxv}
H_{\rm int} &=& H_B + H_F\cr
&=&  -\frac{g}{2\pi^2}\sum^4_{a=2} \partial_x \phi_{Ra}\partial_x\phi_{La}\cr
&& -4g\sum^4_{a\neq b=2}\cos (\theta_a)\cos (\theta_b)\,.
\end{eqnarray}
This Hamiltonian leads to gapped behavior for bosons $\phi_i$,
$i=2,3,4$.  As we show below by a fermionization, the model governing
the gapped behavior is the SO(6) Gross Neveu model. On the other hand
we see that the interaction Hamiltonian does not involve the boson
governing total charge, $\theta_1=\theta_{\rho+}$.  This boson remains
gapless and so takes the form of a Luttinger liquid.  If we were
instead to allow Umklapp interactions (that is if the two chains were
not doped away from half-filling), this boson would be similarly
gapped and the governing model would be SO(8) Gross Neveu.

With the interaction Hamiltonian above, the theory describing the
boson, $\theta_{\rho_+}$, is  a $K=1$ Luttinger liquid.  We will
however consider deviations in $K$ away from 1. Such deviations might
occur either if the RG flow terminated before reaching the above set
of fixed ratios or if differences in the Fermi velocity of the two
bands were taken into account \cite{Fabr93}. Using the case of
half-filled chains (equivalent to SO(8) Gross-Neveu) as a starting
point, it was argued that generically $K < 1$ when doping was
introduced \cite{konik}.

We now recast this Hamiltonian explicitly in the form of the SO(6)
Gross-Neveu model. We refermionize the bosons $\phi_{Pa}$,
$a=1,\cdots,4$ via 
\begin{eqnarray}\label{eIIxvi}
\Psi_{Pa} &=& \kappa_a e^{i\phi_{Pa}}\,, ~~~~ {\rm a = 1,\cdots,3}\,;\nonumber\\[1mm]
\Psi_{P4} &=& P\kappa_4 e^{i\phi_{P4}}\,,\end{eqnarray}
where the Klein factors are given by
\begin{equation}\label{eIIxvii}
\kappa_1 = \kappa_{2\uparrow}, ~~~ \kappa_2 = \kappa_{1\uparrow}, ~~~
\kappa_3 = \kappa_{1\downarrow}, ~~~ \kappa_4 = \kappa_{2\downarrow}\,.
\end{equation}
We then find the free Hamiltonian can be written as
\begin{equation}\label{eIIxviii}
H_0 = \int dx \sum_a 
\bigl [
\Psi^\dagger_{Ra}i\partial_x \Psi_{Ra} -
\Psi^\dagger_{La} i\partial_x \Psi_{La} 
\bigr ]\,,
\end{equation}
where the Fermi velocity, $v_F$, has been set to 1, while
the interaction Hamiltonian can be written as
\begin{equation}\label{eIIxix}
H_{int} =
-2g \big[\sum^4_{a=2} 
(i\Psi^\dagger_{La}\Psi_{Ra} - i\Psi^\dagger_{Ra}\Psi_{La})\big]^2 .
\end{equation}
This is precisely the interaction piece of the SO(6) Gross-Neveu
model. It will sometimes prove convenient to recast the theory in
terms of Majorana fermions, $\eta_{Pa}$.  In terms of the Dirac
fermions,  $\Psi_{Pa}$,
they are given
by
\begin{equation}\label{eIIxx}
\Psi_{Pa} = \frac{1}{\sqrt{2}}(\eta_{2a-2,P} +i \eta_{2a-3,P})\,,
\qquad (a=2,3,4)\,.
\end{equation}
In this basis, $H_{int}$ can be recast as
\begin{equation}\label{eIIxxi}
H_{int} = 2g G^{ab}_R G^{ab}_L\,, \qquad (a\neq b = 1,...,6)\,,
\end{equation}
where $G^{ab}_P = i \eta_{Pa}\eta_{Pb}$ is one of the 15 $SO(6)$
Gross--Neveu currents.

\subsection{Excitation spectrum of SO(6) Gross-Neveu model}
The SO(6) Gross Neveu has a total of 14 different types of excitations
\cite{thun}. 
These excitations are organized into two four dimensional spinor
representations and one six dimensional vector representation.  
The latter corresponds to the Majorana fermions introduced in the
previous section. We denote the fermionic creation operators
for the vector representation by $A^\dagger_a$, $a=1,\ldots,6$. The
excitations corresponding to the spinor representations are kinks and
represent interpolations of the fields $\theta_b$, $b=1,2,3$ between
minima of the cosine potentials in Eqn.(\ref{eIIxv}). 
There are eight different types of kinks, which we will denote by
$A^\dagger_\a$.  Here $\a$ is a ``multi-index'' of the form $\a = (\pm
1/2,\pm 1/2, \pm 1/2)$.  The eight kinks divide themselves into two
different four dimensional irreducible representations, something
discussed in greater detail in Section \ref{sec:FF}. 

$SO(6)$ is a rank 3 algebra and the eigenvalues $N_i$, $i = 1,\ldots
,3$ of the three Cartan generators can be used as quantum numbers for
the excitations. The combination of Majorana fermions
\begin{equation}\label{eIIxxii}
\eta_{2a-1} \pm i\eta_{2a},
\end{equation}
generates excitations with quantum number $N_a=\pm  1$, $N_b=0, b \neq  a$. 
The quantum numbers carried by the kinks $A^\dagger_\alpha$ are
directly encoded in $\alpha$. 
If $\a = (\alpha_1,\alpha_2,\alpha_3)$, $\alpha_i=\pm 1/2$, then
$A^\dagger_\a$ creates a kink with quantum numbers, $N_i = \alpha_i$.  

The quantum numbers, $N_i$, are related to the physical quantum
numbers of the system. In particular these are the z-component of
spin, $S_z$, the difference in z-component of spin between the two
bands, $S_{12}$, and the ``relative band chirality'', $P_{12}$,
defined as $P_{12} = N_{R1} - N_{L1} - N_{R2} + N_{L2}$, where
$N_{Pj}$ is the number electrons in band $j$ with chirality $P$.
In Ref.  \onlinecite{lin}, it was found:
\begin{eqnarray}\label{eIIxxiii}
(N_1=1,0,0) &\leftrightarrow& (S_z = 1, S_{12} = 0, P_{12} = 0);\cr
\cr
(0,N_2=1,0) &\leftrightarrow& (S_z = 0, S_{12} = 1, P_{12} = 0);\cr
\cr
(0,0,N_3=1) &\leftrightarrow& (S_z = 0, S_{12} = 0, P_{12} = 2).
\end{eqnarray}
We note that these three quantum numbers are independent of the total
charge, $Q$. Charge excitations all occur in the gapless sector
governed by the boson $\theta_1$. 

We can see that the vector representation of excitations corresponds 
to states of two electrons (particle-holes pairs) in the original 
formulation.
For example, the fermion $\eta_1 \pm i \eta_2$ carries spin, $S_z =
\pm 1$ and no charge and so is a magnon excitation. 
The spinor representations (the kinks) in turn are closely related to
single particle excitations as their quantum numbers are combinations
of $N_i/2$ (in fact they are single particle excitations modulo their
charge component). 

\subsection{Identification of Physical Operators in terms of SO(6)
Gross-Neveu Fields}

In this section we provide a dictionary between the
$SO(6)$ Gross-Neveu model
and the original fermions of the Hubbard ladders.  
As discussed previously, the kinks are closely related to single particle 
excitations and so the original
electron operators.  There are 16 kinks in total
(counting both left and right movers) and sixteen electron
operators, the $c$'s and $c^{\dagger}$'s (four for each of the 
four Fermi points).  
In terms of the four bosons $\phi_{R/L1} = \phi_{R/L\rho+}$ and
$\phi_{R/Li}$, $i=2,3,4$, the electron operators are given by $c$'s
are as follows: 
\begin{eqnarray}\label{eIIxxiv}\cr
c_{Rb\uparrow} &\sim& e^{\frac{i}{2}\phi_{R1}}
e^{\frac{i}{2}(\phi_{R2} - \phi_{R3} - \phi_{R4})};
\cr
c_{Rb\downarrow} &\sim& e^{\frac{i}{2}\phi_{R1}}
e^{\frac{i}{2}(-\phi_{R2} + \phi_{R3} - \phi_{R4})};
\cr
c_{Rab\uparrow} &\sim& e^{\frac{i}{2}\phi_{R1}}
e^{\frac{i}{2}(\phi_{R2} + \phi_{R3} + \phi_{R4})};
\cr
c_{Rab\downarrow} &\sim& e^{\frac{i}{2}\phi_{R1}}
e^{\frac{i}{2}(-\phi_{R2} - \phi_{R3} + \phi_{R4})};\cr\cr
&& {\rm (even~chirality)}\cr
\cr\cr
c_{Lb\downarrow} &\sim& e^{\frac{i}{2}\phi_{L1}}
e^{\frac{i}{2}(-\phi_{L2} + \phi_{L3} + \phi_{L4})};\cr
c_{Lb\uparrow} &\sim& e^{\frac{i}{2}\phi_{L1}}
e^{\frac{i}{2}(\phi_{L2} - \phi_{L3} + \phi_{L4})};\cr
c_{Lab\downarrow} &\sim& e^{\frac{i}{2}\phi_{L1}}
e^{\frac{i}{2}(-\phi_{L2} - \phi_{L3} - \phi_{L4})};\cr
c_{Lab\uparrow} &\sim&e^{\frac{i}{2}\phi_{L1}}
e^{\frac{i}{2}(\phi_{L2} + \phi_{L3} - \phi_{L4})};\cr\cr\cr
&& {\rm (odd~chirality)} .
\end{eqnarray}
The lattice fermions are all given by a charge piece,
$e^{i\phi_{R1}}$, multiplied by a field creating an SO(6) kink,
$e^{\frac{i}{2}(\pm\phi_{R2} \pm \phi_{R3} \pm \phi_{R4})}$.  
The first four fermions involve even chirality kink fields
while the latter involve odd chirality kinks.  We denote the group
representation of the even chirality kinks by ${\bf 4}$.  
Electrons transforming as even chirality kinks are those above 
in addition to
$(c^\dagger_{Lb\downarrow},c^\dagger_{Lb\uparrow},c^\dagger_{Lab\downarrow},c^\dagger_{Lab\uparrow})$,
that is the Hermitean conjugate of the above odd kinks.
The odd chirality kinks transform under the ${\bf \bar{4}}$ representation.
Again, electrons transforming as such are the above as well as
the Hermitean conjugate of the above even kinks,
$(c^\dagger_{Rb\uparrow},c^\dagger_{Rb\downarrow},c^\dagger_{Rab\uparrow},c^\dagger_{Rab\downarrow})$
The lack of symmetry between left and right movers reflects
the parity sign in the definition of $\phi_{P4}$ in Eqn.(\ref{eIIix}).

To construct the spin-spin correlation function we need to consider
the description of fermion bilinears in the language of the SO(6)
Gross-Neveu model. Fermion bilinears can be of two types: one
involving fermions associated with kinks of different chiralities and
one involving fermions associated with kinks of the same chiralities. 
In the first case we want to consider fermion bilinears transforming under 
${{\bf 4 \otimes \bar{4}} = {\bf 1 \oplus 15}}$.  We will want to focus on 
the fifteen dimensional representation, $\bf{15}$, which is
no more than the set of SO(6) currents.  In the second case we want to consider 
bilinears behaving as ${{\bf 4 \otimes 4} = {\bf 6 \oplus 10}}$ 
(and similarly for ${\bf \bar{4} \bf\otimes \bf\bar{4}}$).  The six dimensional representation,
${\bf 6}$, is associated with particle-hole excitations carrying unit quantum numbers
such as the magnon.  The particle-hole excitations 
corresponding to the $\bf{6}$ are formed by taking symmetric combinations
excluding diagonal terms of the even electrons in Eqn.(\ref{eIIxxiv}):
\begin{equation}\label{eIIxxv}
c_{Rb\uparrow}c^\dagger_{Lab\downarrow} + c_{Rab\uparrow}c^\dagger_{Lb\downarrow}.
\end{equation}
Such {\it symmetric} fermion combinations behave {\it
  antisymmetrically} 
under $SO(6)$.  As will be discussed, this correspondence between the
${\bf 6}$ and symmetric combinations of fermion bilinears is most
easily seen upon bosonization of the fermionic expressions. We will denote
the fields creating these two particle excitations as $\psi^a$, $a=1,\ldots,6$.  
On the other hand, the ten dimensional representation, ${\bf 10}$,
arises from antisymmetric combinations of 
particle-hole excitations,
$$
c_{Rb\uparrow}c^\dagger_{Lab\downarrow} - c_{Rab\uparrow}c^\dagger_{Lb\downarrow},
$$
together with diagonal combinations of fermions
$$
c_{Rb\uparrow}c^\dagger_{Lb\downarrow}.
$$
We will denote the fields creating these excitations by $\psi^{[abc]_A}$ where
$a,b,c=1,\ldots 6$ are anti-symmetrized.  The odd representation for
these fields will facilitate the computation of their associated matrix elements.\cite{note}.

\section{Spin-Spin Correlation Functions of Doped Ladders}
\label{sec:corr}

In this section we utilize the form factors we calculate in Section IV to
determine the low-energy behavior of the dynamical structure factor
of the doped two-leg ladder.  For readers uninterested in the details
of the form factor computations, we have summarized the results in Section IV H.

\subsection{Preliminaries}

In order to compute the spin response, we need to relate the
lattice spin-spin correlation function to correlation functions in the
SO(6) Gross-Neveu model.
The dynamical magnetic susceptibilities of the doped ladder, $\chi
(\omega, q_x,q_y)$, is given in terms of two different spin
correlators: one along a leg of the ladder  
\begin{eqnarray}\label{eIIIi}
\chi_{11} (x,t) &\equiv& \langle {\bf S}_1(x,t)\cdot 
{\bf S}_1(0,0)\rangle_{\rm ret}\nn
&=&\langle {\bf S}_2(x,t)\cdot {\bf S}_2(0,0)\rangle_{\rm
  ret}\equiv\chi_{22} (x,t), 
\end{eqnarray}
and one involving cross leg correlations
\begin{eqnarray}\label{eIIIii}
\chi_{12}(x,t) &\equiv&  \langle {\bf S}_1(x,t)\cdot{\bf
  S}_2(0,0)\rangle_{\rm ret}\nn
&=& \langle {\bf S}_2(x,t)\cdot{\bf   S}_1(0,0)\rangle_{\rm ret}
  \equiv \chi_{21}(x,t). 
\end{eqnarray}
Here $\langle \rangle_{\rm ret}$ denotes the retarded correlation
function.  The dynamical susceptibility of the ladder is then given in
terms of the susceptibilities along the legs and rungs as
\begin{eqnarray}\label{eIIIiii}
\chi (\omega, {\bf k}) &=& \chi_{11}(\omega ,k_x) + \chi_{22}(\omega
,k_x)\nn 
&&+ 2\chi_{12}(\omega ,k_x)\cos(k_y).
\end{eqnarray}
To evaluate these correlators we first express $S^{\pm}$ in terms of
the bonding/anti-bonding fermions ($a=1,2$)
\begin{eqnarray}\label{eIIIiv}
2S^+_a &=& c^\dagger_{\uparrow b} c_{\downarrow b}
+c^\dagger_{\uparrow ab} c_{\downarrow ab}
-(-1)^a \bigl(c^\dagger_{\uparrow b}
c_{\downarrow ab} + c^\dagger_{\uparrow ab} c_{\downarrow b}\bigr)\ ,\nn
2S^-_a &=&
c^\dagger_{\downarrow b} c_{\uparrow b} +
c^\dagger_{\downarrow ab} c_{\uparrow ab}
-(-1)^a\big( c^\dagger_{\downarrow b}
c_{\uparrow ab} +c^\dagger_{\downarrow ab} c_{\uparrow b}\big)\ ,\nn
\end{eqnarray}
The correlators we need can then be written (using SU(2) invariance) as
\begin{eqnarray}\label{eIIIv}
-\chi_{ab}(\omega,k) & = & \int_{-\infty}^\infty d\tau dx e^{i(-i\omega+\epsilon)
 \tau-ikx} \cr
&& \hskip .5in \times \langle T{\bf S}_a(x,\tau)\cdot {\bf S}_b(0)\rangle\nonumber\\
&\equiv&\langle {\bf  S}_a\cdot {\bf S}_b\rangle (\omega, k).
\end{eqnarray}
The dynamical spin-spin correlation functions are expressed as
\begin{eqnarray}\label{eIIIvi}
\frac{2}{3}\langle {\bf S}_1\cdot {\bf S}_1\rangle (\omega, k) &=&
\langle S^+_1 S^-_1 \rangle (\omega, k) = \frac{1}{4}(I_1 + I_2 + I_3);\cr
\frac{2}{3}\langle {\bf S}_1\cdot {\bf S}_2 \rangle (\omega, k) &=&
\langle S^+_1 S^-_2 \rangle (\omega, k) = \frac{1}{4}(I_1 + I_2 - I_3),\cr &&
\end{eqnarray}
where the $I_i$ are given by
\begin{widetext}
\begin{eqnarray}\label{eIIIvii}
I_1(\omega, k) &=&\int d\tau dx e^{i(-i\omega+\epsilon) \tau - i k x} \langle
T c^\dagger_{b\uparrow}(x,\tau)c_{b\downarrow}(x,\tau)
c^\dagger_{b\downarrow}(0)c_{b\uparrow}(0)\rangle,\nn
I_2(\omega, k) &=& \int d\tau dx e^{i(-i\omega+\epsilon) \tau - i k x} \langle
T c^\dagger_{ab\uparrow}(x,\tau)c_{ab\downarrow}(x,\tau)
c^\dagger_{ab\downarrow}(0)c_{ab\uparrow}(0)\rangle,\nn
I_3(\omega, k)&=&\int d\tau dx e^{i(-i\omega+\epsilon) \tau - i k x} 
\langle T\big[c^\dagger_{ab\uparrow}(x,\tau)c_{b\downarrow}(x,\tau) +
c^\dagger_{b\uparrow}(x,\tau)c_{ab\downarrow}(x,\tau)\big]
\big[c^\dagger_{b\downarrow}(0)c_{ab\uparrow}(0) +
c^\dagger_{ab\downarrow}(0)c_{b\uparrow}(0)\big]\rangle. 
\end{eqnarray}
In the next step we take the low-energy limit, in which the fermion
operators are expanded according to (\ref{eIIiii}). This gives, for
example 
\begin{eqnarray}\label{eIIIix}
c^\dagger_{b\uparrow}(x,\tau)c_{b\downarrow}(x,\tau)&\sim&
c^\dagger_{Rb\uparrow}(x,\tau)c_{Rb\downarrow}(x,\tau)
+c^\dagger_{Lb\uparrow}(x,\tau)c_{Lb\downarrow}(x,\tau)\nn
&&+e^{-2iK_{Fb}x}c^\dagger_{Rb\uparrow}(x,\tau)c_{Lb\downarrow}(x,\tau)
+e^{2iK_{Fb}x}c^\dagger_{Lb\uparrow}(x,\tau)c_{Rb\downarrow}(x,\tau)\ .
\end{eqnarray}
Substituting (\ref{eIIIix}) and its analogs into the
expressions for $I_j(\omega,k)$ we obtain sums of terms, each of
which can be calculated using a form factor expansion as sketched in
the introduction. For example in the imaginary piece of 
$I_1(\omega,k)$ we have a contribution of the form
\begin{eqnarray}\label{eIIIx}
\sum^\infty_{n=0}\sum_{\{a_n\}} 
\int \frac{d\theta_1\ldots d\theta_n}{(2\pi)^{n-2}n!}
\bigl|\langle 0|c^\dagger_{Rb\uparrow}(0)c_{Rb\downarrow}(0)
|\theta_1\ldots\theta_n\rangle_{a_1\ldots a_n}\bigr|^2
\delta\Bigl(\omega-\sum^n_{i=1}m_{a_i}\cosh\theta_i\Bigr)\
\delta\Bigl(k-\sum^n_{i=1}\frac{m_{a_i}}{v_F}\sinh\theta_i\Bigr),
\end{eqnarray}
\end{widetext}

\noindent where the first sum $\sum_n$ runs over the number of
particles in a particular term of the form factor expansion and the
second sum $\sum_{a_i}$ runs over the different possible particle types. 
The complete set of intermediate states used in the spectral
representation (\ref{eIIIix}) is defined by
\begin{equation}\label{eIIIxi}
|\theta_1\ldots\theta_n\rangle_{a_1\ldots a_n}=
A^\dagger_{a_n}(\theta_n) \cdots A^\dagger_{a_1}(\theta_1) |0\rangle\ .
\end{equation}
Carrying out the expansion (\ref{eIIiii}) for all terms in (\ref{eIIIvii})
we obtain 
\be\label{eIIIxii}
I_1 (\omega, k) = 
\begin{cases}
A_{11} J_1(\omega, k) & \text{if $k\approx 0$}\\
A_{12} J_2(\omega,k\mp 2K_{Fb}) &  \text{if $k\approx \pm 2K_{Fb}$},\\
\end{cases}
\ee
\be\label{eIIIxiii}
I_2 (\omega, k) = 
\begin{cases}
A_{21} J_1(\omega, k) & \text{if $k\approx 0$}\\
A_{22} J_2(\omega,k\mp 2K_{Fab}) &  \text{if $k\approx \pm 2K_{Fab}$},\\
\end{cases}
\ee
\bea\label{eIIIxiv}
I_3 (\omega, k) &=& 
\begin{cases}
A_{31}J_1(\omega,  k\mp K_{F-}) &\text{if $k\approx\pm K_{F-}$} \\
A_{32}J_3(\omega,  k\mp K_{F+}) &\text{if $k\approx\pm K_{F+}$}.\\
\end{cases}\nn
\eea
Here we have defined
\be
K_{F\pm}=K_{Fab}\pm K_{Fb}\ .
\ee
The $J_i$'s correspond to the lowest (one and two) particle (bound state and 
kink) contributions to the spin response.  Their explicit forms
are derived in Appendix \ref{app:spinspin}. At low energies, as
explained in the introduction, these terms provide the sole
contributions. The various $A's$ in the above are normalization constants.
To fix them, we compare the spectral weight in the spin response obtained
using our low energy field theoretic reduction to that obtained using a random
phase approximation (RPA).  While the RPA is a crude approximation, we believe
that it will provide a rough estimate for the integrated spin response characterized
by
\begin{equation}\label{eIIIxv}
\int d\omega \frac{{\rm Im} \chi(\omega , k)}{w}.
\end{equation}
Thus by equating the weight given by the two approaches, we are able
to fix the values of the $A's$.  The details may be found in Appendix \ref{app:spinspin}.

\renewcommand{\thesubfigure}{}
\renewcommand{\subfigcapskip}{0pt}
\begin{figure*}
\begin{center}
\vskip -.2in
\subfigure[$k_y=0$]{
\epsfysize=1.25\textwidth
\includegraphics[height=5cm]{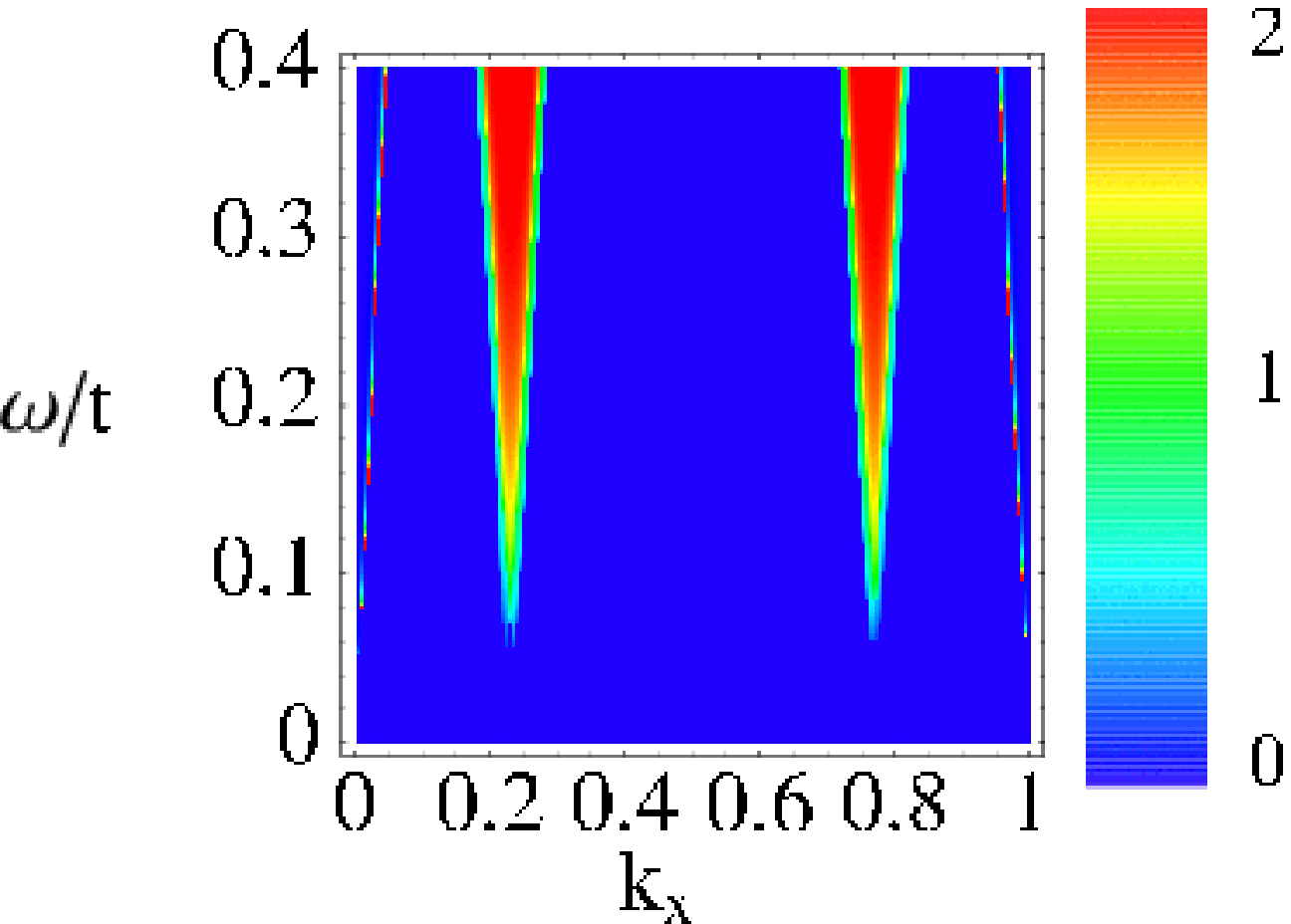}}
\subfigure[$k_y=\pi$]{
\epsfysize=1.25\textwidth
\includegraphics[height=5cm]{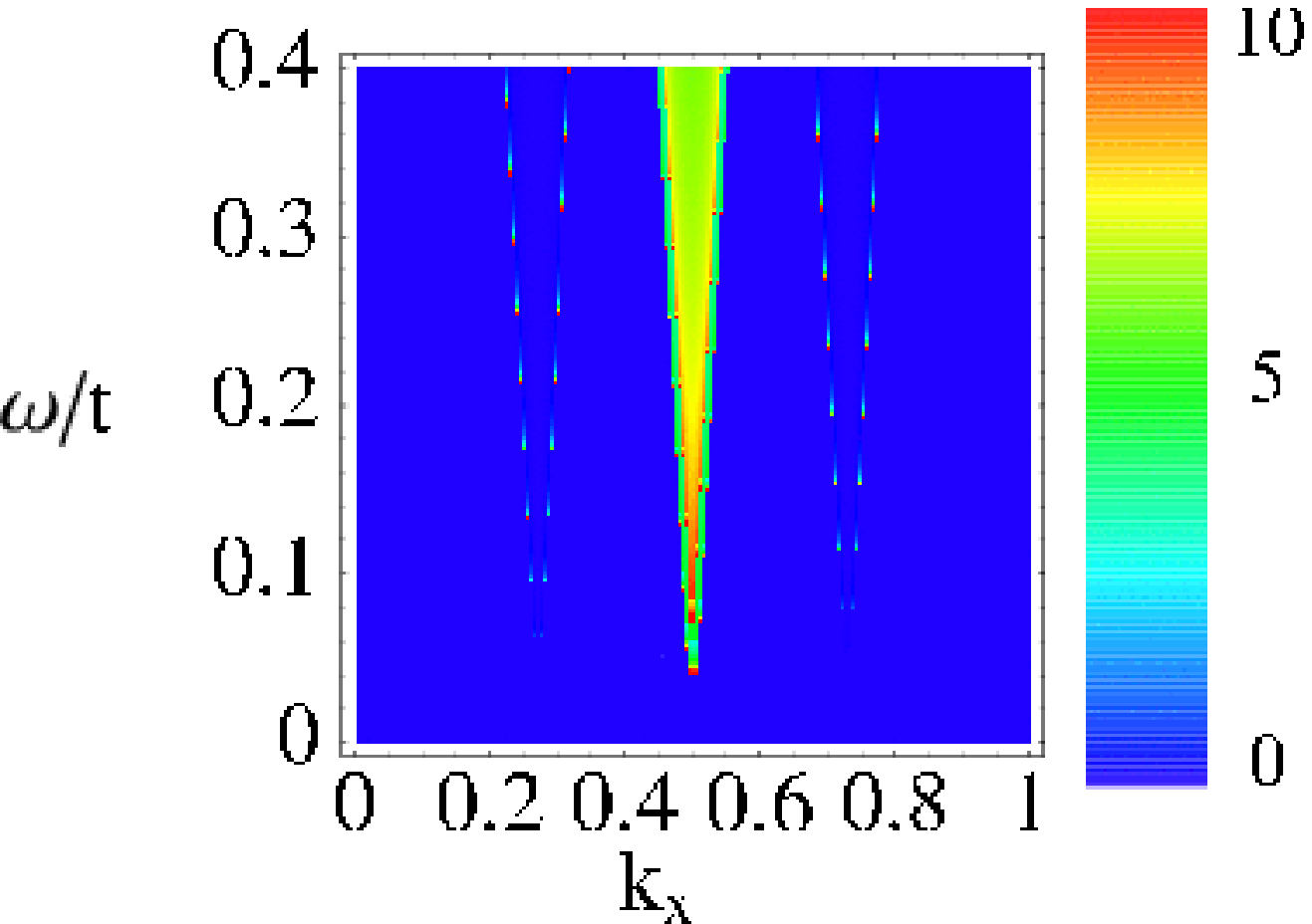}}
\end{center}
\vskip -.2in
\centerline{\hbox{SO(6) GN}}
\vskip .4in
\begin{center}
\subfigure[$k_y=0$]{
\epsfysize=1.25\textwidth
\includegraphics[height=5cm]{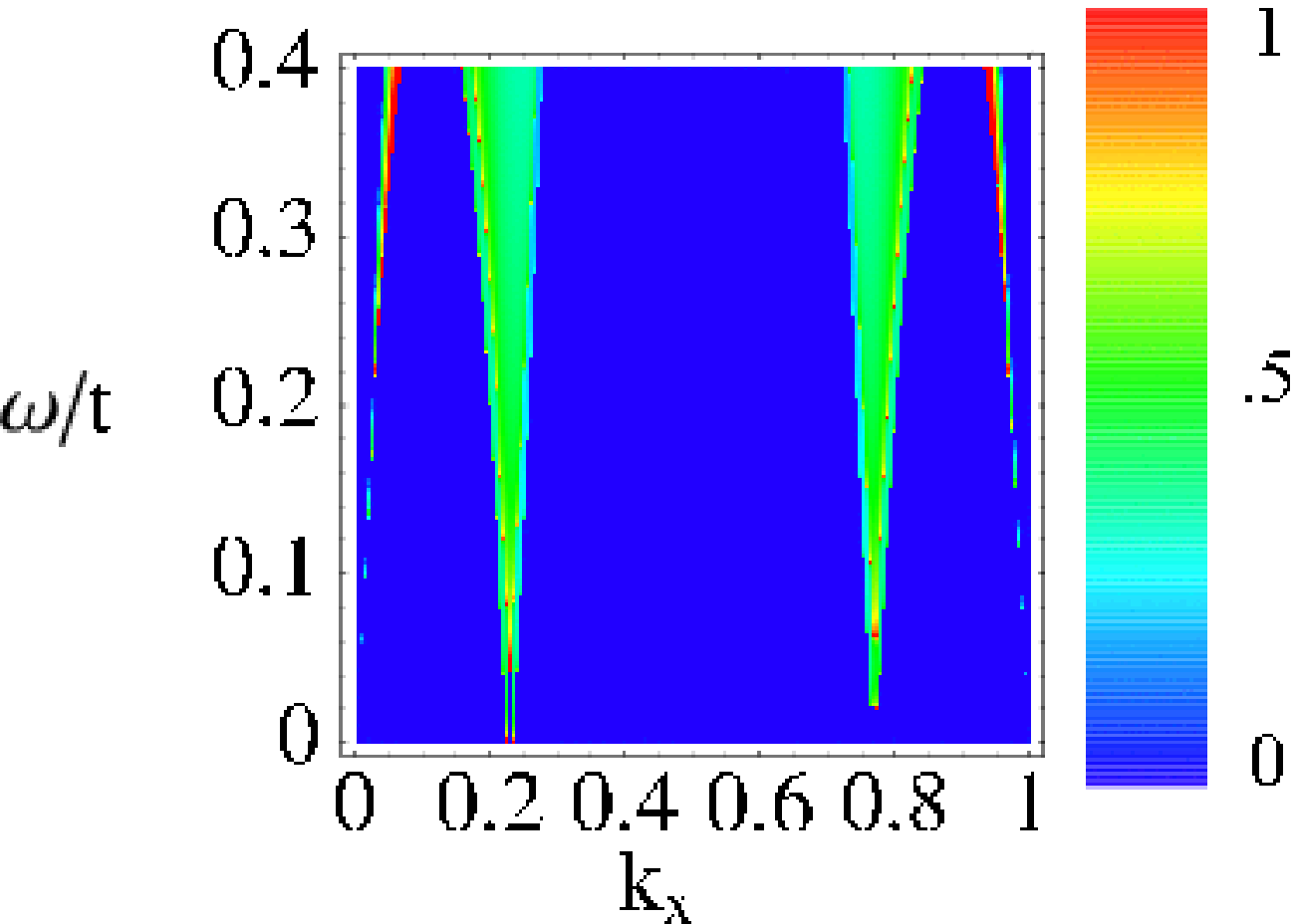}}
\subfigure[$k_y=\pi$]{
\epsfysize=1.25\textwidth
\includegraphics[height=5cm]{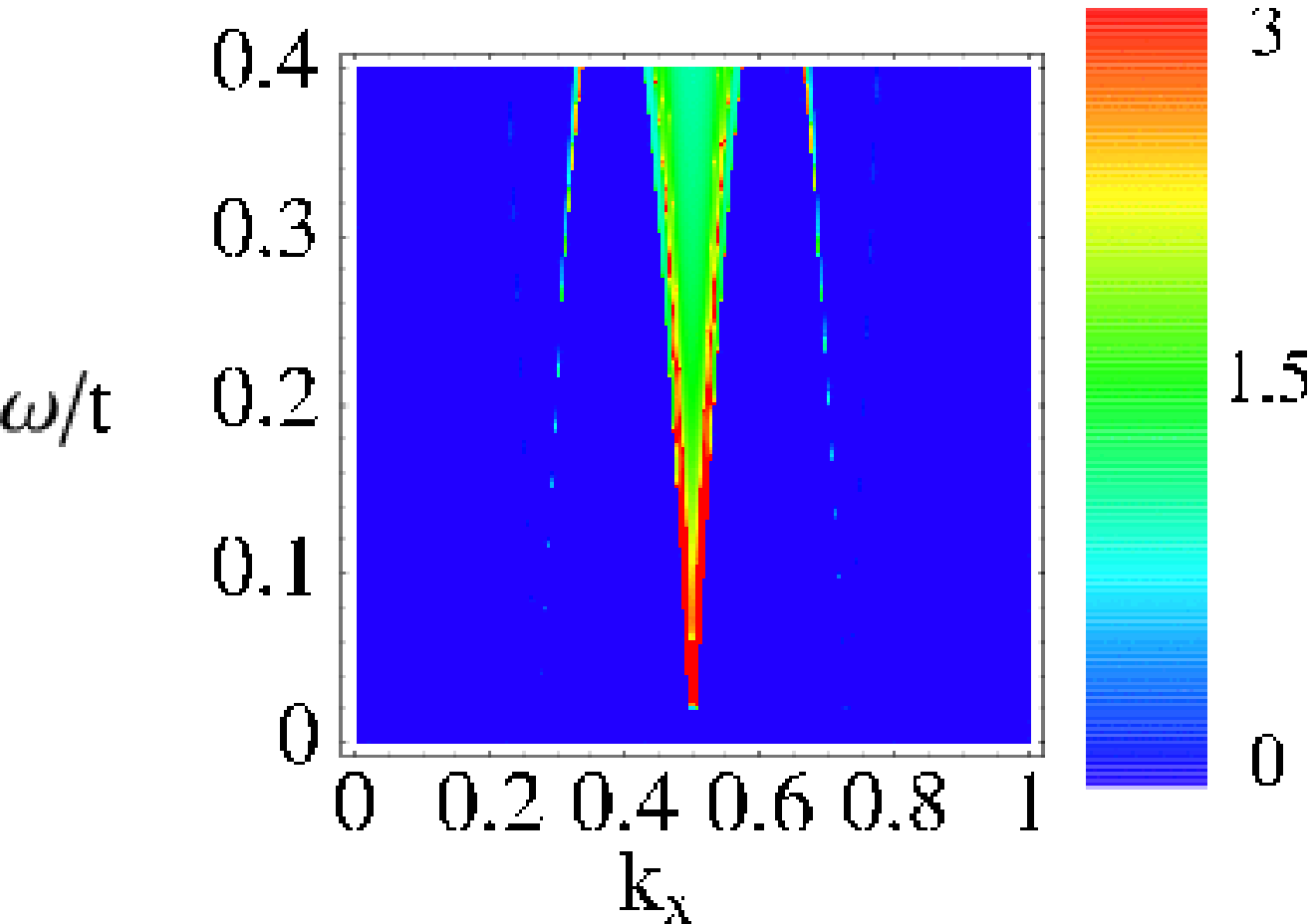}}
\vskip 0in
\end{center}
\vskip -.2in
\centerline{\hbox{RPA}}
\vskip .4in
\begin{center}
\subfigure[$k_y=0$]{
\epsfysize=1.25\textwidth
\includegraphics[height=5cm]{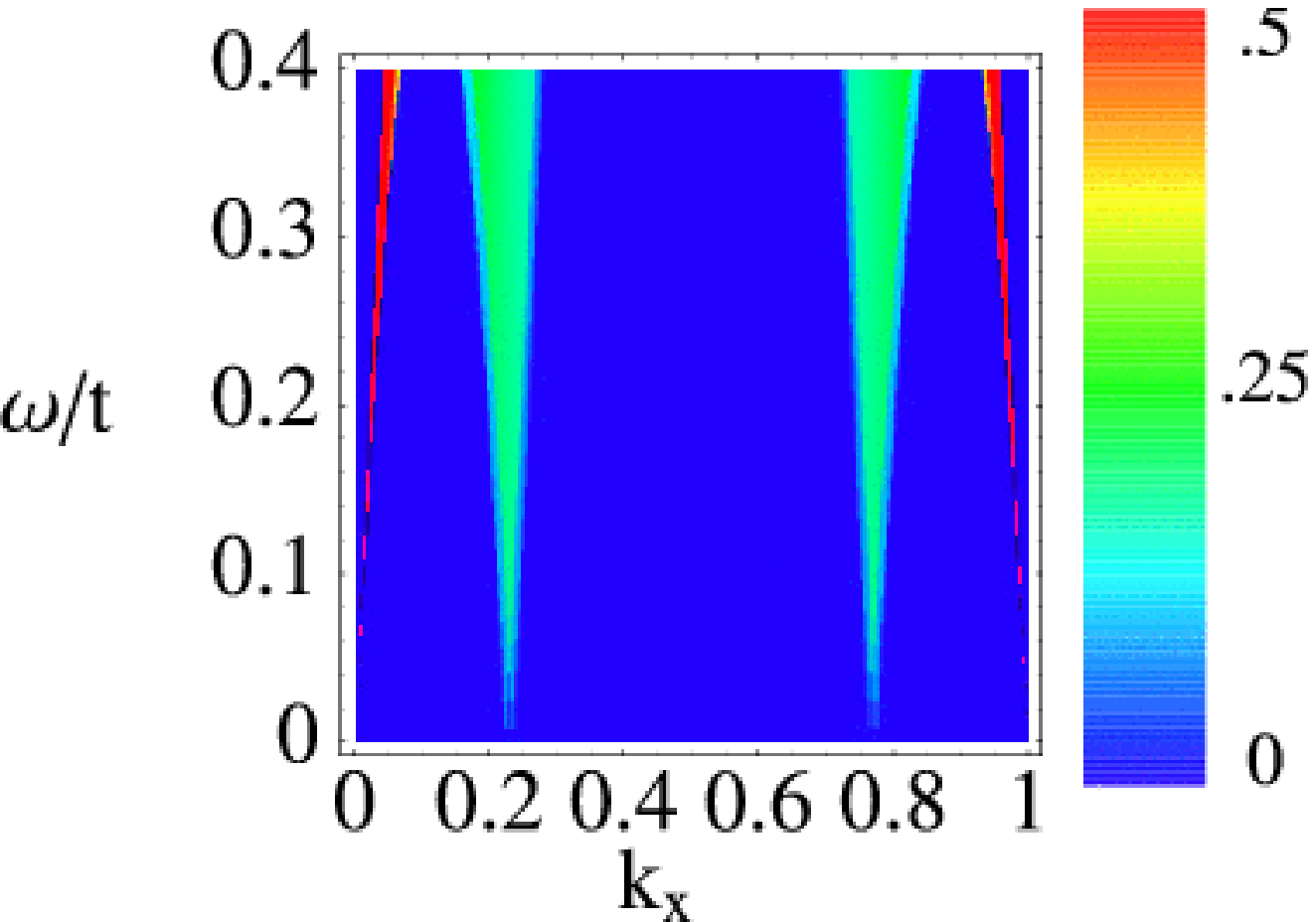}}
\subfigure[$k_y=pi$]{
\epsfysize=1.25\textwidth
\includegraphics[height=5cm]{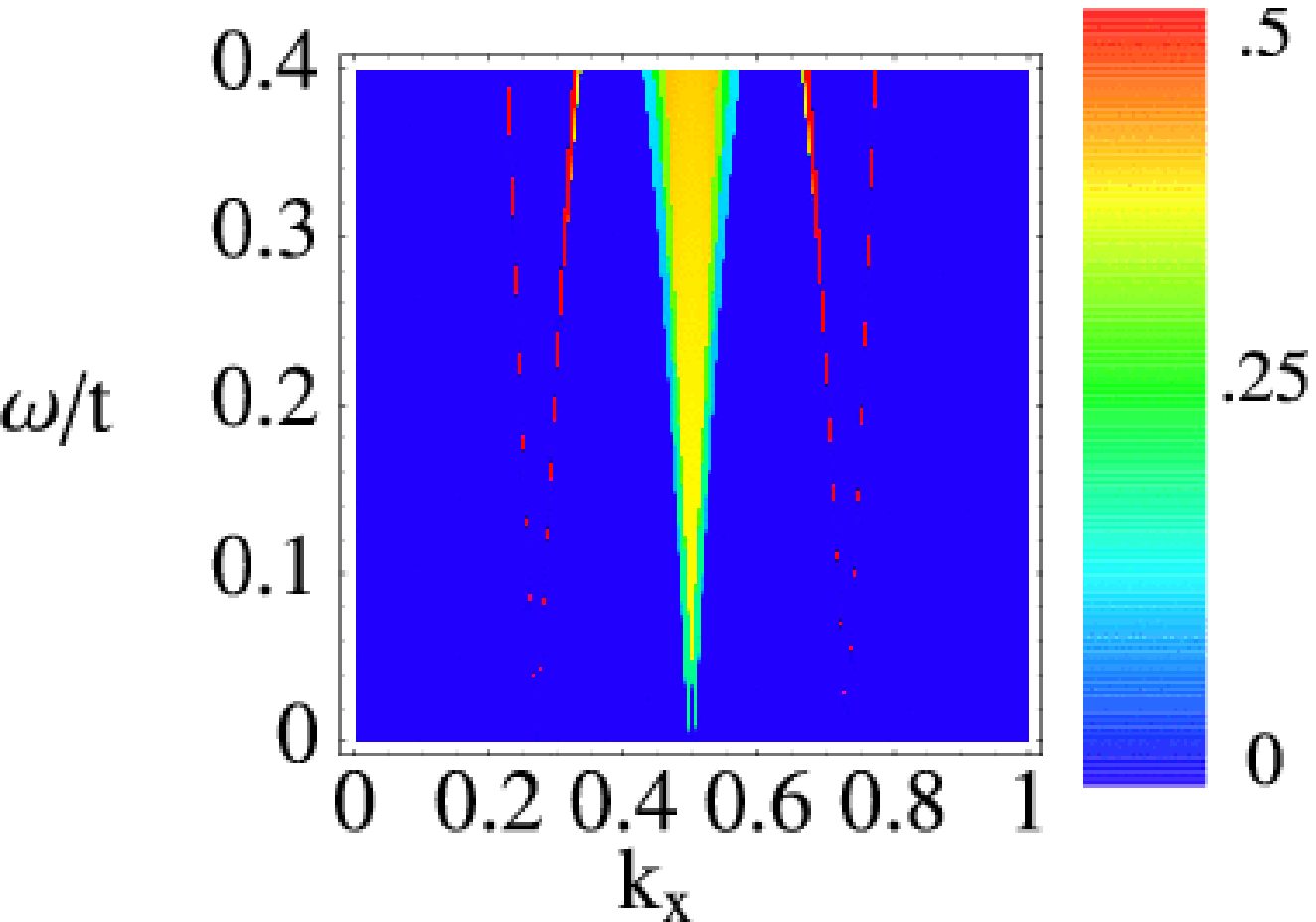}}
\end{center}
\vskip -.2in
\centerline{\hbox{Non-interacting}}
\vskip .15in
\caption{Plots of the spin response function computed in three different fashions: 
using SO(6) Gross-Neveu as a low-energy effective theory; RPA; and taking the ladders as non-interacting.
The left-hand panels give the response at $k_y=0$ while the right-hand panels correspond to $k_y=\pi$.
In making these plots, we have assumed 1\% doping and $U=t$.  We have plotted $k_x$ in 
rationalized units of $2\pi$.}
\end{figure*}

\subsection{Spin Response at $U=t$}

In this section we discuss the spin response at the relatively moderate value of $U=t$.  
At the top of Figure 1, we present an intensity plot from the field
theoretic computation for the spin response function at 1\% doping,
both at $k_y=0$ (the left panel) and $k_y=\pi$ (the right panel).  The low energy spectral weight
is found near values of $k_x$ related to $K_{Fab}$ and $K_{Fb}$, 
the bonding and anti-bonding wavevectors.  
As a function of doping, $\mu$, away from half-filling, the values of $K_{Fab}$ and $K_{Fb}$ are
\begin{eqnarray}\label{eIIIxvi}
K_{Fb} &=& \cos^{-1}\bigg[-\frac{t_\perp}{2t}\cr\cr
&& \hskip .2in + \frac{1}{2}
\big(\frac{t_\perp^2}{t^2}-2\frac{t_\perp^2/t^2-\sin^2(\pi\mu)}{1+\cos(\pi\mu)}\big)^{1/2}\bigg]\cr\cr
K_{Fab} &=& \pi(1-\mu) - K_{Fb}. 
\end{eqnarray}
At $k_y=\pi$, the low energy weight
is present near $k_x= K_{Fab}+K_{Fb},K_{Fb}-K_{Fab},2\pi-K_{Fab}-K_{Fb},$ and $2\pi-K_{Fb}+K_{Fab}$.
We see that the greatest intensity
is found at $k_x=K_{Fab}+K_{Fb}$ and $k_x=2\pi-K_{Fb}-K_{Fab}$.  Exactly at half-filling, 
these wave-vectors equal $\pi$, i.e. $K_{Fab}+K_{Fb}=\pi$.  Thus as the system is doped the $(\pi,\pi )$ 
peaks splits into
two incommensurate peaks.  
However at 1\% doping, the splitting is barely resolved in Figure 1.
For $k_y = \pi$, the spectral weight at $k_x = K_{Fb}-K_{Fab}$ is considerably smaller.  

Spectral weight is also found for $k_y=0$ at $k_x = 2K_{Fab},$ $2\pi-2K_{Fab},$ 
$2K_{Fb},$ $2\pi-2K_{Fb},$ and finally at $k_x=0,2\pi$.  
As we can see from the intensity scales, it is considerably
less than that at $k_y = \pi$.  We also see that the energy at which weight is first found
is higher.  This reflects the fact that the first contribution to the spectral weight is a 
two-particle contribution
and so is at an energy a $\sqrt{2}$ times higher than that at $k_y=\pi$ 
where the initial contribution
is made by a single coherent bound state.

\renewcommand{\thesubfigure}{}
\renewcommand{\subfigcapskip}{-40pt}
\begin{figure*}
\begin{center}
\vskip -.2in
\subfigure[]{
\epsfysize=1.25\textwidth
\includegraphics[height=6.5cm]{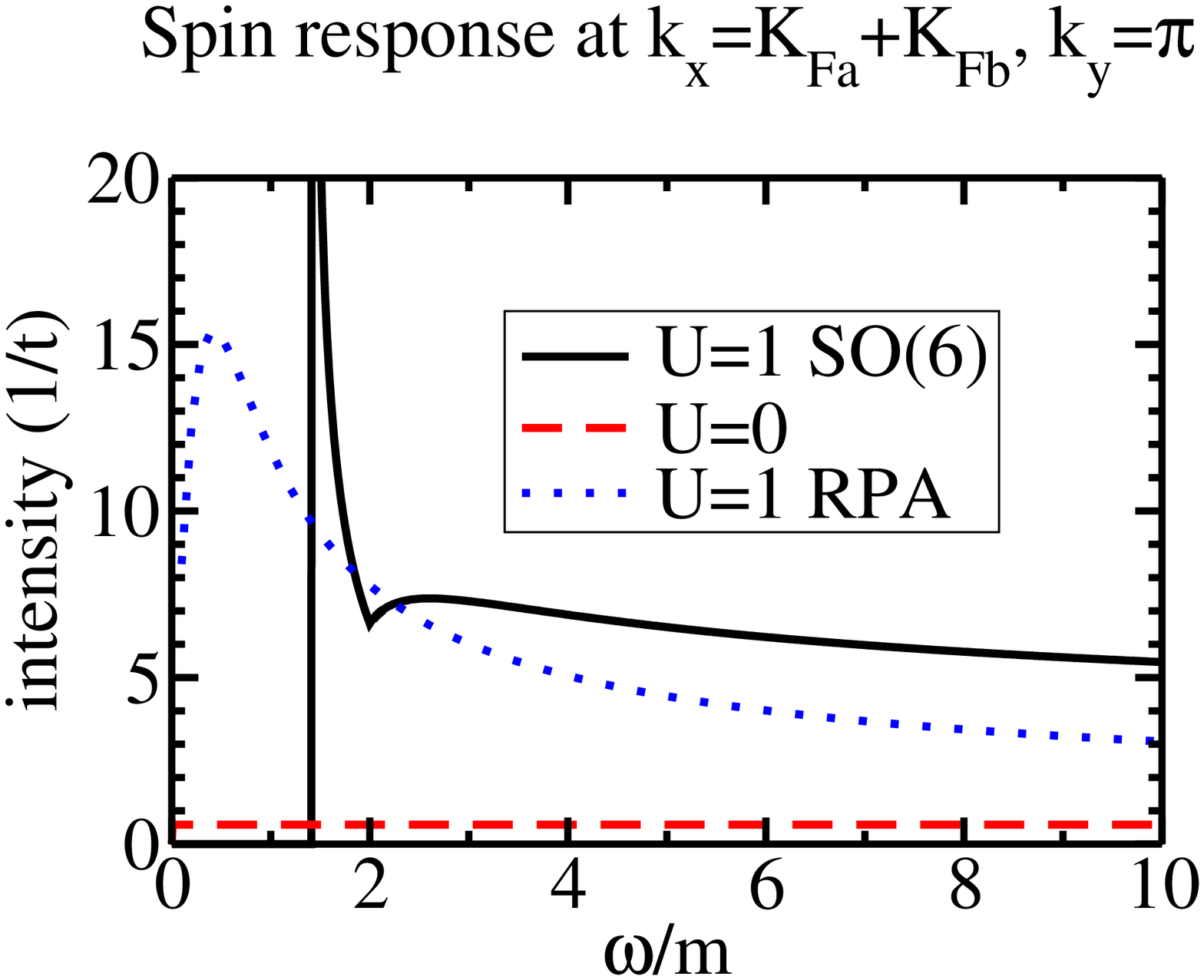}}
\epsfysize=1.25\textwidth
\subfigure[]{
\includegraphics[height=6.5cm]{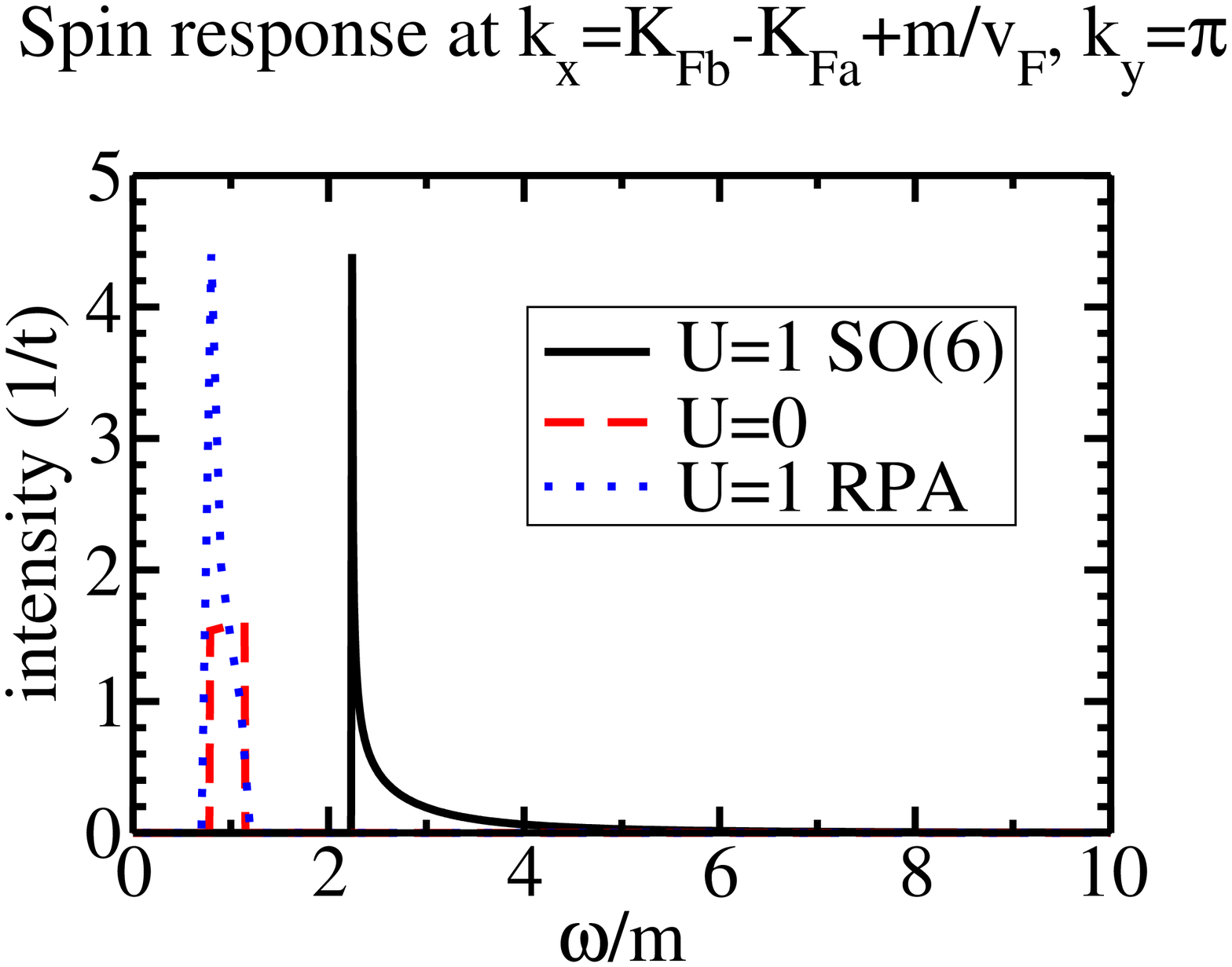}}
\vskip 0in
\end{center}
\vskip -.2in
\vskip -.5in
\centerline{\hbox{}}
\begin{center}
\subfigure[]{
\epsfysize=1.25\textwidth
\includegraphics[height=6.5cm]{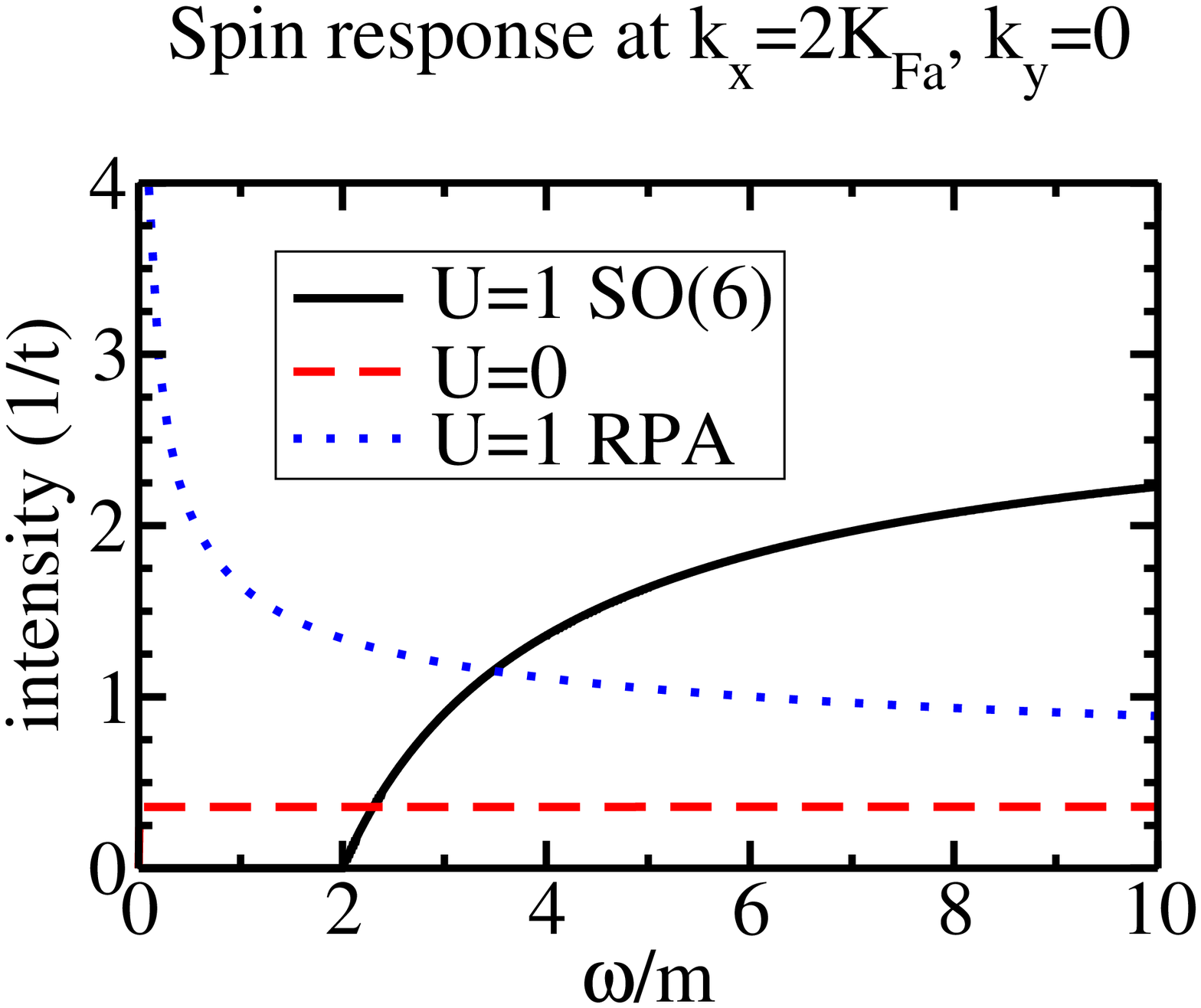}}
\subfigure[]{
\epsfysize=1.25\textwidth
\includegraphics[height=6.5cm]{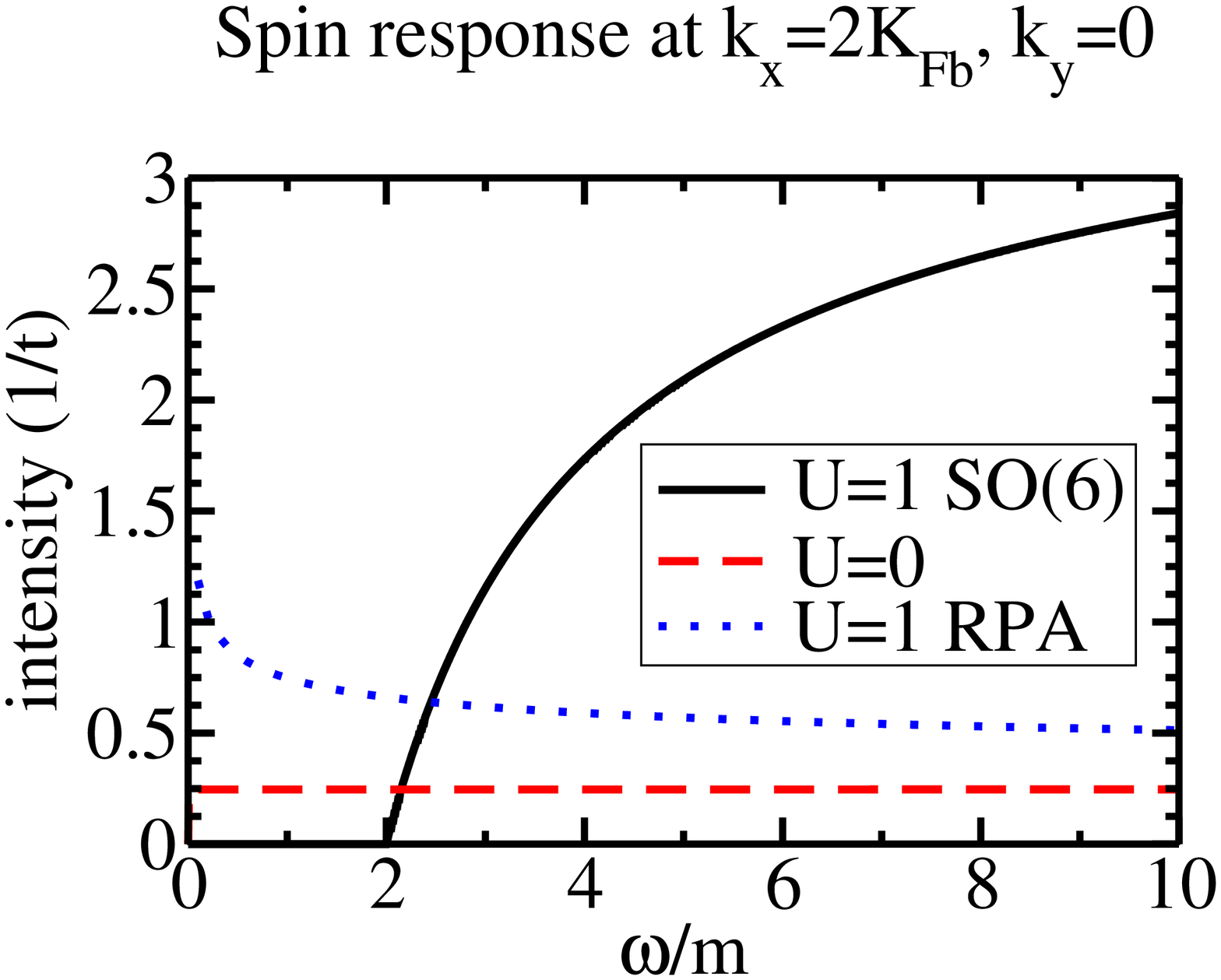}}
\vskip 0in
\end{center}
\vskip -.2in
\vskip -.5in
\centerline{\hbox{}}
\begin{center}
\subfigure[]{
\epsfysize=1.25\textwidth
\includegraphics[height=6.5cm]{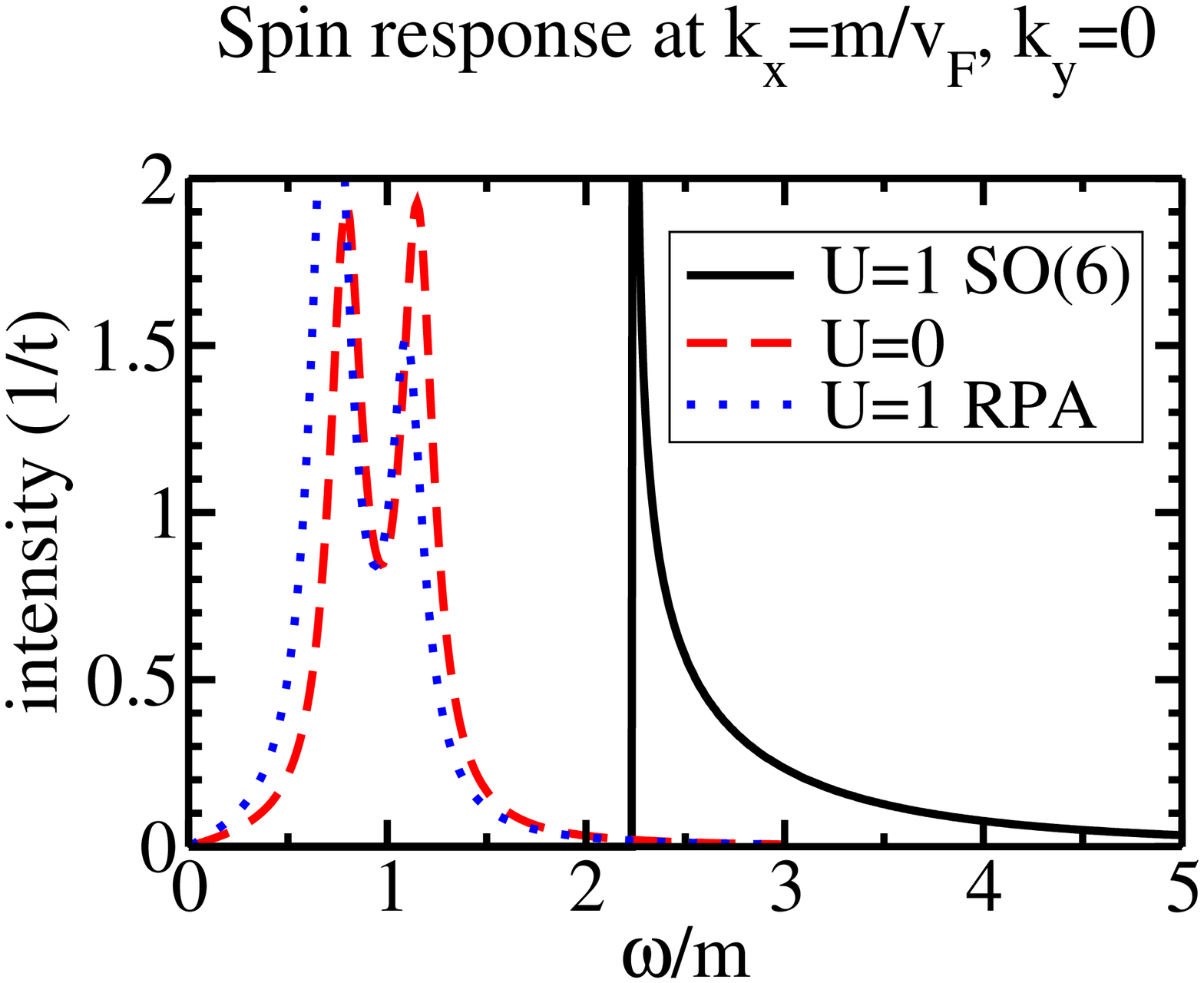}}
\end{center}
\vskip -.2in
\caption{A number of constant wavevector cuts of the spin response function for 10\% doping.
We compare the computations done in the SO(6) Gross-Neveu framework with both non-interacting ladders and
an RPA treatment.}
\end{figure*}

In the lower and middle panels of Figure 1, we compare our field theoretic results for the spin 
response
function both to the results for non-interacting ladders (see Appendix C) and to interacting ladders treated using an 
RPA approximation (see Appendix A 4).  The RPA approximation sees a more diffuse distribution of spectral weight than
that of the field theoretic treatment.  In particular the RPA approximation does a poor job at 
deducing
the existence of the contribution of a single bound state to the spin response near $(\pi,\pi)$.
But both the RPA and field theoretic treatment predict a much larger low-energy response 
than that of non-interacting ladders.  In this sense the ladders mimic the behavior of individual
chains where the presence of interactions shifts spectral weight from high to low energies\cite{joe}.
The presence of low energy spectral weight in a single chain is captured 
by the M\"uller ansatz\cite{muller}.
It would be interesting if a similar ansatz might be developed for the spin response of doped
ladders using our field theoretic treatment as a starting point.

To obtain a more quantitative understanding of the spin response, 
we have plotted constant wavevector cuts of
the spectral function in Figure 2.  Here we have done so at 10\% doping so that 
the low energy spectral weight present
near any of the values of $k_x$ related to $K_{Fab}$ or $K_{Fb}$ does not overlap with 
weight arising from 
a different such value of $k_x$.  We first consider the cut at $K_{Fab}+K_{Fb}$.  We 
see explicitly the onset of
weight at the spin gap, $\Delta_s = \sqrt{2}m$, the excitation energy of the
bound state.
We note that contribution of the bound state does
not lead to a (theoretically) infinitely narrow feature in the spin response.  
While this would be so at half-filling, 
the gapless nature of the charge mode at finite doping smooths the feature out giving 
it a finite spread.  Nonetheless there is a divergence at $(\omega^2-k^2/v_F^2)\rightarrow 0$
going as $(\omega^2-k^2/v_F^2)^{K/2-1}$.
In the energy interval $(0,10m)$ we can also estimate the amount of spectral weight
due to the single bound state.  We find that it accounts for 37\% of the total spectral weight
(with a Luttinger parameter, $K=0.93$).
Of this 37\%, 42\% of it occurs in the interval $(0,2m)$, i.e. before the incoherent
two kink contribution
to the spectral function begins to be felt.

\renewcommand{\thesubfigure}{}
\renewcommand{\subfigcapskip}{0pt}
\begin{figure*}
\begin{center}
\vskip -.2in
\subfigure[$k_y=0$]{
\epsfysize=1.25\textwidth
\includegraphics[height=5cm]{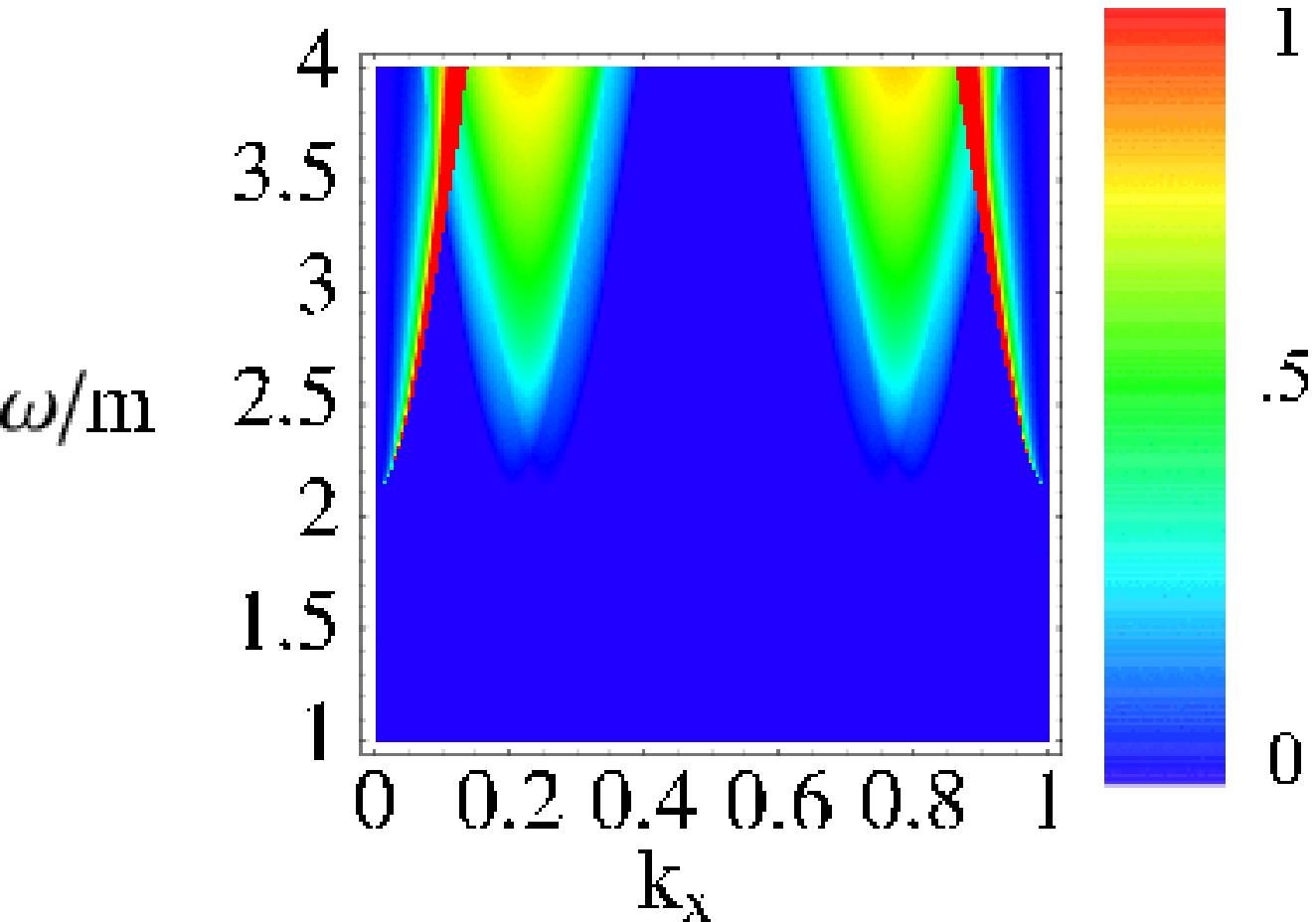}}
\subfigure[$k_y=\pi$]{
\epsfysize=1.25\textwidth
\includegraphics[height=5cm]{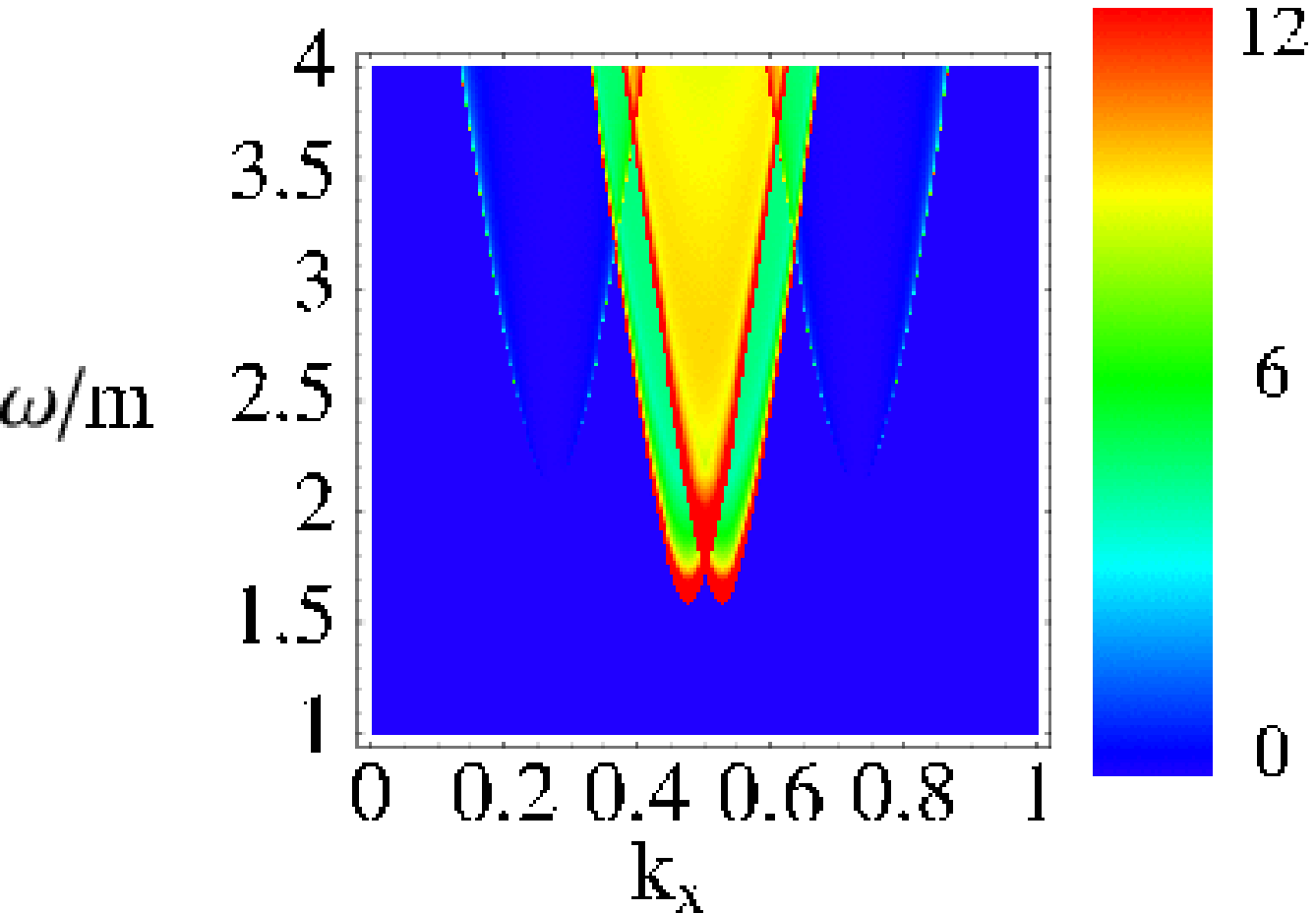}}
\end{center}
\vskip -.2in
\centerline{\hbox{5\% doping}}
\vskip -.5in
\begin{center}
\subfigure[$k_y=0$]{
\epsfysize=1.25\textwidth
\includegraphics[height=5cm]{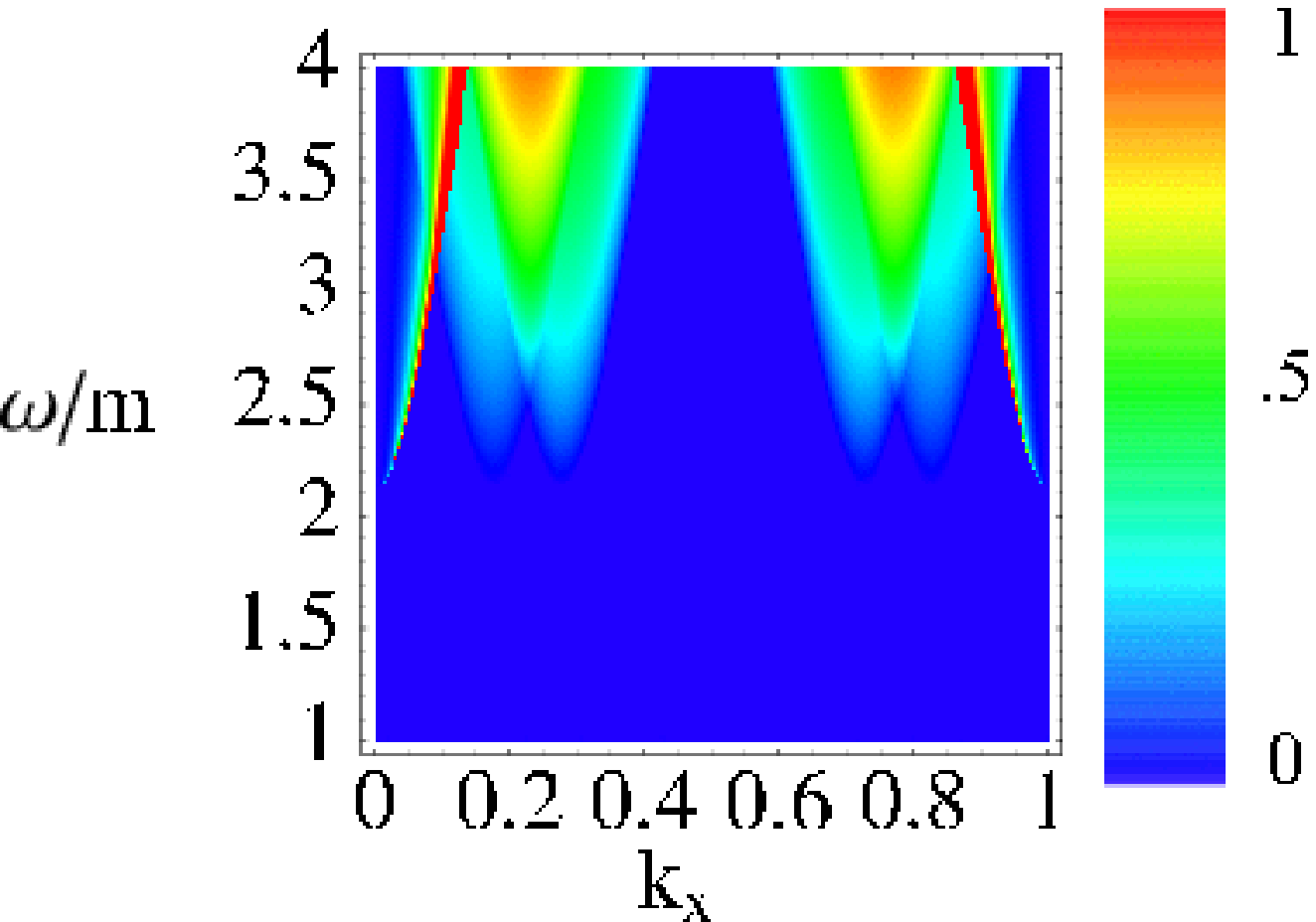}}
\subfigure[$k_y=\pi$]{
\epsfysize=1.25\textwidth
\includegraphics[height=5cm]{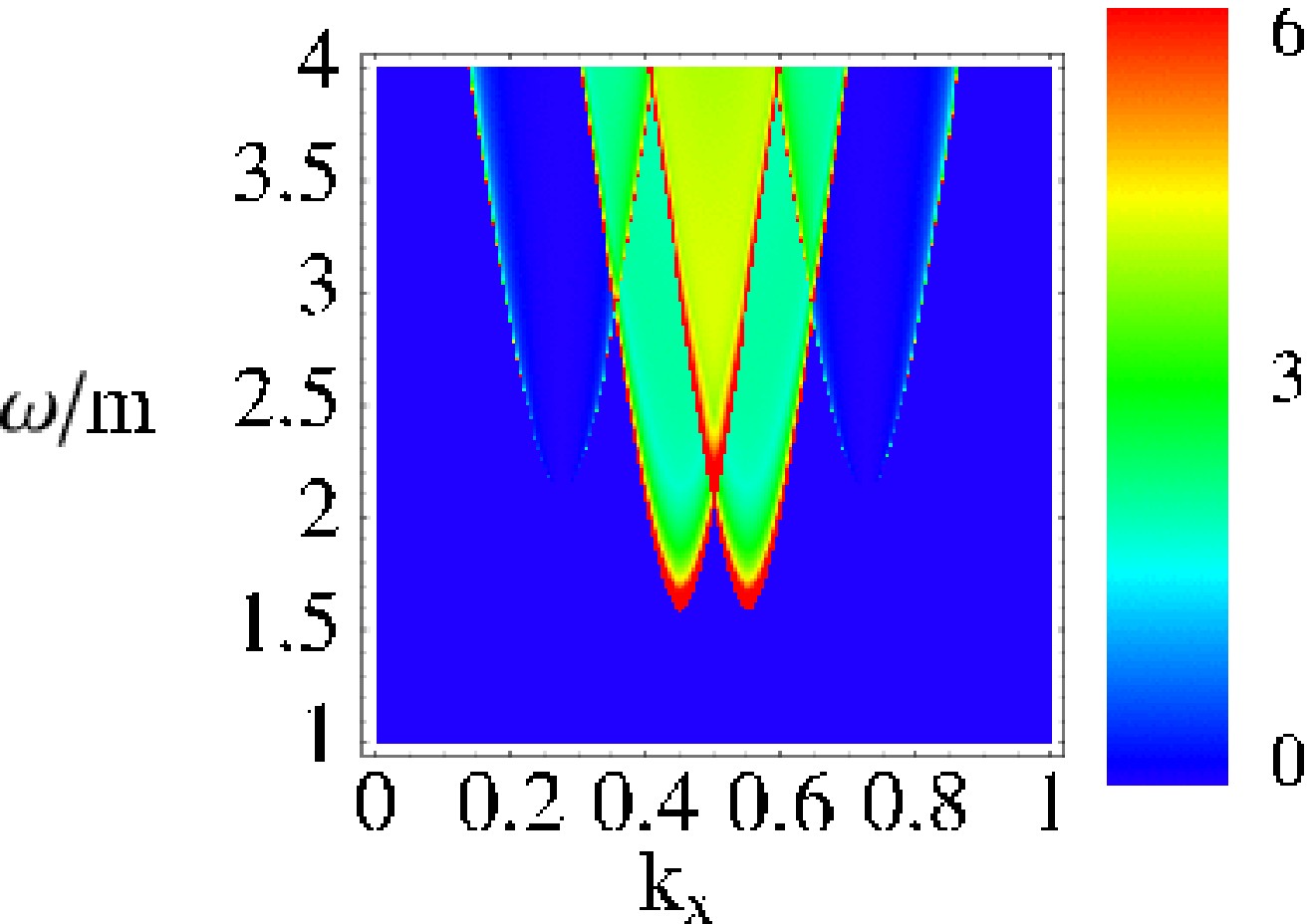}}
\vskip 0in
\end{center}
\vskip -.2in
\centerline{\hbox{10\% doping}}
\vskip -.5in
\begin{center}
\subfigure[$k_y=0$]{
\epsfysize=1.25\textwidth
\includegraphics[height=5cm]{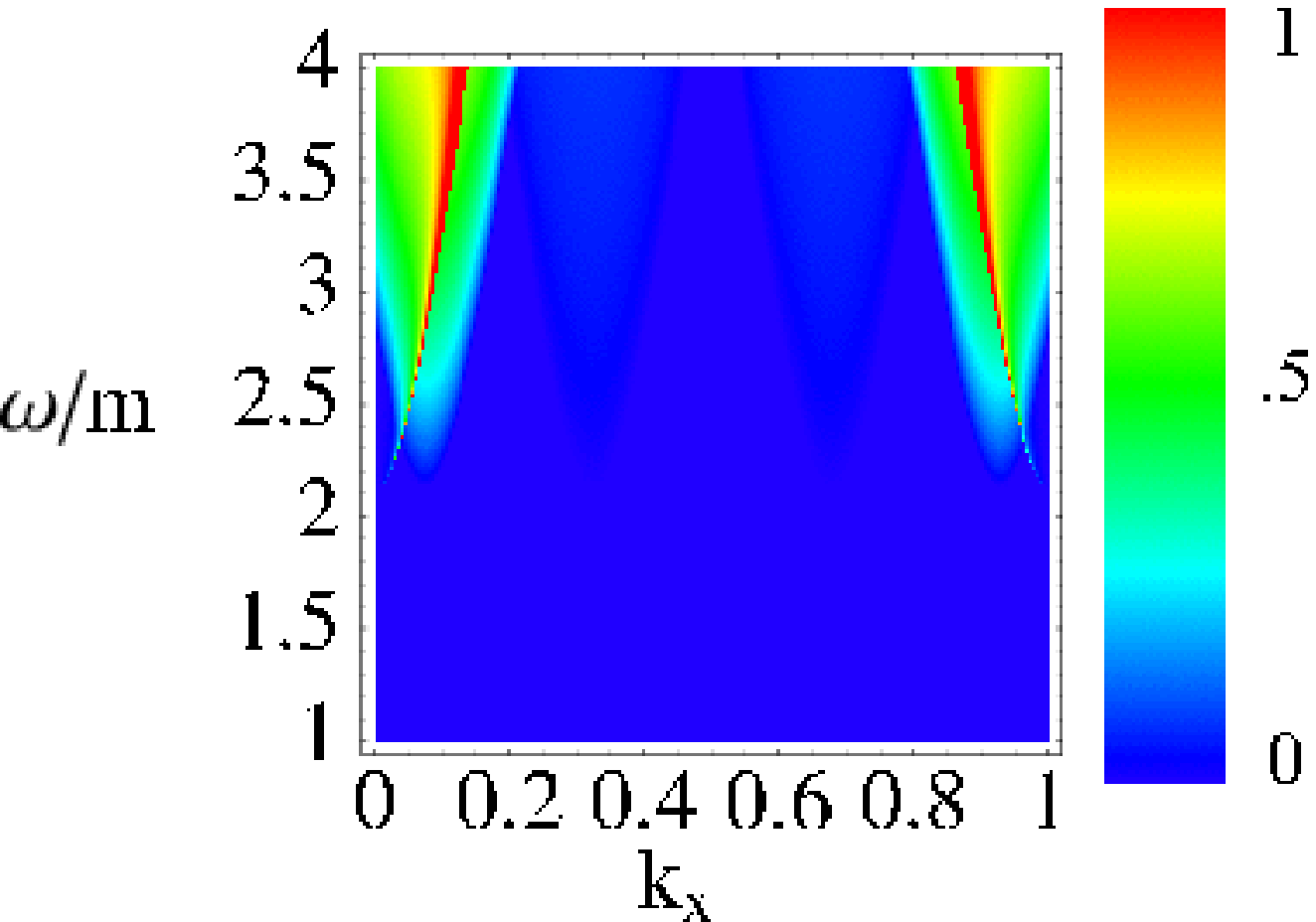}}
\subfigure[$k_y=\pi$]{
\epsfysize=1.25\textwidth
\includegraphics[height=5cm]{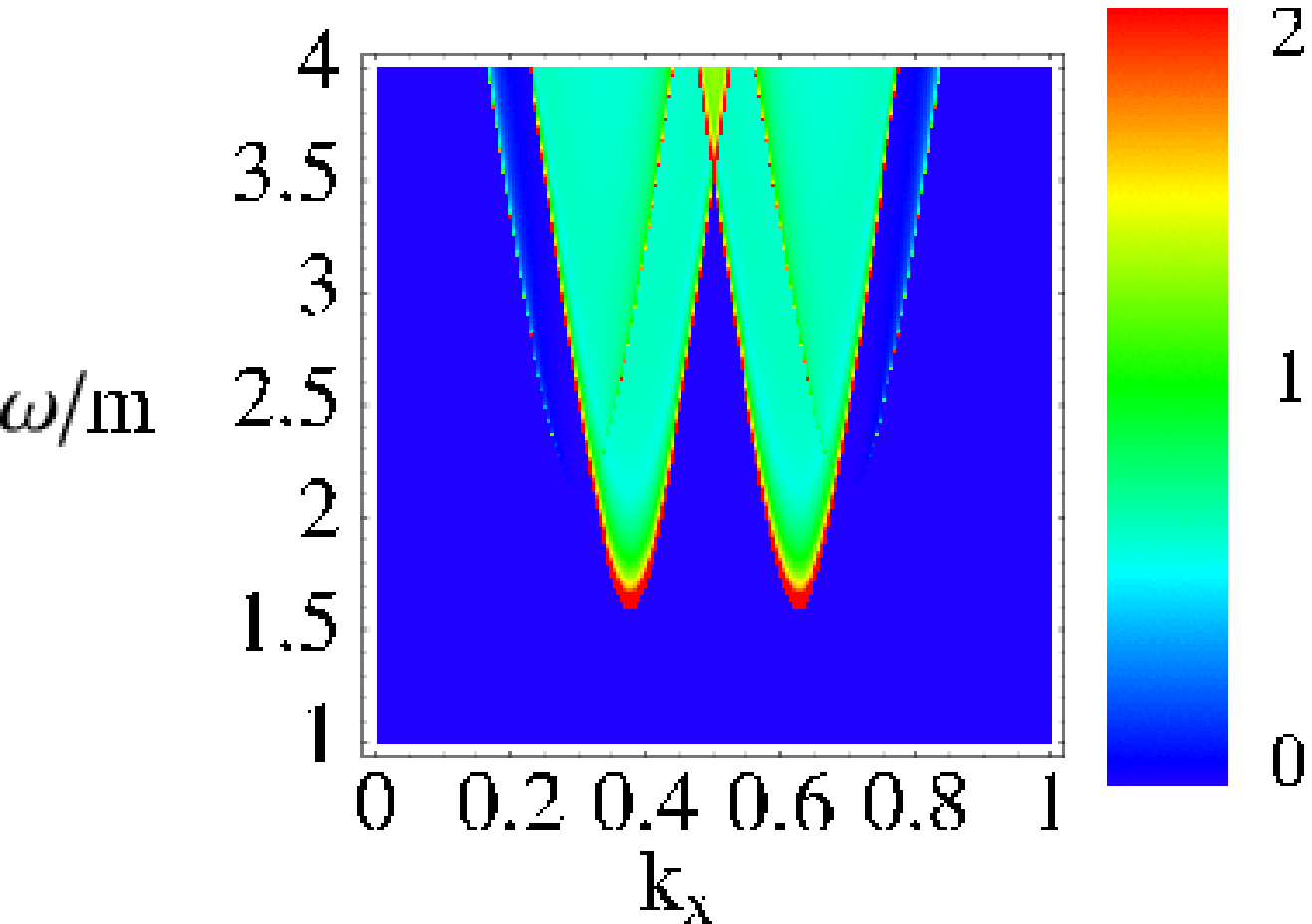}}
\vskip 0in
\end{center}
\centerline{\hbox{25\% doping}}
\caption{Plots of the spin response function for three different dopings, 5\%, 10\%, and 25\%.
The left-hand panels give the response at $k_y=0$ while the right-hand panels correspond to $k_y=\pi$.
These plots are made assuming a strongly renormalized Fermi velocity,
$v_F/m = 4$.}
\end{figure*}

\subsection{Spin Response at Larger U}

As U becomes larger we expect that the Fermi velocity will renormalize significantly
but that our effective low energy description will remain valid.  This view is supported
by the two loop RG computations of Fabrizio\cite{Fabr93}.  There it was shown that
the two-loop RG preserved the flow to SO(6) but that unlike one-loop RG equations,
the Fermi velocity was renormalized downward.

To study the spin response at large U we thus suppose that the ratio $v_F/m$ is much
smaller than at weak coupling.  The RPA breaks down at $U$ in excess of $t$, and so we
cannot directly estimate the corresponding normalization constants, $A$'s.  Instead we simply
continue to use the values calculated for $U=t$.  We point out that while at best any comparison
of low energy spectral weight at different wavevectors will then be qualitative, the 
spectral weight at any one value of $k$ will be unaffected by this procedure. 

In Figure 3 we plot the spin response for a value of $U$ corresponding to $v_F/m=4$ and
for three different values of doping, 5\%, 10\%, and 25\%.  In comparison with the spin
response in Figure 1 where $v_F/m \gg 1$, the response at a given energy occurs over a much
wider band of $k_x$ values.  

From Figure 3, we can observe how the spin response evolves as a function of doping.  Two
primary effects are observed.  The first is increasingly significant incommensuration effects 
as doping is increased.
Incommensuration effects are most noticeable near $(\pi,\pi)$, the region of greatest
spin response.  At small doping the sum
of bonding and anti-bonding wavevectors remain close to $\pi$ (see Eqn.(\ref{eIIIxvi})).  Consequently
at small dopings the spin response near $(\pi,\pi)$ appears akin to that at half-filling
with only a single peak of intensity apparent.  As doping is increased, however, two regions
of intensity, one at $K_{Fab}+K_{Fb}$ and one at $2\pi - K_{Fab}-K_{Fb}$, become resolved.
At 5\% doping, the two regions are becoming individually visible.  As we increase
doping further, the regions separate and become distinct over a wider and wider range
of energy.  
In Figure 4, we plot the spin response in a narrow region about $(\pi,\pi)$ for the three
different values of dopings.  In this figure, the splitting of the $(\pi,\pi)$ peak
as function of doping is more readily apparent.

An obvious signature of the incommensuration in a neutron scattering experiment would 
be found in a series of constant k-scans as a function of energy.  In comparing such scans at
$(\pi,\pi)$ and $(K_{Fab}+K_{Fb},\pi)$, a greater
intensity at the latter value of $k_x$ would mark the presence of incommensuration.

Incommensuration effects will be less obvious at $k_y=0$ simply because the intensity present
at this value of $k_y=0$ is markedly less than at $k_y=\pi$.  But the $k_x$ points at which
low energy intensity is found, i.e. $k_x= 2 K_{Fab}$, $2 K_{Fb}$, and $0$ will shift relative to 
one another.  This effect is convoluted with changing relative intensity at $k_x = 2 K_{Fb}$ and
$k_x = 2 K_{Fab}$.  For smaller dopings (i.e. 5\% and 10\%), the intensities found near these
two points is roughly equal.  For larger dopings, however, the intensity at $2 K_{Fab}$
becomes notably larger.  In the Figure 3 panel for $k_y=0$ at 25\% doping, the contribution at 
$k_x = 2 K_{Fab}$
and $k_x = 0$ overlap, while the intensity at $k_x = 2K_{Fb}$ is not observable in the plot at hand.

The second notable effect of doping is
that with increasing doping one sees
a decrease in maximum intensity at $k_y=\pi$ relative to that at $k_y=0$.
At 5\% doping, the relative intensity at $k_y=\pi$ is twelve times
that at $k_y=0$.  At 25\% doping, this ratio has been reduced from
twelve to two.

\subsection{Effects of Integrability Breaking Perturbations}

In this section we consider the effects that integrability breaking perturbations will have upon our
results.  As we discussed in the introduction, this is important to consider because 
the renormalization
group flow leaves a generic system more symmetric but does not necessarily
render the low energy sector
exactly equivalent to the SO(6) Gross-Neveu model.  As was done for the SO(8) Gross-Neveu
description of half-filled ladders\cite{so8}, 
we consider this question in broad terms so as to demonstrate
that small perturbations away from the SO(6) Gross-Neveu model do not drastically change the results.
(For a more detailed analysis for general ladder models see Ref. (\onlinecite{controzzi2}).)

With this in mind we examine general possible perturbations of a U(1) 
Luttinger liquid $\otimes$ 
SO(6) Gross-Neveu model.  Possible perturbations can be read off 
from Eqn.(\ref{eIIxiii}), the Hamiltonian
of the doped ladders with generic (bare) interactions.  These interactions 
all take the form of
\begin{equation}\label{eIIIxvii}
H_{\rm pert} = \lambda \cos(\theta_i)\cos(\theta_j), ~i\neq j
\end{equation}
or 
\begin{equation}\label{eIIIxviii}
H_{\rm pert} = \lambda \cos(\varphi_3)\cos(\theta_j).
\end{equation}
In general these interaction preserve the charge, spin, and
relative band chirality but not the difference in the z-component of spin between bands
(owing to the presence
of $\cos(\varphi_3)$).

To determine the manner in
which the above type of perturbations will affect the spectrum of the model is straightforward.
We can immediately note that the various multiplets of SO(6) Gross Neveu will be split
by the perturbation, but the single particle multiplets themselves will not be mixed.  We expect
also that the notion of single particle excitations will remain robust -- no
perturbation in Eqn.(\ref{eIIIxvi}) will lead to kink confinement.
Moreover no perturbation will gap out the U(1) Luttinger
liquid.

To determine the splitting of 
any given multiplet
of particles $\{ A_i(\theta) \}$, we employ stationary state perturbation
theory.   In this sense we operate in the same spirit as Ref. (\onlinecite{mussardo})
in treating the off-critical Ising model in a magnetic field or Ref. (\onlinecite{davide}) in 
treating perturbations of the O(3) non-linear sigma model.  At leading order this amounts then to 
diagonalizing the matrix
\begin{equation}\label{eIIIxix}
M_{ij} = \frac{\langle A_i (\theta ) H_{\rm pert} A_j^\dagger (\theta )\rangle}
{(\langle A_i(\theta)A_i^\dagger(\theta)\rangle\langle A_j(\theta)A_j^\dagger(\theta)\rangle)^{1/2}}.
\end{equation}
The computation of these matrix elements is straightforward in the context of integrability.
Each matrix element is no more than a form factor which can easily be computed in the same fashion as
done in Section IV.  Moreover because the theory is massive, perturbation theory is controlled and so
small perturbations should only introduce small quantitative changes to our results.

Deviations in the mass spectrum will have a two fold effect.  The bound state producing
the peak in the spin response at $\omega = \sqrt{2}m$ and $k = K_{Fab}+K_{Fb}$ will, 
under the perturbation,
be mixed into other members of the
vector multiplet, leading to a number of particles that will couple
to the spin operator.  Consequently the peak will split.  
This splitting however will be obscured by
the gapless charge excitations with their tendency to smear any sharp spectral features.  
This smearing
will be enhanced by finite temperature.  Unless the splitting is then large, it is unlikely
the effects of the perturbation will be detectable experimentally.

Like the splitting of the multiplet of bound states, the two particle threshold will similarly split.
Rather than occur at $\omega = 2m$, a number of thresholds will 
appear about this energy.  This effect will be muted at
points in k-space where the threshold 
opens smoothly from zero rather
than appearing as a singularity.  But the effect will be more
pronounced at $k_x=0, k_y=0$ and $k_x=K_{Fb}-K_{Fab}, k_y=\pi$, where
the two particle threshold is marked by a van-Hove like square root singularity.  
However the relative
weight of spectral intensity is small at these points (in comparison to $k_x=K_{Fab}+K_{Fb},k_y=\pi$)
and so even here any splitting will presumably be difficult to detect.

We can also consider the effects of integrable breaking perturbations upon matrix elements.
The behavior of matrix elements near threshold determine whether a threshold opens up smoothly
(for example the two particle matrix element vanishing at threshold, i.e.
$(\omega^2-k^2/v_F^2)\rightarrow 0)$, extinguishing a van-Hove singularity) 
or whether it is marked by a discontinuity (the matrix element is finite at threshold).  
Small perturbations
may then lead a vanishing matrix element to be finite forcing a qualitative change in the physics.
However, we now argue that this does not happen at leading order in the perturbation.

To compute the leading order correction of a matrix element of 
some operator, ${\cal O}(x=0,t=0)$, we employ
the relation (arrived at from Keldysh perturbation theory)
\begin{eqnarray}\label{eIIIxx}
\hskip -.6in \delta \langle 0 | {\cal O}(0,0) | i \rangle &=& \cr\cr
&& \hskip -.9in  -i \int^0_{-\infty} dt \int^\infty_{-\infty} dx 
\langle 0 | [ {\cal O}(0), H_{\rm pert}(x,t) ]|i\rangle.
\end{eqnarray}
Here $|0\rangle$ is the vacuum state and $|i\rangle$ is some other state with some arbitrary number
of particles.
If we employ a form-factor expansion to evaluate the expectation values involving operator bilinears
in the above expression we obtain
\begin{eqnarray}\label{eIIIxxi}
\hskip 0in \delta \langle 0 | {\cal O}(0,0) | i \rangle &=& \cr\cr
&& \hskip -.9in  -i \int^0_{-\infty} dt \int^\infty_{-\infty} dx
\sum_j \bigg( \langle 0 |{\cal O}(0) | j \rangle \langle j | H_{\rm pert}(x,t)|i\rangle \cr\cr
&& \hskip -0in - \langle 0 |H_{\rm pert}(x,t)|j\rangle \langle j|{\cal O}(0)|i\rangle\bigg),
\end{eqnarray}
where $\sum_j$ represents schematically a sum over all states appearing in a resolution of the identity.
We argue below that this correction will uniformly vanish for matrix elements, $|i\rangle$, involving
two kinks or anti-kinks with minimal energy, $2m$, and for which the matrix element,
$\langle 0 | {\cal O}(0,0) | i \rangle$,
itself vanishes.
This implies that the leading correction 
leaves unaffected the behavior of all relevant matrix elements at threshold appearing in the spin response.

In the relevant cases where the matrix element itself vanishes,
the vanishing is a consequence of the structure of the low energy S-matrix.
The S-matrix at vanishing energies of two kinks is $-{\bf P}$, where ${\bf P}$ is the permutation matrix.
Not only then does $\langle 0 | {\cal O}(0,0) | i \rangle$ vanish, but any matrix element
involving $|i\rangle$ vanishes, for example
$\langle j | H_{\rm pert}(x,t)|i\rangle$ and $\langle 0 | H_{\rm pert}(x,t)|i\rangle$.
The vanishing comes about through an application of the scattering axiom (see Section IV B).  
This axiom implies the following relation,
\begin{equation}\label{eIIIxxii}
\langle j | H_{\rm pert}(x,t)|i\rangle = - \langle j | H_{\rm pert}(x,t)|i\rangle.
\end{equation}
This then necessarily implies the matrix element vanishes regardless of the nature of $|j\rangle$
or $H_{\rm pert}$.

\renewcommand{\thesubfigure}{}
\renewcommand{\subfigcapskip}{0pt}
\begin{figure*}
\begin{center}
\subfigure[5\% doping]{
\epsfysize=1\textwidth
\includegraphics[height=4.cm]{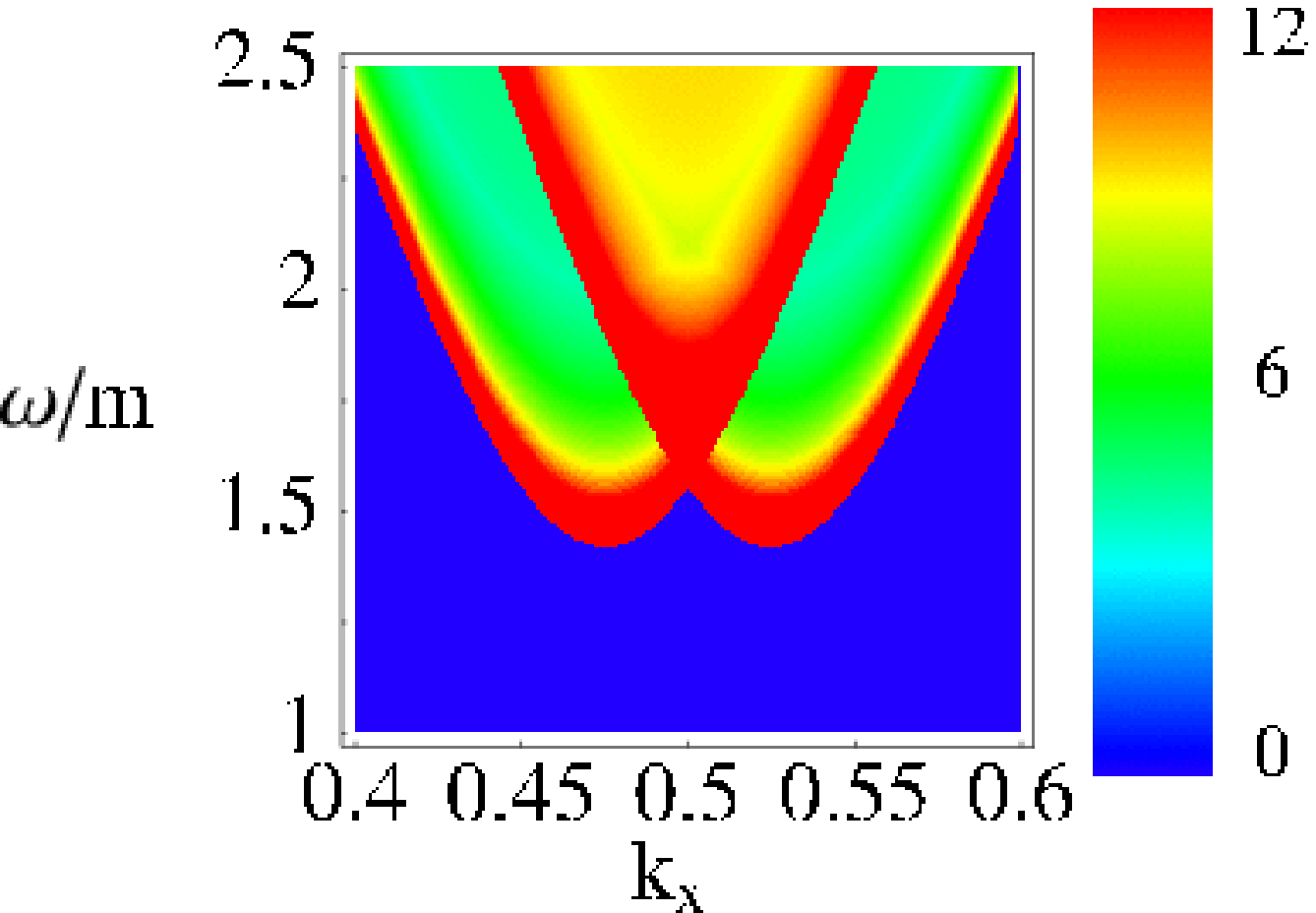}}
\subfigure[10\% doping]{
\epsfysize=1\textwidth
\includegraphics[height=4.cm]{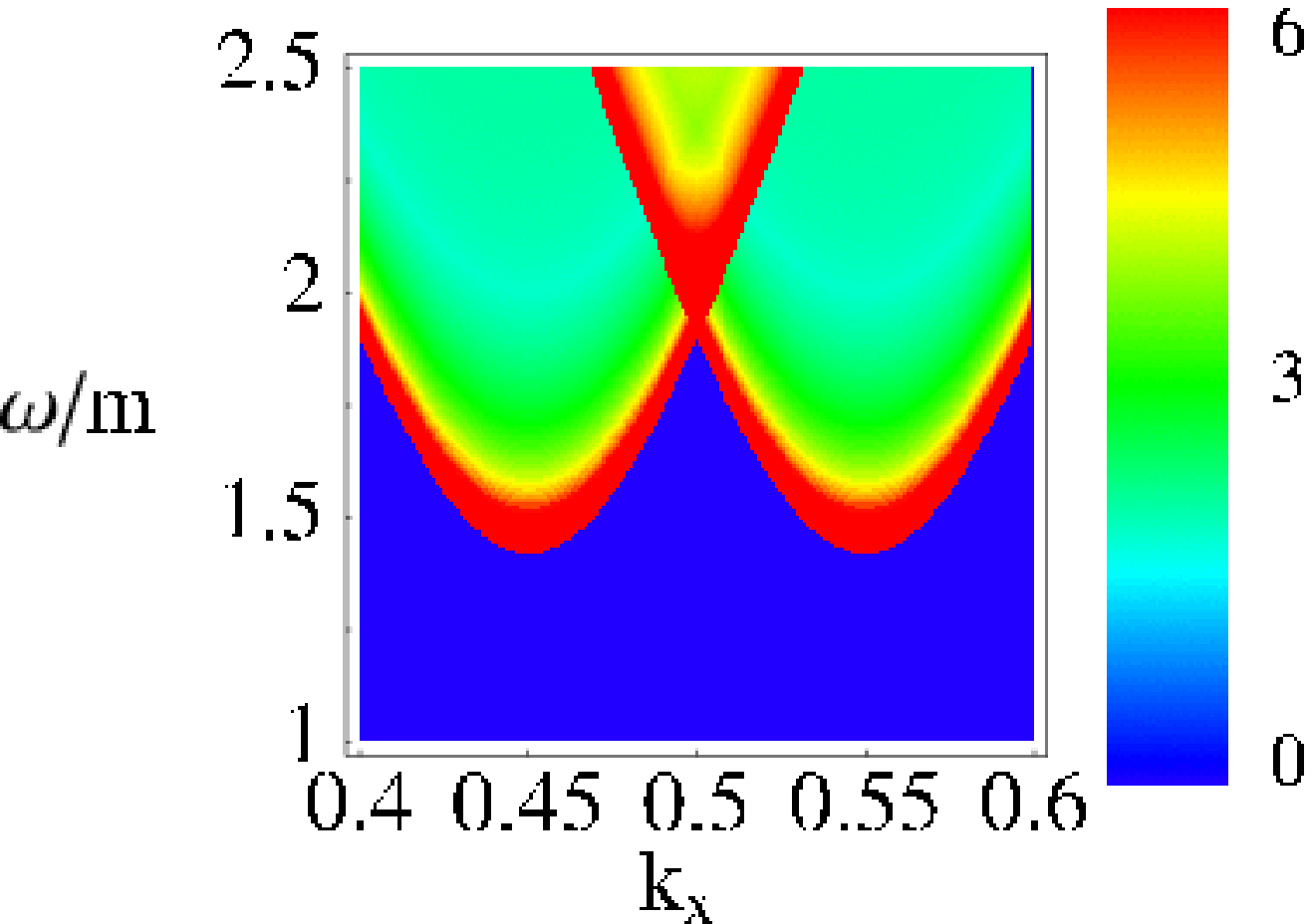}}
\subfigure[25\% doping]{
\epsfysize=1\textwidth
\includegraphics[height=4.cm]{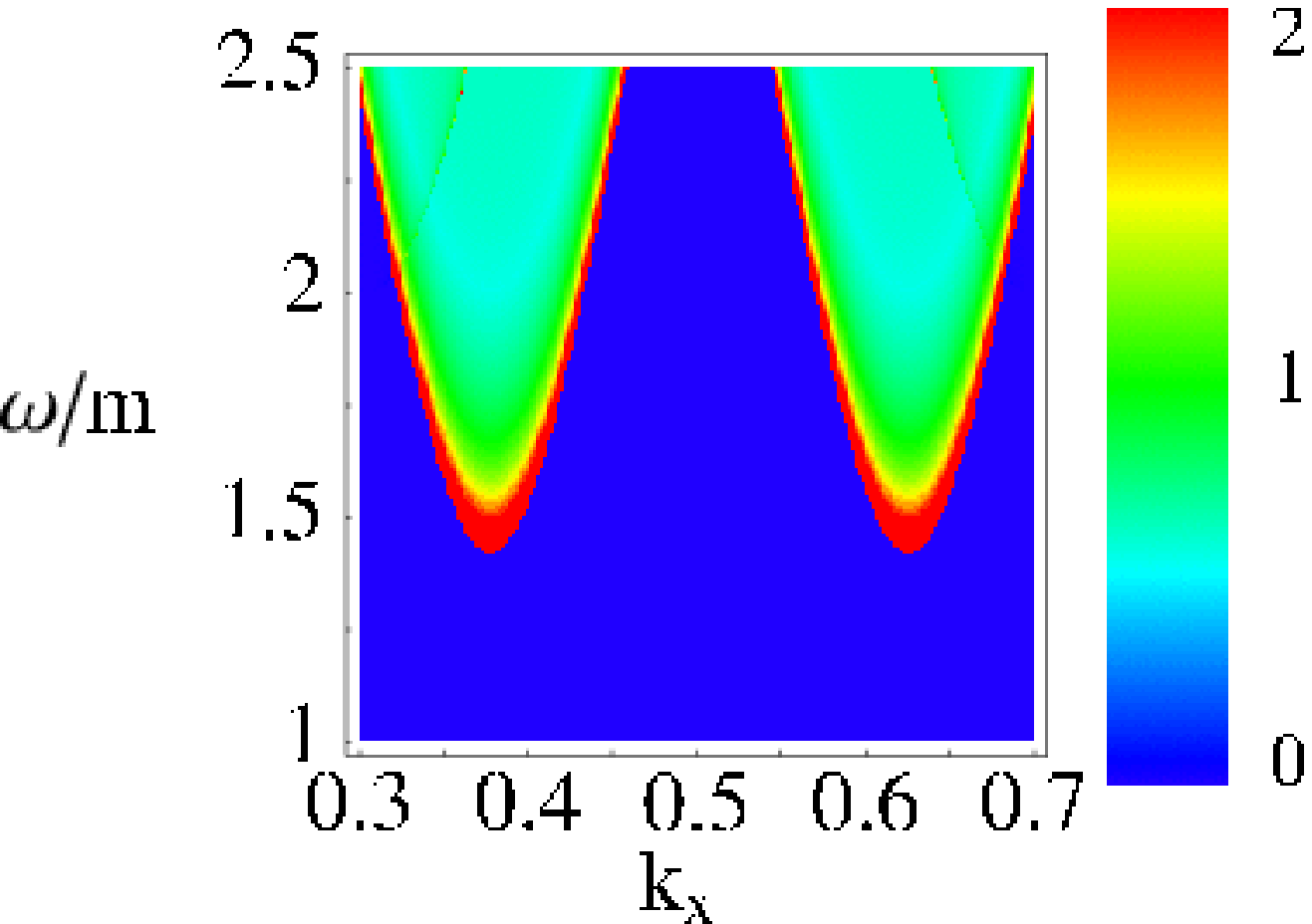}}
\end{center}
\vskip -.2in
\caption{Plots of the spin response function for three different dopings, 5\%, 10\%, and 25\%
for $k_y=\pi$ and a narrow range of $k_x$ chosen to emphasize the region where the
maximum intensity is found.  These plots are made assuming a strongly renormalized Fermi velocity,
$v_F/m = 4$.}
\end{figure*}

\section{Form Factors for the $SO(6)$ Gross-Neveu Model}
\label{sec:FF}

\subsection{S-matrices}
\newcommand{\so}{\sigma_1}
\newcommand{\st}{\sigma_2}
\newcommand{\sth}{\sigma_3}
\newcommand{\ith}{i\theta/2\pi}
\newcommand{\sos}{$SO(6)$ }
\newcommand{\son}{$SO(6)$}
\newcommand{\tht}{\theta}

In this section we lay out the S-matrices for scattering between the kinks.
From these, we will be able to develop formulae for the various
form-factors needed to compute the spin-spin ladder correlation functions.

\subsubsection{$SO(6)$ Spinor Representations}
The spinor representations are expressed in terms of \sos $\gamma$-matrices.
The $\gamma$-matrices in turn are built out of the two-dimensional Pauli
matrices, $\sigma^i$'s.  As we are interested in \son, we consider three
copies of the $\sigma^i$'s,
\begin{equation}\label{eIVi}
\sigma^i_1, \sigma^i_2, \sigma^i_3,
\end{equation}
each acting on a different two-dimensional space, i.e.:
\begin{equation}\label{eIVii}
\sigma^i_2 |\alpha_1,\alpha_2,\alpha_3\rangle =
|\alpha_1,\sigma^i_2 \alpha_2,\alpha_3\rangle  .
\end{equation}
A particular basis vector, $|\alpha\rangle$,
in the corresponding 8 dimensional vector space
can be labeled by a series of three numbers $\pm 1/2$, i.e.
\begin{equation}\label{eIViii}
|\alpha\rangle \equiv |\alpha_1, \alpha_2, \alpha_3\rangle = |\pm 1/2, \pm 1/2, \pm 1/2\rangle .
\end{equation}
Physically, the kink associated with $|\alpha\rangle$ then 
carries quantum numbers $N_i=\pm 1/2$ for the three U(1) symmetries
corresponding to the Cartan elements of $SO(6)$, discussed in the
previous section.

In terms of the $\sigma^i$'s, the $\gamma$-matrices are defined
by (following the conventions of Ref. \onlinecite{thun}),
\begin{eqnarray}\label{eIViv}
\gamma_{2n-1} &=& \sigma^1_n \otimes \prod^{n-1}_{k=1} \sigma^3_k ;\cr\cr
\gamma_{2n} &=& \sigma^2_n \otimes \prod^{n-1}_{k=1} \sigma^3_k ;
\end{eqnarray}
for $1 \leq n \leq 3$.  These matrices satisfy the necessary Clifford
algebra:
\begin{equation}\label{eIVv}
\{\gamma_a, \gamma_b\} = 2 \delta_{ab}.
\end{equation}
In this representation the Clifford-algebra generators,
$\gamma_n$, are imaginary and antisymmetric for n even, 
while for n odd, they are real and symmetric. The $SO(6)$-generators
are represented by
\begin{equation}\label{eIVvi}
\sigma^{ab} = \frac{1}{2}(\gamma^a\gamma^b - \gamma^b\gamma^a),
\end{equation}
in analogy with the more familiar $SO(4)$ case.

The 8-dimensional space of the $\gamma$'s decomposes
into two 4-dimensional
spaces, each of which forms one of the two
irreducible $SO(6)$ spinor representations.
The decomposition is achieved explicitly via the hermitian 
chirality operator, 
$\Gamma$,
\begin{equation}\label{eIVvii}
\Gamma = (-i)^3\prod^{6}_{a=1} \gamma^a = \sigma^3_1\otimes
\sigma^3_2\otimes \sigma^3_3. 
\end{equation}
$\Gamma$ is such that it commutes with all \sos generators and is diagonal
with eigenvalues $\pm 1$.  If $|\alpha^\pm\rangle$
is a state with an even (odd)
number of negative components (i.e. states with positive or negative
isotopic chirality), then
\begin{equation}\label{eIVviii}
\Gamma|\alpha^\pm\rangle  = \pm|\alpha^\pm\rangle .
\end{equation}
Thus the operators $(1\pm\Gamma )/2$ project onto the two irreducible
subspaces.  When it is necessary to make a distinction, we will denote
kinks with an even number of negative components by $\alpha$ and those
with an odd number by $\bar\alpha$. 

The last item to be presented in this section is
the charge conjugation matrix, $C$.  In terms of the $\gamma$'s, $C$
is given by
\begin{equation}\label{eIVix}
C = \gamma_1\gamma_3\gamma_5 .
\end{equation}
$C$ is completely off-diagonal (as expected).  Moreover we note that
$C$ is skew symmetric ($C^T = -C$, $C^2 = -1$).  
This fact will have important consequences in
the determination of the form factors.  

If $\alpha = |\alpha_1,\alpha_2,\alpha_3\rangle$ is an even kink its
corresponding anti-particle is denoted by $\bar\alpha =
|-\alpha_1,-\alpha_2,-\alpha_3\rangle$.  By the skew-symmetry of $C$,
we thus have  
\begin{equation}\label{eIVx}
|\bar\alpha\rangle = (-1)^{\alpha_2-1/2}C_{\alpha\beta}|\beta\rangle.
\end{equation}

\subsubsection{S-Matrices for the Kinks}
\newcommand{\Aa}{A_a^\dagger (\theta )}
\newcommand{\Aal}{A_\alpha^\dagger (\theta )}
\newcommand{\Ab}{A_b^\dagger (\theta )}
\newcommand{\Abe}{A_\beta^\dagger (\theta )}
\newcommand{\AB}{A_B^\dagger (\theta )}

In order to describe factorizable scattering, we introduce Faddeev - Zamolodchikov (FZ) operators,
$\Aal$, that create the elementary kinks.  $\tht$ is the rapidity
which parameterizes a particle's energy and momentum:
\begin{equation}\label{eIVxi}
P(\theta) = m \sinh (\tht ) ; ~~~~~~~ E(\theta) = m \cosh (\tht ) .
\end{equation}
Because the \sos Gross-Neveu model is integrable, 
scattering is completely encoded in the two-body S-matrix.  This
S-matrix, in turn, is parameterized by the commutation relations of
the Faddeev-Zamolodchikov operators:
\begin{equation}\label{eIVxii}
A_1^\dagger (\tht_1 ) A_2^\dagger (\tht_2 ) = S^{34}_{12}(\tht_{12}) 
A_4^\dagger (\tht_2 ) A_3^\dagger (\tht_1 ).
\end{equation}
$S^{34}_{12}(\tht_{12})$ is the amplitude of a process
by which particles 1 and 2
scatter into 3 and 4.  It is a function
of $\tht_{12} \equiv \tht_1 - \tht_2$ by reason of Lorentz invariance.

Kink-kink scattering takes the form
\begin{equation}\label{eIVxiii}
S_{\alpha\beta}^{\gamma\delta}(\tht) = \frac{1}{8} \sum^{6}_{r=0}
\frac{u_r (\tht )}{r!}
\sigma^{(r)}_{\gamma\beta} \sigma^{(r)}_{\delta\alpha},
\end{equation}
where $\sigma^{(r)}$ is a rank-r antisymmetric tensor:
\begin{equation}\label{eIVxiv}
\sigma^{(r)} \equiv \sigma^{a_1 \cdots a_r}
= (\gamma^{a_1} \cdots \gamma^{a_r})_A .
\end{equation}
Here A represents a complete anti-symmetrization of the gamma matrices.
By $\sigma^{(r)}_{\gamma\beta} \sigma^{(r)}_{\delta\alpha}$ we mean a
trace over all possible rank-r antisymmetric tensors:
\begin{equation}\label{eIVxv}
\sigma^{(r)}_{\gamma\beta} \sigma^{(r)}_{\delta\alpha} =
\sum_{a_1 \cdots a_r} \sigma^{a_1 \cdots a_r}_{\gamma\beta}
\sigma^{a_1 \cdots a_r}_{\delta\alpha}.
\end{equation}
The generic form of $S_{\alpha\beta}^{\gamma\delta}$ was determined
in Ref.  \onlinecite{witt}, 
while the specific forms of the $u$'s were given in Ref. \onlinecite{thun}.
There it was found
\begin{eqnarray}\label{eIVxvi}
u_{3+r} (\tht ) &=& (-1)^r u_{3-r}(\tht ) ;\cr\cr
u_{r+2} (\tht ) &=& u_r (\tht ) { (1 - r/2) - (1 + i\tht/\pi ) \over
(1 - r/2) + (1 + i \tht/\pi )} ;\cr\cr
u_2 (\tht ) &=& {\Gamma(1-\tht/2i\pi)\Gamma(\tht/2i\pi-1/4) \over \Gamma(3/4-\tht/2i\pi)\Gamma(\tht/2i\pi)};\cr\cr
u_1 (\tht ) &=&  {3/2-\tht/i\pi \over 1-\tht/i\pi}{\Gamma(1/2+\tht/2i\pi) \over \Gamma(1/2-\tht/2i\pi)}\cr
&& \hskip .6in \times{\Gamma(1/4-\tht/2i\pi) \over \Gamma(1/4+\tht/2i\pi)}.
\end{eqnarray}
We note that there is a sign difference in the definition of $u_1$ between Ref. \onlinecite{thun}
and Eqn.(\ref{eIVxvi}).  This sign difference arises as we give here the S-matrices of the physical
particles (those with fractional statistics).  
$S_{\alpha\beta}^{\gamma\delta}(\tht )$ satisfies a Yang-Baxter equation,
\begin{eqnarray}\label{eIVxvii}
S^{\gamma_1\gamma_2}_{\alpha_1\alpha_2}(\tht_{12})S^{\beta_1\gamma_3}_{\gamma_1\alpha_3}(\tht_{13})
S^{\beta_2\beta_3}_{\gamma_2\gamma_3}(\tht_{23}) &=& \cr\cr
&& \hskip -1.7in S^{\gamma_2\gamma_3}_{\alpha_2\alpha_3}(\tht_{23})
S^{\gamma_1\beta_3}_{\alpha_1\gamma_3}(\tht_{13}) S^{\beta_1\beta_2}_{\gamma_1\gamma_2}(\tht_{12}).
\end{eqnarray}
Physically, the Yang-Baxter equation encodes the equivalence of
different ways of representing three-body interactions in terms of
two-body amplitudes.  Formally, it expresses the associativity of
the Faddeev-Zamolodchikov algebra.
The S-matrix, $S_{\alpha\beta}^{\gamma\delta}(\tht )$, also satisfies both a crossing,
\begin{equation}\label{eIVxviii}
S^{\alpha\beta}_{\gamma\delta}(i\pi - \theta) = 
S^{\bar{\gamma}\beta}_{\bar{\alpha}\delta}(\theta ) C_{\bar{\gamma}\gamma} C_{\alpha\bar{\alpha}}.
\end{equation}
and unitarity relation,
\begin{equation}\label{eIVxix}
S^{\gamma\delta}_{\alpha\beta}(\tht )S^{\alpha'\beta'}_{\gamma\delta}(-\tht ) 
= \delta_\alpha^{\alpha'}\delta_\beta^{\beta'}.
\end{equation}
We note that for the above kink-kink S-matrices, crossing is satisfied, as expected,
with $C=\gamma_1\gamma_3\gamma_5$.

These constraints determine $S^{\gamma\delta}_{\alpha\beta}$ up to a `CDD'-factor.
Such factors allow additional poles to be added, if needed, in the physical strip,
${\rm Re} (\tht ) = 0$,
$0 < {\rm Im}{\tht} < 2\pi$, to the scattering matrix
and so are indicative of bound states.  Here we expect to find a pole
in $u_n$, n even, at $\tht = i\pi/2$ reflecting the
fact that two (same chirality) kinks can form a bound state (the fundamental fermions
of the model) of
mass $\sqrt{2}m$.  
The overall sign of the S-matrix is determined by examining the residue
of the pole in $S^{\alpha\beta}_{\alpha\beta}$ at $\tht = i\pi/2$.  This pole is
indicative of the formation of a mass $\sqrt{2}m$ bound state in the
s-channel and so should have positive imaginary residue.

The $SO(6)$ Gross-Neveu 
model has an isotopic chirality conservation law \cite{thun}.
Thus opposite chirality kink scattering is determined solely 
by $u_n$, $n$ odd,
while same chirality kink scattering is
determined solely by $u_n$, $n$ even.  
For example, it can be shown that for even chirality kinks, the S-matrix reduces to
\begin{eqnarray}\label{eIVxx}
S_{\alpha\beta}^{\gamma\delta}(\tht ) &=& -u_2(\tht)\delta_{\gamma\alpha}\delta_{\delta\beta} \cr\cr
&&\hskip .25in + {1\over 4}(u_0(\tht)+u_2(\tht ))\delta_{\gamma\beta}\delta_{\delta\alpha}.
\end{eqnarray}


\subsection{Basic Properties: Two Particle Form Factors}
\newcommand{\fp}{f^\psi}

The two particle form factors of a field $\psi (x)$ are defined as the
matrix elements,
\begin{equation}\label{eIVxxi}
f^\psi_{12} (\tht_1 , \tht_2 ) = 
\langle 0|\psi (0) A_2^\dagger(\tht_2) 
A_1^\dagger(\tht_1)|0 \rangle .
\end{equation}
The form of $\fp_{12}(\tht_1 ,\tht_2 )$ is constrained by integrability,
braiding relations, Lorentz invariance, and hermiticity.

The constraint coming from integrability
arises from the scattering of Faddeev-Zamolodchikov
operators.  For Eqn.(\ref{eIVxxi}) to be consistent with Eqn.(\ref{eIVxii}), we must have
\begin{equation}\label{eIVxxii}
\fp_{21}(\tht_2 ,\tht_1 ) = S^{34}_{12}(\tht_{12})\fp_{34}(\tht_1,\tht_2).
\end{equation}
The second constraint can be thought of as a periodicity axiom.
It reads
\begin{equation}\label{eIVxxiii}
\fp_{21}(\tht_2,\tht_1) = Q_{12} R_{\psi_1\psi}\fp_{12}(\tht_1-2\pi i,\tht_2).
\end{equation}
Both $Q_{12}$ and $R_{\psi_1\psi}$ are phases. $Q_{12}$ arises from
the charge conjugation matrix in SO(6) Gross-Neveu being skew-symmetric
\cite{smirbook}.  Its determination is discussed in some detail
in Appendix B. 
The phase factor $R_{\psi_1\psi}$, on the other hand, arises from the
``semi-locality'' of the fields in the \sos Gross-Neveu model. 
Working in Euclidean space, fields can be said to be semi-local if
their product sees a phase change when the fields are taken around one
another in the plane via the analytic continuation $z \rightarrow
ze^{i2\pi}, \bar{z} \rightarrow ze^{-i2\pi}$. More specifically we have
\begin{equation}\label{eIVxxiv}
\psi_1 (ze^{i2\pi},\bar{z}e^{-i2\pi})\psi (0) = R_{\psi_1\psi}
\psi_1(z,\bar{z})\psi (0). 
\end{equation}
Here $\psi_1$ is the field that is associated with the particle
$A_1^\dagger (\tht_1)$. The phase $R_{\psi_1\psi}$ can be calculated from the
operator product expansions of the operator $\psi$ with bosonic vertex
operator representations of the kink fields.  If $A_1$ is a kink,
$|\alpha\rangle = (\pm 1/2,\pm 1/2, \pm 1/2)$, the corresponding
representation of the field is
\begin{equation}\label{eIVxxv}
\psi_1 = e^{\frac{i}{4}(\pm\phi_1+\pm\phi_2+\phi_3)}
\end{equation}
In contrast to the representation of the left and right components of the lattice
fermions (Eqn.(\ref{eIIxxiv})), these fields are non-chiral. 
The form factor, $\fp_{12}$, must also satisfy Lorentz covariance.  If
$\psi$ has Lorentz spin, $s$, $\fp_{12}$ will take the form
(at least in the cases at hand),
\begin{equation}\label{eIVxxvi}
\fp_{12}(\tht_1,\tht_2 ) = e^{s(\tht_1+\tht_2)/2}f(\tht_{12}),
\end{equation}
where by virtue of $\tht_{12} \equiv \tht_1 - \tht_2$, $f$ is a Lorentz
scalar.

The constraints Eqn.(\ref{eIVxxii}), Eqn.(\ref{eIVxxiii}), and
Eqn.(\ref{eIVxxvi}), do not uniquely specify $\fp_{12}$.
It is easily seen that if $\fp_{12}$ satisfies these axioms then so does
$\fp_{12}\ g(\cosh (\tht_{12}))$, where $g(x)$ is some rational function.  
The strategy is then as follows.  One first determines the minimal
solution to the constraints, minimal in the sense that is has the
minimum number of zeroes and poles in the physical strip,
${\rm Re}(\tht ) = 0, 0 < {\rm Im}(\tht ) < 2\pi$.  Then one adds poles 
according to the theory's bound state structure.  If the S-matrix
element describing scattering of particles 1,2 has a pole at $\tht =
iu$, then $\fp_{12}(\tht_1,\tht_2)$ has a pole at
\begin{equation}\label{eIVxxvii}
\tht_2 = \tht_1 + iu .
\end{equation}
Insisting that $\fp_{12}$ has such poles and only such poles fixes 
$g(x)$ up to a constant.

The phase of this constant can be readily determined.  Appealing
to hermiticity gives us
\begin{eqnarray}\label{eIVxxviii}
\langle 0| \psi (0) A_2^\dagger(\tht_2) 
A_1^\dagger(\tht_1)|0 \rangle^*  &=&
\langle 0|A_1(\tht_1) A_2(\tht_2) \psi^\dagger (0)|0 \rangle\cr\cr
&& \hskip -1.45in= 
\bar{C}_{\bar{1}1}\bar{C}_{\bar{2}2} 
\langle 0|\psi^\dagger (0) A^\dagger_{\bar 1}(\tht_1-i\pi) 
A^\dagger_{\bar 2}(\tht_2-i\pi)|0\rangle,
\end{eqnarray}
where the last line follows by crossing and so
\begin{eqnarray}\label{eIVxxix}
f^\psi_{12}(\tht_1 ,\tht_2)^* =
\bar{C}_{\bar{1}1}\bar{C}_{\bar{2}2}
f^{\psi^\dagger}_{\bar{2}\bar{1}}(\tht_2-i\pi,\tht_1-i\pi). 
\end{eqnarray}
For the kinks the matrix $\bar{C}$ equals the standard charge
conjugation matrix, $C=\gamma_1\gamma_3\gamma_5$, while for the
Majorana fermions $\bar{C} = 1$.  (We note however that $C$ for the
Majorana fermions is not identically 1.  This difference between
$\bar{C}$ and $C$ for the Majorana fermions arises because under
Hermitean conjugation these fermions are invariant while under charge
conjugation they undergo sign changes.)
For the form factors we will examine, hermiticity will be enough to
fix their overall phase.

As a final consideration we examine the large rapidity ($\tht
\rightarrow \infty$) asymptotics of a form factor.  These asymptotics
are governed by the (chiral) conformal dimension, $\Delta$, of the
operator appearing in the form factor \cite{delone}.  In particular,
if we write  
\begin{equation}\label{eIVxxx}
\lim_{|\tht_i| \rightarrow \infty} \fp (\tht_1,\tht_2) \sim
e^{y_\psi|\tht_i|}, 
\end{equation}
the quantity, $y_\psi$ is bounded from above by $\Delta$.  This bound
will prove useful in checking the end result.

\subsection{Basic Properties: One Particle Form Factors}
One particle form factors are in a sense trivial;  Lorentz covariance
completely determines their form.  If $\psi(x)$ has Lorentz spin, $s$, 
then
\begin{equation}\label{eIVxxxi}
\fp_1(\tht) = \langle 0| \psi (0)A_1^\dagger (\tht)|0\rangle 
= c e^{s\tht} ,
\end{equation}
where $c$ is some constant.  To determine $c$ we use the theory's
bound state structure.

If a particle $A_3$ is a bound state of two particles 
$A_1$ and $A_2$ we can represent it formally as
product of two Faddeev-Zamolodchikov operators: 
\begin{equation}\label{eIVxxxii}
i g^3_{12}A_3^\dagger(\tht ) =  {\rm res}_{\delta = 0}~
A_1^\dagger(\tht + \delta + i \bar{u}^{\bar{2}}_{1\bar{3}})
A_2^\dagger(\tht - i \bar{u}^{\bar{1}}_{2\bar{3}}) .
\end{equation}
Here $\text{res}_{\delta = 0}$ denotes the residue at $\delta = 0$ and
$\bar 1$ is the charge conjugate particle of $1$.  Implicit to this
relation is the particle normalization $\langle \tht | \tht '\rangle =
2\pi \delta (\tht - \tht')$.  The $\bar u$'s 
are given by $\bar u^{\bar 2}_{1{\bar 3}} = \pi - u^{\bar 2}_{1{\bar
    3}}$ where $u^3_{12}$ is the location of the bound state pole of
the S-matrix for particles $1,2$ indicative of particle
$3$. $g^3_{12}$ is the amplitude to create particle $1$ from particles
$2$ and $3$.  $g^3_{12}$ is defined by 
\begin{equation}\label{eIVxxxiii}
S^{12}_{12} (\theta ) \sim
i {g^3_{12} g_3^{12}
\over \theta - i u^3_{12}} .
\end{equation}
Then we have
\begin{equation}\label{eIVxxxiv}
i g^3_{12}\fp_3(\tht ) =  {\rm res}_{\delta = 0}~
\fp_{12}(\tht - i\bar{u}^{\bar{1}}_{2\bar{3}},
\tht + \delta + i \bar{u}^{\bar{2}}_{1\bar{3}}).
\end{equation}
Thus knowledge of the two particle form factors completely determines
their one-particle counterparts.

\subsection{Two particle form factors for the SO(6) currents}

\newcommand{\fcf}{_\mu f^{ab}_{cd}}
\newcommand{\fcft}{_\mu f^{ab}_{cd}(\tht_1,\tht_2)}
\newcommand{\fck}{_\mu f^{ab}_{\alpha\bar{\beta}}}
\newcommand{\fckt}{_\mu f^{ab}_{\alpha{\bar\beta}}(\tht_1,\tht_2)}
\newcommand{\fcka}{_\mu f^{ab}_{\bar{\alpha}\beta}}
\newcommand{\fckta}{_\mu f^{ab}_{\bar{\alpha}\beta}(\tht_1,\tht_2)}
\newcommand{\cop}{G^{ab}_\mu}

There are two possible two particle form factors for the $SO(6)$ currents,
one involving two fermions,
\begin{equation}\label{eIVxxxv}
\fcft = \langle \cop(0) A^\dagger_d(\tht_2 )A^\dagger_c(\tht_1)\rangle ,
\end{equation}
and one involving two kinks,
\begin{equation}\label{eIVxxxvi}
\fckt = \langle \cop(0) A^\dagger_{\bar\beta} (\tht_2 )A^\dagger_\alpha (\tht_1)\rangle ,
\end{equation}
where the two kinks have opposite chirality.  The contribution of the
former to the spectral function is only felt at energies $\omega >
2^{3/2}\Delta$ while the contribution of the second is seen at $\omega
> 2\Delta$.  We thus focus on the latter.

\newcommand{\cs}{(C \sigma^{ab})_{\alpha\bar\beta}}
\newcommand{\sco}{(\sigma^{ab}C)_{\alpha\bar\beta}}
\newcommand{\csa}{(C \sigma^{ab})_{\bar\alpha\beta}}
\newcommand{\scoa}{(\sigma^{ab}C)_{\bar\alpha\beta}}

We begin by identifying the group theoretical structure of
$\fck$:
\begin{equation}\label{eIVxxxvii}
\fckt = \cs f_\mu (\tht_1,\tht_2) ,
\end{equation}
where $C$ is the charge conjugation matrix introduced previously.
$\cs$ is antisymmetric in $a,b$.
Now $\cs$
is not the only obvious choice of tensor.  One also has
$\sco$, antisymmetric combinations built up out of $\gamma^a C \gamma^b$
or $C\gamma^a C \gamma^b C$, or some combination of all three (but not
$\sigma^{ab}$ as this choice violates obvious U(1) conservation).  Perhaps
the most natural choice is to make $\fck$ explicitly symmetric in C:
\begin{equation}\label{eIVxxxviii}
\fckt = (\cs + \sco ) f_\mu (\tht_1,\tht_2).
\end{equation}
However this forces $_\mu f^{(2a-1)(2a)}_{\alpha\bar{\beta}}$ to zero, something we do
not necessarily expect.
We instead arrive at the choice as given above by explicitly checking the other possibilities for invariance 
under $SO(6)$.  Of the choices only $\cs$ is consistent as such.

Having specified the coupling of a kink+anti-kink to $\cop$, we use $C$-symmetry to specify
the corresponding anti-kink+kink form-factor $\fckta$.  Under the action of the charge
conjugation matrix, $C$, the currents $\cop$ transform as
\begin{equation}\label{eIVxxxix}
C\cop C^\dagger = (-1)^{a+b}\cop.
\end{equation}
We thus expect
\begin{equation}\label{eIVxl}
\fckta = -(-1)^{a+b}(-1)^{\bar{\alpha}_2+\beta_2}~\fckt.
\end{equation}\label{eIVxli}
An ansatz for $\fckta$ satisfying this condition is
\begin{equation}
\fckta = \csa f_\mu (\tht_1,\tht_2).
\end{equation}
In particular we note the absence of any phase in the expression for $\fckta$ relative to that of $\fckt$.

As before Lorentz invariance demands
\begin{equation}\label{eIVxlii}
f_\mu (\tht_1,\tht_2) =
(e^{(\tht_1+\tht_2)/2} - (-1)^\mu e^{-(\tht_1+\tht_2)/2})f(\tht_{12}) .
\end{equation}
Then by the scattering axiom, $f(\tht_{12})$ must satisfy
\begin{eqnarray}\label{eIVxliii}
f(-\tht ) &=& u_1 (\tht) {1 - \tht/i\pi \over 3/2 - \tht/i\pi} f(\tht), \cr\cr
&=& \exp\bigg[\int^\infty_0 {dx \over x} G_c(x) \sinh({\tht x \over i\pi})\bigg]f(\tht );\cr\cr
G_c(x) &=& {e^{x/2}-1 \over \sinh(x) }.
\end{eqnarray}

The periodicity axiom takes the form
\begin{equation}\label{eIVxliv}
_\mu f^{ab}_{\bar{\beta}\alpha}(\tht_2,\tht_1 ) = -_\mu f^{ab}_{\alpha{\bar\beta}}(\tht_1 - 2\pi i,\tht_2) .
\end{equation}
The minus sign arises from $Q_{\alpha\bar\beta}R_{\alpha,G^{ab}} = -1$.
As shown in Appendix B, $Q_{\alpha\bar\beta}=-1$ and
$R_{\alpha,G^{ab}}=1$ for neutral currents and vice-versa for currents
carrying charge.

In terms of the scalarized form
factor, $f$, the periodicity axiom then reads
\begin{equation}\label{eIVxlv}
f(-\tht ) = f(\tht - 2\pi i).
\end{equation}
Two kinks of opposite chirality do not form a bound state.
Thus the form factor
should have no poles in the physical strip.  Thus with Eqns.(\ref{eIVxlii}),(\ref{eIVxliii}), 
and (\ref{eIVxlv}),
the form factor is 
\begin{eqnarray}\label{eIVxlvi}
\fckt &=& \cr\cr
&& \hskip -.8in = A_c \cs \bigl (
e^{(\tht_1+\tht_2)/2} - (-1)^\mu e^{-(\tht_1+\tht_2)/2}
\bigr )\cr\cr
&& \hskip -1.in 
\times \exp \bigg[\int^\infty_0 {dx \over x} {G_c(x) \over s(x)}
\sin^2({x\over 2\pi}(i\pi + \tht_{12}))\bigg] .
\end{eqnarray}
The hermiticity condition for this current form-factor reads
\begin{eqnarray}\label{eIVxlvii}
f^{ab}_{\alpha\bar{\beta}}(\theta_1,\theta_2)^* = f^{ab}_{\beta\bar{\alpha}}(\theta_2-i\pi,\theta_1-i\pi) C_{\beta\bar{\beta}}
C_{\bar{\alpha}\alpha}
\end{eqnarray}
Applying this constraint shows that
$A_c$ is some purely imaginary constant.

\subsection{Two particle form factors for $SO(6)$ 
symmetric tensor fields}
Here we will compute the two particle form factor coupling to an operator
that transforms as a scalar under Lorentz transformations and as the
10 dimensional symmetric representation under $SO(6)$ (we indicate
this operator as $\psi^{[abc]_A}$ where $a,b,c$ are anti-symmetrized).
We also assume that the operator's conformal dimension is $\Delta =
3/8$.  Such an operator arises in the computation of the spin response
for the ladders when considering fermion bilinears of the form
$c^\dagger_{Rb\uparrow}c_{Lb\downarrow}$ or antisymmetric combinations
$c^\dagger_{Rb\uparrow}c_{Lab\downarrow} -
c^\dagger_{Rab\uparrow}c_{Lb\downarrow}$. 
Each of the
fermions transforms under the ${\bf 4}$ of $SO(6)$ ($SU(4)$).  
These bilinears are then in the ${\bf 10}$
appearing in ${\bf 4_R}\otimes {\bf 4_L}$.
Here the elements of ${\bf 4_R}$ are 
$(c_{Rb\uparrow},c_{Rb\downarrow},c_{Rab\uparrow},c_{Rab\downarrow})$
while the elements of ${\bf 4_L}$ are
$(c^\dagger_{Lb\downarrow},c^\dagger_{Lb\uparrow},c^\dagger_{Lab\downarrow},c^\dagger_{Lab\uparrow})$.
We can see, for example, that $c^\dagger_{Rb\uparrow}c_{Lb\downarrow}$ lies
in the ${\bf 10}$ as in its bosonized form,
$$ 
c^\dagger_{Lb\uparrow}c_{Rb\downarrow} \sim e^{i\frac{1}{2}(\theta_{+c} - \theta_{-c} - \phi_{+s}+\phi_{-s})}
= e^{i\frac{1}{2}(\theta_1 - \phi_2 + \phi_3 - \phi_4)}.
$$
it carries all three of the quantum numbers in $SO(6)$.  In considering $\psi^{[abc]_A}$,
we rewrite the $SO(6)$ fields appearing in the bosonization
as 
$$
e^{i\pm\phi/2} \sim \chi_2 \pm i \chi_1,
$$
and so work in a Majorana like basis.  In this basis we need to know
how to construct the ${\bf 10}$ out of the ${\bf 6}$.  From
Ref. \onlinecite{slansky}, we know the antisymmetric product of three
${\bf 6}$'s yields the ${\bf 10}$.  We thus denote this field as
$\psi^{[abc]_A}$ where the $A$ indicates anti-symmetrization.

At the two particle level, two kinks of the
same chirality will couple to such an operator.  We will consider for
the moment $\psi^{[abc]_A}$ coupling to two even chirality kinks.  
This then leads to form factors of the type
\begin{equation}\label{eIVxlviii}
f^{[abc]_A}_{\alpha\beta}(\tht_1,\tht_2) = \langle \psi^{[abc]_A} 
A^\dagger_\beta(\tht_2)A^\dagger_\alpha(\tht_1)\rangle .
\end{equation}
$SO(6)$ covariance suggests that $f^{[abc]_A}_{\alpha\beta}(\tht_1,\tht_2)$ take the
form
\begin{eqnarray}\label{eIVxlix}
f^{[abc]_A}_{\alpha\beta}(\tht_1,\tht_2) &=& \bigg( c_1(C\gamma^{[a}\gamma^b\gamma^{c]})_{\alpha\beta} \cr\cr
&& \hskip -.85in + c_2 (\gamma^{[a}C\gamma^b\gamma^{c]})_{\alpha\beta}
 + c_3 (\gamma^{[a}\gamma^b C \gamma^{c]})_{\alpha\beta}\cr\cr
&& \hskip -.85in + c_4 (\gamma^{[a}\gamma^b \gamma^{c]} C)_{\alpha\beta}\bigg) f(\tht_{12}).
\end{eqnarray}
Here the $f^{[abc]_A}$ is a function of $\tht_1-\tht_2$ because $\psi^{[abc]_A}$ is a Lorentz scalar.

We point out that the $SO(6)$ tensors in the above expression are symmetric
in $\alpha$ and $\beta$.
To fix the values of the various coefficients we have two arguments available to us.
By checking explicitly for $SO(6)$ covariance we see that all of the coefficients but
$c_1$ are zero.  
We can also 
fix the values of $c_i$ relying on U(1) charge conservation alone.
Suppose $|\alpha\rangle = (1/2,\pm 1/2,\pm 1/2)$ and
$|\beta\rangle = (1/2,\pm 1/2,\pm 1/2)$.  Then by rewriting the Majorana
fields as Dirac fields, one readily finds that
\begin{equation}\label{eIVl}
f^{[1bc]_A}_{\alpha\beta} =  -i f^{[2bc]_A}_{\alpha\beta} ,
\end{equation}
provided $b,c = 3,4,5,$ or $6$.
As can be checked, this yields $c_2 = c_3 = c_4 = 0$.  We thus obtain in the end
\begin{eqnarray}\label{eIVli}
f^{[abc]_A}_{\alpha\beta}(\tht_1,\tht_2) &=& (C\gamma^{[a}\gamma^b\gamma^{c]})_{\alpha\beta} f(\theta_1-\theta_2).
\end{eqnarray}
where we have absorbed a constant into $f(\tht )$.

To determine $f(\tht )$ we apply the scattering and periodicity
axioms.  The scattering axiom constrains $f(\tht )$ as follows:
\begin{eqnarray}\label{eIVlii}
f(-\tht ) &=& (\frac{1}{4}u_0(\tht) - \frac{3}{4}u_2(\tht))f(\tht )\cr\cr
&=& -{\Gamma(3/4+\tht/2i\pi)\Gamma(1-\tht/2i\pi) \over \Gamma(3/4-\tht/2i\pi)\Gamma(1+\tht/2i\pi)}
f(\tht) \cr\cr
&=& -\exp\bigg[\int^\infty_0 
{dx \over x} G_s(x) \sinh(x\tht/i\pi)\bigg] f(\tht);\cr\cr
G_s(x) &=& -2e^{-3x/4}{\sinh (x/4)\over \sinh (x)}.
\end{eqnarray}
The periodicity axioms reads 
\begin{equation}\label{eIVliii}
f(-\tht) = Q_{\alpha\beta} R_{\alpha \psi } f(\tht - 2\pi i) .
\end{equation}
Both phases, $Q_{\alpha\beta}$ and $R_{\alpha\psi}$, are equal to 
1 (Appendix B). 
With this we arrive at the following expression for $f(\tht )$
\begin{equation}\label{eIVliv}
f(\tht ) = \sinh(\frac{\tht}{2})
\exp \bigg[\int^\infty_0 {dx \over x} {G_s(x) \over s(x)}
\sin^2({x\over 2\pi}(i\pi + \tht))\bigg] .
\end{equation}
Although two kinks can form a bound state, they do so in the antisymmetric (vector)
channel and so do not couple to $\psi^{[abc]_A}$.  Thus there is no need to add
additional poles to the form factor.
We then can put everything together leading to
\begin{eqnarray}\label{eIVlv}
f^{[abc]_A}_{\alpha\beta}(\tht_1,\tht_2) 
&=& A_s (\gamma^{[a}\gamma^b \gamma^{c]} C)_{\alpha\beta} \sinh(\frac{\tht_{12}}{2})\cr\cr
&& \hskip -1.in \times 
\exp \bigg[\int^\infty_0 {dx \over x} {G_s(x) \over s(x)}
\sin^2({x\over 2\pi}(i\pi + \tht_{12}))\bigg],
\end{eqnarray}
where $A_s$ is a real constant (as determined by hermiticity).

The UV asymptotics of this form factor are easily found to be
\begin{equation}\label{eIVlvi}
\lim_{\tht_1\rightarrow \infty} f^{[abc]_A}_{\alpha\beta}(\tht_1,\tht_2) \sim e^{\frac{3}{8}\tht_1}.
\end{equation}
Given that the conformal dimension of $f^{[abc]_A}$ is $\Delta=3/8$, we see that
the UV behavior of this form factor saturates its upper bound.

We now return to the situation of two odd kinks coupling to $f^{[abc]_A}$.  We thus want to consider
the form factors 
\begin{equation}\label{eIVlvii}
f^{[abc]_A}_{\bar{\alpha}\bar{\beta}}(\tht_1,\tht_2) = \langle \psi^{[abc]_A} A_{\bar{\beta}}(\tht_2)
A_{\bar{\alpha}(\tht_1)}\rangle .
\end{equation}
Under charge conjugation, the fields, $f^{[abc]_A}$, transform as
\begin{equation}\label{eIVlviii}
f^{[abc]_A} = -(-1)^{a+b+c} f^{[abc]_A}
\end{equation}
We thus expect
\begin{equation}\label{eIVlix}
f^{[abc]_A}_{\bar{\alpha}\bar{\beta}}(\tht_1,\tht_2) = (-1)^{a+b+c}(-1)^{\alpha_2+\beta_2}
f^{[abc]_A}_{\alpha\beta}(\tht_1,\tht_2)
\end{equation}
We may satisfy this constraint by choosing
$f^{[abc]_A}_{\bar{\alpha}\bar{\beta}}(\tht_1,\tht_2)$ to have the
same form as $f^{[abc]_A}_{\alpha\beta}(\tht_1,\tht_2)$, i.e.
\begin{eqnarray}\label{eIVlx}
f^{[abc]_A}_{\bar{\alpha}\bar{\beta}}(\tht_1,\tht_2) &=& 
(C\gamma^{[a}\gamma^b\gamma^{c]})_{\bar{\alpha}\bar{\beta}} f(\tht_{12}).
\end{eqnarray}
Here $f(\tht_{12})$ is the same as before as all the other axioms
governing the form factor remain unchanged.

\newcommand{\fff}{\psi^{abc}}
\newcommand{\ffkk}{f^{abc}_{\alpha\beta}}
\newcommand{\ffkkt}{f^{abc}_{\alpha\beta}(\tht_1,\tht_2)}

\subsection{Two particle form factors for $SO(6)$ vector fields}
\newcommand{\ff}{\psi^a}
\newcommand{\fkk}{f^a_{\alpha\beta}}
\newcommand{\fkkt}{f^a_{e\alpha\beta}(\tht_1,\tht_2)}
\newcommand{\fokkt}{f^a_{o\bar{\alpha}\bar{\beta}}(\tht_1,\tht_2)}
\newcommand{\cg}{(C\gamma^a)_{\alpha\beta}}
\newcommand{\gc}{(\gamma^aC)_{\alpha\beta}}
\newcommand{\fpm}{{f_e(\tht_1,\tht_2)}}
\newcommand{\fo}{f}

In this section we are interested in computing the two particle form
factors for an operator, $\ff$, transforming as both a Lorentz scalar and a vector
under SO(6).  For specificity we consider such an operator arising from
the following symmetric combinations
of fermion bilinears
$c^{\dagger}_{Rb\uparrow}c_{Lab\downarrow} +
c^{\dagger}_{Rab\uparrow}c_{Lb\downarrow}$.
This operator has conformal dimension, $\Delta = 3/8$.
At the two particle level, two kinks of the same chirality will again
couple to such an operator.  We first consider the case of two kinks of even chirality
coupling to this operator.

We thus consider form factors,
$\fkkt$, defined by
\begin{equation}\label{eIVlxi}
\fkkt = \langle \psi^a (0) A_\beta(\tht_2)A_\alpha(\tht_1)\rangle .
\end{equation}
Covariance under $SO(8)$ suggests that $\fkk$ takes the form
\begin{equation}\label{eIVlxii}
\fkkt = [ c_1 \gc + c_2 \cg ]\fpm ,
\end{equation}
where $c_1$ and $c_2$ are constants.  (That $c_1$ and $c_2$ are not more
generally independent functions of $\tht_1$ and $\tht_2$ is easily seen;
the constraints Eqn.(\ref{eIVxxii}) and Eqn.(\ref{eIVxxiii}) do not allow it.)
$c_1$ and $c_2$ can be fixed easily.  Suppose
$|\alpha\rangle = (1/2,1/2,1/2)$ and
$|\beta\rangle = (1/2,-1/2,-1/2)$.  Then by rewriting the Majorana
fields as Dirac fields, one readily finds that
\begin{equation}\label{eIVlxiii}
f^1_{e\alpha\beta} =  -i f^2_{e\alpha\beta} .
\end{equation}
This in turn forces $c_1 =0$.  We then set $c_2 = 1$ as we 
are uninterested at this point in an overall normalization.

As the operator is a Lorentz scalar, we must have
\begin{equation}\label{eIVlxiv}
f_e(\theta_1,\theta_2) = f_e (\tht_{12}) .
\end{equation}
$f_e (\tht)$ is then constrained by the scattering axiom.  Using the form of
the kink-kink S-matrix for kinks of the same chirality together
with the anti-symmetry of $\cg$ in $\alpha$ and $\beta$ we obtain
\begin{eqnarray}\label{eIVlxv}
f_e (-\tht ) &=& {\Gamma(1-\tht/2i\pi)\Gamma(\tht/2i\pi + 3/4) 
\over \Gamma(1+\tht/2i\pi)\Gamma(3/4-\tht/2i\pi)}\cr\cr
&& \hskip 1in \times {1/4+\tht/2i\pi \over -1/4+\tht/2i\pi}
f_e (\tht ) \cr\cr
&=& -\exp\bigg[\int^\infty_0 {dx\over x}G_v(x) \sinh({x\tht \over
    i\pi})\bigg] f_e (\tht);\cr\cr  
G_v(x) &=& \frac{2}{1-e^{-2x}}\bigg( e^{-2x}(1-e^{x/2})\cr\cr
&& \hskip .85in -e^{-5x/2}(1-e^{2x})\bigg).
\end{eqnarray}
We now apply the periodicity axiom. It takes the form
\begin{equation}\label{eIVlxvi}
f^a_{e\beta\alpha}(-\tht) = Q_{\alpha\beta} R_{\alpha \psi } f^a_{e\alpha\beta}(\tht - 2\pi i) .
\end{equation}
As discussed in Appendix B, we find that for even kinks the product of
phases $Q_{\alpha\beta} R_{\alpha \psi }$ is equal to 1.  With this the
periodicity axiom reduces to
\begin{equation}\label{eIVlxvii}
f_e(-\tht) = -f_e(\tht - 2\pi i).
\end{equation}
The presence of a minus sign (compare Eqn.(\ref{eIVlviii})) results from the antisymmetry of $C\gamma^a$.
The periodicity and scattering axioms then imply that $f_e(\tht )$ must have the minimal form
\begin{equation}\label{eIVlxviii}
f_e (\tht ) = \sinh (\tht )\exp\bigg[\int^\infty_0 {dx \over x} {G_v(x) \over \sinh(x)}
\sin^2 ({x\over 2\pi}(i\pi + \tht))\bigg].
\end{equation}
As two kinks of the same chirality form a fermionic bound
state, $f^a_{e\alpha\beta}$ should have a pole at
$\tht_2 = \tht_1 + i u^a_{\alpha\beta} = \tht_1 + i\pi /2$.
Thus $f^a_{e\alpha\beta}$ becomes,
\begin{eqnarray}\label{eIVlxix}
\fkkt &=&  A_v \cg {\sinh(\tht_{12}) \over \cosh(\tht_{12} ) }
\times\cr\cr\cr
&& \hskip -.75in \exp\bigg[\int^\infty_0 {dx \over x} {G_v(x) \over \sinh(x)}
\sin^2 ({x\over 2\pi}(i\pi + \theta_{12}))\bigg],
\end{eqnarray}
Here $A_v$ is the some normalization with mass dimension $[m]^{3/8}$.  
We now consider the situation of two odd kinks coupling to $\psi^a$.
We thus want to consider form factors of the type
\begin{equation}\label{eIVlxx}
f^{a}_{o\bar{\alpha}\bar{\beta}}(\tht_1,\tht_2) = \langle \psi^{a}
A_{\bar{\beta}}(\tht_2) A_{\bar{\alpha}(\tht_1)}\rangle .
\end{equation}
All but the periodicity axiom is of same form as for the two even
kinks. For the two odd kinks, the periodicity axiom changes to
\begin{equation}
f_o(-\tht) = f_o(\tht - 2\pi i).
\end{equation}
The difference with Eqn.(\ref{eIVlxvii}) arises because the OPE of the 
odd kinks with the operator produces an additional sign.
This sign changes the form factor involving two odd kinks to be
\begin{eqnarray}\label{eIVlxxii}
\fokkt &=& \frac{A_v}{\sin(\pi/4)} \cg {\sinh(\tht_{12}/2 ) 
\over \cosh(\tht_{12} ) }\times\cr\cr\cr
&& \hskip -.75in \exp\bigg[\int^\infty_0 {dx \over x} {G_v(x) 
\over \sinh(x)}\sin^2 ({x\over 2\pi}(i\pi + \theta_{12}))\bigg],
\end{eqnarray}
Here $A_v$ is the same normalization as for the even kinks.
We can fix
the normalization of $\fokkt$ relative to $\fkkt$ because both form
factors must have an identical bound state pole structure.  This then
mandates the relative factor, $\sqrt{2}$. To determine the phase of
$A_v$ we again employ the hermiticity condition:
\begin{equation}\label{eIVlxxiii}
\fkkt^* = C_{\bar{\beta}\beta} C_{\bar{\alpha}\alpha} 
f^a_{e\bar{\beta} \bar{\alpha}}(\tht_2 - i\pi , \tht_1 - i\pi).
\end{equation}
This implies $A_v$ is real.  

We can compute the asymptotics of these form factors.  We find
that for the even kinks, $\lim_{\tht_i \rightarrow \infty}\fkkt$
behaves as $\exp(3\tht_i/8)$. Thus the bound is saturated as we found
in the case of two kinks coupling to a field transforming as the
$SO(6)$ 10 dimensional symmetric representation. However for the odd
kinks we find instead that $\lim_{\tht_i \rightarrow \infty}\fokkt =
\exp(-1/8\tht_i)$. 

If we had considered an operator different than 
$c^{\dagger}_{Rb\uparrow}c_{Lab\downarrow} +
c^{\dagger}_{Rab\uparrow}c_{Lb\downarrow}$, but still
transforming as a SO(6) vector, we would arrive at the same results
with the possible caveat that the form factors of the even and odd kinks
may be swapped.  For practical purposes this is a distinction without
a difference.  In any sum over form factors that would appear in a computation
of a correlation function, the form factors involving both even and odd kinks
would appear with equal weight.

\subsection{One particle form factor of $SO(6)$ vector fields}
The $SO(6)$ vector field will couple to a single Gross-Neveu fermion, $a$.
So we consider the form factor,
\begin{equation}\label{eIVlxxiv}
f^a_b(\tht)  = \langle \psi^a (0) A^\dagger_b (\tht )\rangle ,
\end{equation}
which must take the form
\begin{equation}\label{eIVlxxv}
f^a_b(\tht) = c \delta^a_b.
\end{equation}
To determine the constant, we use the two particle form
factor $f^a_{\alpha\beta}$.  Let $|\alpha\rangle = (1/2,1/2,1/2)$ and
$|\beta\rangle = (1/2,-1/2,-1/2)$.  $A_a$, $a=1$, can be written
in terms of these two even chirality kinks:
\begin{equation}\label{eIVlxxvi}
i g^a_{\alpha\beta} A_a (\tht) = {\rm res}_{\delta = 0}
A_{\alpha}(\tht + \delta + i \bar{u}^{\bar\alpha}_{\beta a})
A_{a}(\tht - i \bar{u}^{\bar{\beta}}_{\alpha a}).
\end{equation}
Again the $u$'s
mark out poles in the S-matrix indicative of bound states.  Here they are given
by
\begin{equation}\label{eIVlxxvii}
\bar{u}^{\bar{\alpha}}_{\beta a} = \bar{u}^{\bar{\beta}}_{\alpha a}
= \pi - u^{\bar{\beta}}_{\alpha a} = \pi/4.
\end{equation}
$g^a_{\alpha\beta}$ can be determined up to a phase
from the kink-kink S-matrix
as before to be
\begin{equation}\label{eIVlxxviii}
g^a_{\alpha\beta} = 
\bigg(\frac{2}{3}\sqrt{\pi} \frac{\Gamma (7/4)}{\Gamma (5/4)}\bigg)^{1/2} .
\end{equation}
Given that,
\begin{equation}\label{eIVlxxix}
i g^a_{\alpha\beta} f^a_a(\tht) = {\rm res}_{\delta = 0} ~ {_\pm}
f^a_{e\alpha \beta} (\tht - i\pi/4 , \tht + \delta + i\pi/4) ,
\end{equation}
together with the hermiticity constraint,
\begin{equation}\label{eIVlxxx}
{(f^a_b)}^*(\tht) =  f^a_b(\tht),
\end{equation}
we find (up to a sign) the constant in Eqn.(\ref{eIVlxxv}) to be
\begin{eqnarray}\label{eIVlxxxi}
c &=& \frac{A_v}{g^a_{\alpha\beta}} \exp\bigg[-\int^\infty_0 \frac{dx}{x} \frac{G_v(x)}{\sinh(x)}\sinh^2(\frac{x}{4})\bigg].
\end{eqnarray}
If we repeat the above procedure using the two particle form-factor involving two odd chirality kinks, 
$f^a_{\bar{\alpha}\bar{\beta}}$, we obtain the same result.

\subsection{Summary of form factor results}

In this section we summarize the results of this section for the various
form factors.

\vskip 10pt

{\bf Two particle form factors:}

\vskip 10pt

The form factors of the SO(6) currents, $G^{ab}_\mu$, are:

\begin{eqnarray}\label{eIVlxxxii}
\fckt &=& \cr\cr
&& \hskip -.8in = A_c \cs \bigl (
e^{(\tht_1+\tht_2)/2} - (-1)^\mu e^{-(\tht_1+\tht_2)/2}
\bigr )\cr\cr
&& \hskip -1.in 
\times \exp \bigg[\int^\infty_0 {dx \over x} {G_c(x) \over s(x)}
\sin^2({x\over 2\pi}(i\pi + \tht_{12}))\bigg].
\end{eqnarray}

\vskip 10pt

The form factors of SO(6) symmetric tensor fields, $\psi^{[abc]_A}$, are:

\begin{eqnarray}\label{eIVlxxxiii}
f^{[abc]_A}_{\alpha\beta}(\tht_1,\tht_2) 
&=& f^{[abc]_A}_{\bar{\alpha}\bar{\beta}}(\tht_1,\tht_2) 
A_s (\gamma^{[a}\gamma^b \gamma^{c]} C)_{\alpha\beta} \sinh(\frac{\tht_{12}}{2})\cr\cr
&& \hskip -1.in \times 
\exp \bigg[\int^\infty_0 {dx \over x} {G_s(x) \over s(x)}
\sin^2({x\over 2\pi}(i\pi + \tht_{12}))\bigg].
\end{eqnarray}

\vskip 10pt

The form factors of SO(6) vector fields, $\psi^a$, involving
two even kinks are

\begin{eqnarray}\label{eIVlxxxiv}
\fkkt &=&  A_v \cg {\sinh(\tht_{12}) \over \cosh(\tht_{12} ) }
\times\cr\cr\cr
&& \hskip -.75in \exp\bigg[\int^\infty_0 {dx \over x} {G_v(x) \over \sinh(x)}
\sin^2 ({x\over 2\pi}(i\pi + \theta_{12}))\bigg],
\end{eqnarray}
and for two odd kinks change to
\begin{eqnarray}\label{eIVlxxxv}
\fokkt &=& \frac{A_v}{\sin(\pi/4)} \cg {\sinh(\tht_{12}/2 ) 
\over \cosh(\tht_{12} ) }\times\cr\cr\cr
&& \hskip -.75in \exp\bigg[\int^\infty_0 {dx \over x} {G_v(x) 
\over \sinh(x)}\sin^2 ({x\over 2\pi}(i\pi + \theta_{12}))\bigg].
\end{eqnarray}
We note that depending on the particular vector field considered the
form factors of even and odd kinks may be swapped.

\vskip 10pt

{\bf One particle form factors:}

\vskip 10pt

The one particle form factor for SO(6) vector fields is

\begin{equation}\label{eIVlxxxvi}
f^a_b(\tht) = c \delta^a_b,
\end{equation}
where $c$ is given by
\begin{eqnarray}\label{eIVlxxxvii}
c &=& \frac{A_v}{g^a_{\alpha\beta}} 
\exp\bigg[-\int^\infty_0 \frac{dx}{x} \frac{G_v(x)}{\sinh(x)}\sinh^2(\frac{x}{4})\bigg].
\end{eqnarray}

\section{Summary and Conclusions}
\label{sec:summ}
In this work we have calculated the dynamical magnetic
susceptibilities of doped Hubbard-like ladders in the framework of a
low-energy description by means of the SO(6) Gross-Neveu model. The
most prominent feature of the low-energy spin response is a narrow
peak at wave-vectors (along the leg direction) $K_{Fab}+ K_{Fb}$ and
$-K_{Fab}- K_{Fb}$. This peak corresponds to a fermionic bound state
of two SO(6) kinks, which is broadened by the gapless bonding charge
mode. This narrow peak sits on top of an incoherent scattering
continuum of two SO(6) kinks. The low-energy dynamic magnetic response
in the vicinity of the wave numbers $\pm 2K_{Fb}$ and $\pm 2K_{Fab}$ is
rather featureless. The response at low momenta and in the vicinity of
$\pm(K_{Fab}- K_{Fb})$ exhibits incoherent scattering continua with a
threshold singularity. 

Other dynamical correlation functions such as the single-particle
Green's function, density-density or superconducting correlators can
be determined along the same lines.

\acknowledgments
We are grateful to M. Karowski, A.A. Nersesyan, F. Smirnov and
A.M. Tsvelik for helpful discussions and communications. This work was
supported by the EPSRC under grant GR/R83712/01 (FHLE), the DOE under
contract DE-AC02-98 CH 10886 (RMK) and the Theory Institute for Strongly Correlated and
Complex Systems at Brookhaven National Laboratory (FHLE).
We thank the ICTP Trieste, where part of this work was carried out,
for hospitality.
\appendix

\section{Calculating the Dynamical Spin-Spin Correlators}
\label{app:spinspin}
In this appendix we present the detailed derivations of the functions
$J_{1,2,3}(\omega, k)$ in Eqns.(\ref{eIIIxi})-(\ref{eIIIxiii})
that determine the dynamical susceptibility. In subsection
{\bf A 4} we estimate the amplitudes $A_{ij}$ that determine the
relative spectral weights of the various contributions to the
structure factor.

\subsection{Calculation of $J_1(\omega,k)$}
The contribution $J_1(\omega,k)$ involves fields arising from
right/left symmetric combinations of fermion bilinears of the form 
$c^\dagger_{Rb\uparrow}c_{Rb\downarrow} +
c^\dagger_{Lb\uparrow}c_{Lb\downarrow}$.  These fields in turn are
expressible purely in terms of zeroth Lorentz component of the the
SO(6) Gross-Neveu currents.  For example 
\begin{eqnarray}\label{eAi}
c^\dagger_{Rb\uparrow}c_{Rb\downarrow} + c^\dagger_{Lb\uparrow}c_{Lb\downarrow} 
&=& \psi^\dagger_{2R}\psi_{3R} + \psi^\dagger_{2L}\psi_{3L}\cr\cr
&& \hskip -1in = \frac{1}{2}(-iG^0_{46} + G^0_{45} - G^0_{36} -iG^0_{35}).
\end{eqnarray}
We can then write down $J_1$ as
\begin{eqnarray}\label{eAii}
J_1(\omega,k) \!\!\!&=&\!\!\! \int d\tau dx e^{i(-i\omega+\epsilon)\tau - ikx}
\langle T G^0_{cd}(x,\tau)G^0_{cd}(0)\rangle .\cr
&&
\end{eqnarray}
Here $cd$ gives one SO(6) current.  Given that the different currents
that arise in expanding $c^\dagger_{Rb\uparrow}c_{Rb\downarrow} +
c^\dagger_{Lb\uparrow}c_{Lb\downarrow}$ give identical contributions
(up to a constant) we simply write $J_1$ in terms of a single current,
understanding that any degeneracy is fixed through absorption into the
associated pre-factor, $A_{j1}$, $j=1,2,3$.  Under a form factor
expansion, the lowest energy excitations contributing to $J_1$ are two
kinks of opposite chirality.  We first focus on the imaginary piece of $J_1$.
For the real piece it is easiest to obtain via a Kramers-Kronig transformation.

\begin{figure}
\begin{center}
\epsfxsize=0.45\textwidth
\epsfbox{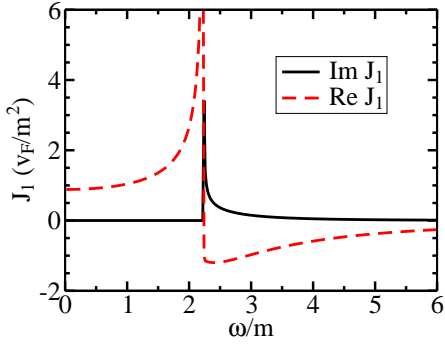}
\end{center}
\caption{The real and imaginary parts of the function $J_1(\omega,m/v_F)$ for low frequencies. For
general values of $q$ there is a square root singularity in the imaginary part above the
threshold, $\omega=\sqrt{v_F^2q^2+4m^2}$.}
\label{Fig:j1}
\end{figure}

The imaginary piece takes the form,
\begin{widetext}
\begin{eqnarray}\label{eAiii}
{\rm Im}\ J_1(\omega, k) &=& 
\pi^2 \sum_{\alpha\beta} \int {d\theta_1d\theta_2 \over (2\pi)^2} 
|f^{0cd}_{\bar{\alpha}\beta}(\theta_1,\theta_2)|^2
\delta(\omega-E(\theta_1,\theta_2))\
\delta(k-P(\theta_1,\theta_2))\ ,
\end{eqnarray}
\end{widetext}
where $_{R/L}f^{cd}_{\alpha\beta}(\theta_1,\theta_2)$ is the kink-kink
form factor for the current operator computed in Section 4 and
\begin{eqnarray}\label{eAiv}
E(\theta_1,\theta_2)&=&m\left[\cosh\theta_1+\cosh\theta_2\right],\nn
P(\theta_1,\theta_2)&=&\frac{m}{v_F}\left[\sinh\theta_1+\sinh\theta_2\right].
\end{eqnarray}
We mean the indices, $\bar{\alpha}$ and $\beta$, of  
the sum, $\sum_{\bar{\alpha}\beta}$, to run over both kinks and anti-kinks.
Performing the sum, we obtain
\begin{eqnarray}\label{eAv}
{\rm Im} J_1(\omega,k) &=& 8\pi^2\int {d\theta_1
d\theta_2 \over (2\pi)^2}  |f^0(\theta_1,\theta_2)|^2\nn
&&\hskip -.2in \times\delta(\omega-E(\theta_1,\theta_2))\
\delta(k-P(\theta_1,\theta_2)) ,
\end{eqnarray}
where (up to a constant which will be absorbed into the appropriate $A_{i1}$),
\begin{eqnarray}\label{eAvi}
|f^0(\theta_1,\theta_2)|^2 &=& \sinh^2\bigl(\frac{\theta_1+\theta_2}{2}\bigr) \cr\cr
&& \hskip -1.1in 
\times \exp\bigg[\int^\infty_0 \frac{dx}{x} \frac{G_c(x)}{s(x)}(1-c(x)\cos(\frac{\theta_{12}x}{\pi})\bigg],
\end{eqnarray}
and $G_c(x)$ is given in Eqn.(\ref{eIVxlvi}).  
We then perform the integrations over $\theta_1$ and $\theta_2$ obtaining
\begin{eqnarray}\label{eAvii}
&& {\rm Im } J_1(\omega,k) = \frac{8v_F\tilde{k}^2}{(\omega^2-\tilde{k}^2)^{3/2}}
\frac{\theta(\omega-\sqrt{\tilde{k}^2+4m^2})}{(\omega^2-\tilde{k}^2-4m^2)^{1/2}}\cr\cr
&&\ \times\ \exp\bigg[\int^\infty_0 \frac{dx}{x}
  \frac{G_c(x)}{s(x)}(1-c(x)\cos(\frac{\theta_{12}x}{\pi})\bigg],
\end{eqnarray}
where $\theta_{12}$ is given by 
\begin{equation}\label{eAviii}
\theta_{12} = \cosh^{-1}(\frac{\omega^2-\tilde{k}^2-2m^2}{2m^2}),
\end{equation}
and $\tilde{k} = v_F k$. We plot ${\rm Im} J_1(\omega,m/v_F)$ in
Fig.\ref{Fig:j1}.

\subsection{Calculation of $J_2(\omega,k)$}

We now consider the computation of $J_2$.  This contribution arises
from considering fermion bilinears of the form
$c^\dagger_{Rb\uparrow}c_{Lb\downarrow}$.  These bilinears are
expressible as products of the gapless charge degrees of freedom with
fields transforming under a 10 dimensional symmetric representation of
$SO(6)$.  For example 
\begin{eqnarray}\label{eAix}
c^\dagger_{Lb\uparrow}c_{Rb\downarrow} &\sim& e^{\frac{i}{2}\theta_{+c}}e^{-\frac{i}{2}(\phi_{+s}-\phi_{-s}-\theta_{-c})}\cr\cr
&\sim& e^{\frac{i}{2}\theta_{+c}} \sum_{abc}c_{abc} \psi^{[abc]_A}
\end{eqnarray}
Here the fields $\psi^{[abc]_A}$ are defined and discussed both in
Sections 2 and 4.  The coefficients $c_{abc}$ do not need to be
determined exactly as the contribution of each $\psi^{[abc]_A}$ to
$J_2(\omega, k)$ will be identical up to some constant.  Any
undetermined constant will be fixed in the end by appealing to a sum rule.

\begin{figure}
\begin{center}
\epsfxsize=0.45\textwidth
\epsfbox{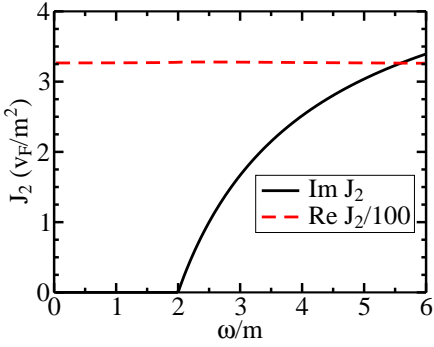}
\end{center}
\caption{The real and imaginary part of the
function, $J_2(\omega,0)$, for low frequencies and $K=0.95$.}
\label{Fig:j2}
\end{figure}

$J_2(x, \tau)$ is then the product of two correlation functions,
one coming from the gapless charge correlations and one coming from
the gapped $SO(6)$ sector.  We write
\begin{equation}\label{eAx}
J_2(x, \tau ) = J_{2g}(x, \tau)
J_{c}(x, \tau),
\end{equation}
where
\begin{eqnarray}\label{eAxi}
J_{c}(x,\tau ) &=& 
\langle T e^{-\frac{i}{2}\theta_{c+}(x,\tau)} e^{\frac{i}{2}\theta_{c+}(0,0)}\rangle\cr\cr
&=& A_c (x^2+\tau^2)^{K/4};\cr\cr
J_{2g}(x, \tau) &=& \langle
T \psi^{[abc]_A}(x,\tau)\psi^{[abc]_A}(0)\rangle.\nn
\end{eqnarray}
Here $K$ is the Luttinger parameter of the gapless charge excitations
and $A_c$ is some constant which we will absorb where appropriate into
the $A$'s of Eqns.(\ref{eIIIxi}-\ref{eIIIxiii}). We evaluate $J_{2g}$ in the same
fashion as $J_1$, namely we use employ a form factor expansion that is
truncated at the two kink level.  We thus obtain 
\begin{eqnarray}\label{eAxii}
J_{2g}(x, \tau) &=& \frac{1}{2} \sum_{\alpha\beta} \int
{d\theta_1d\theta_2 \over (2\pi)^2}  
|f^{{abc}_A}_{\alpha\beta}(\theta_1,\theta_2)|^2\nn
&&\hskip -.2in \times e^{i\tau(-i\omega+\epsilon)-|\tau|E(\theta_1,\theta_2)}
e^{ix(k-{\rm sign}(\tau) P(\theta_1,\theta_2))}\cr\cr
&=& \frac{1}{2\pi^2} \sum_{\alpha\beta}\int d\theta_- 
|f^{{abc}_A}_{\alpha\beta}(2\theta_-)|^2 \cr
&& \hskip -.3in \times K_0 (2m\cosh(\theta_-)((x/v_F)^2+\tau^2)^{1/2}).
\end{eqnarray}
Here $\alpha$ and $\beta$ are kinks of the same chirality.  In the second line of the equation
we have a made a change of variables to $\theta_\pm = (\theta_1\pm\theta_2)/2$ and performed the
$\theta_+$ integral.
Using the form factors of Section 4.4, we can write  $|f^{{abc}_A}_{\alpha\beta}(\theta_1,\theta_2)|^2$
as
\begin{eqnarray}\label{eAxiii}
|f^{{abc}_A}_{\alpha\beta}(\theta_1,\theta_2)|^2 &=&
\sinh(\frac{\theta_{12}}{2})^2\cr
&& \hskip -1.2in \times 
\exp\bigg[\int^\infty_0\frac{dx}{x}\frac{G_s(x)}{s(x)}(1-c(x)\cos(\frac{\theta_{12} x}{\pi}))\bigg].
\end{eqnarray}
Substituting Eqn.(\ref{eAxiii}) and Eqn.(\ref{eAxii}) into Eqn.(\ref{eAxi}), we can carry out
the $x-\tau$ integrals following Refs.\onlinecite{controzzi} and \onlinecite{book}.
Up to some constant (which we absorb into $A_{12}/A_{22}$), we obtain
\begin{widetext}
\begin{equation}\label{eAxiv}
J_2(\omega,k)= \frac{v_F}{m^2}\int_{-\infty}^\infty d\theta 
\frac{\sinh^2(\theta)}{\left(\cosh\theta\right)^\frac{4-K}{2}}
F\bigl(1-\frac{K}{4},1-\frac{K}{4},1,
\frac{\omega^2+\epsilon{\rm sign}(w)-\tilde{k}^2}{4m^2c^2(\theta)}\bigr)
\exp\bigg[\int^\infty_0\frac{dx}{x}\frac{G_s(x)}{s(x)}(1-c(x)\cos(\frac{2\theta x}{\pi}))\bigg].
\end{equation}
\end{widetext}
We plot the imaginary part of $J_2(\omega,0)$ in Fig.\ref{Fig:j2} for the special case $K=1$.

\subsection{Calculation of $J_3(\omega,k)$}

Finally we consider the calculation of $J_3 (\omega , k)$.  $J_3$
comes about from considering the correlations of fermion bilinears of the form 
$c^\dagger_{Rab\uparrow}c_{Lb\downarrow}+c^\dagger_{Rb\uparrow}c_{Lab\downarrow}$. 
In particular we have
\begin{eqnarray}\label{eAxv}
J_3 (\omega, k) &=&
\int d\tau dx e^{i(-i\omega+\epsilon) \tau - ikx}\langle 
{\cal  O}_3(x,\tau){\cal O}^\dagger_3(0,0)\rangle\ ,\nn
\end{eqnarray}
where 
\bea\label{eAxvi}
{\cal O}_3(x,\tau)=
c^\dagger_{Rab\uparrow}(x,\tau)c_{Lb\downarrow}(x,\tau)+b\leftrightarrow
ab\ .
\end{eqnarray}
The above {\it symmetric} combination of fermions is equivalent to the
{\it antisymmetric} field $\psi^a$ discussed in Section 4.  This
becomes clear under bosonization of these fermions 
\begin{equation}\label{eAxvii}
{\cal O}_3(x,\tau)=
2e^{-\frac{i}{2}(\theta_1+\phi_2)}\cos(\frac{1}{2}(\theta_3+\theta_4)).
\end{equation}
The model is symmetric under the $Z_2$ transformation
$\theta_{3,4}\rightarrow -\theta_{3,4}$.  It is clear that this field
couples to the the Gross-Neveu fermion, $\exp(i\phi_{R/L 2})$, whereas the
antisymmetric combination of Fermi fields,
$c^\dagger_{Rab\uparrow}c_{Lb\downarrow}-c^\dagger_{Rb\uparrow}c_{Lab\downarrow}$,
proportional to $\sin(\frac{1}{2}(\theta_3+\theta_4))$, does
not. The ${\bf 6}$ is then identified with symmetric combinations of
fermion bilinears. 

$J_{3}$ has a structure similar to that of $J_2$:
it is a product of correlation functions of both the gapped and gapless degrees of freedom,
\begin{equation}\label{eAxviii}
J_{3}(x, \tau) = J_{3g}(x, \tau') J_{c}(x,\tau).
\end{equation}
However the gapped part of the product is different in that it has contributions
at low energies from both single and two particle form-factors:
\begin{widetext}
\begin{eqnarray}\label{eAxix}
J_{3g}(x, \tau) &=&  \sum_a |c_a|^2 \int \frac{d\theta_1}{2\pi} \bigl|
\langle \psi^a(0)|A^\dagger_{a}(\theta_1)\rangle \bigr|^2
e^{-|\tau|\sqrt{2} m c(\theta_1)+ix{\rm sign}(\tau)\frac{\sqrt{2}m}{v_F}s(\theta_1)}\cr\cr
&+&\frac{1}{2}\sum_a |c_a|^2 \sum_{\alpha\beta} \int {d\theta_1d\theta_2 \over (2\pi)^2} 
\bigr|\langle
\psi^a(0)|A^\dagger_{\alpha}(\theta_1)A^\dagger_{\beta}(\theta_2)
\rangle\bigr|^2 
e^{-|\tau|E(\theta_1,\theta_2)+ix{\rm sign}(\tau)P(\theta_1,\theta_2)}\cr\cr
&=& J_{3g,1}(x, \tau) + J_{3g,2}(x,\tau)\ .
\end{eqnarray}
\end{widetext}
The contribution of the one-particle form-factor marks the most
significant difference between $J_{3}$ and $J_2$.  It indicates a
propagating coherent mode contributes to the spectral function. We do
note that this modes broadens out because of the gapless charge
excitations. The contribution of this mode has been previously studied
in Ref. \onlinecite{orignac}.
We now drop the sum, $\sum_{a}|c^a|^2$, as each term
gives an identical contribution up to some constant.
Using the form factors of Section 4.5 we can then write
\begin{eqnarray}\label{eAxx}
J_{3g,1}(x, \tau) &=& A^2_A \frac{d}{\pi} K_0(\sqrt{2}m(\tau^2+(x/v_F)^2))\cr\cr
d &=& \frac{1}{g^2}\exp[-2\int^\infty_0 \frac{dx}{x}\frac{G_v(x)}{s(x)}s^2(x/4)];\cr\cr
J_{3g,2}(x,\tau) &=& A^2_A \frac{4}{\pi^2}\int d\theta_- (\frac{\tanh(2\theta_-)^2}{2} +
\frac{\sinh(\theta_-)^2}{\cosh(2\theta_-)^2})\cr\cr
&& \hskip -.8in \times
K_0 (2m\cosh(\theta_-)((x/v_F)^2+\tau^2)^{1/2})\cr\cr
&& \hskip -.8in \times
\exp\bigg[\int^\infty_0\frac{dx}{x}\frac{G_v(x)}{s(x)}(1-c(x)\cos(\frac{2\theta_- x}{\pi}))\bigg].
\end{eqnarray}

\begin{figure}
\begin{center}
\epsfxsize=0.45\textwidth
\epsfbox{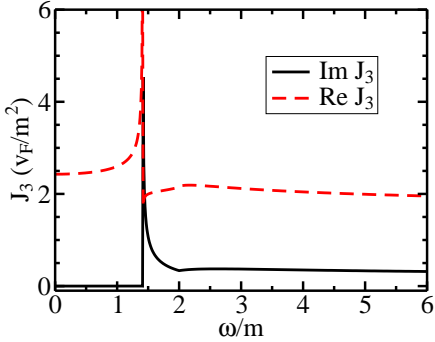}
\end{center}
\caption{The real and imaginary part of the 
function $J_3(\omega,0)$ for low frequencies and $K=0.95$. 
There is a singularity above the threshold at $\omega=\sqrt{2}m$. 
The kink-kink continuum contributes for $\omega>2m$. } 
\label{Fig:j3}
\end{figure}

Here the relative weights of $J_{3g,1}$ and $J_{3g,2}$ are fixed using the bootstrap as
described in Section 4.6.  $g$ is given by 
$$
g^2 = \frac{2\sqrt{\pi}\Gamma(7/4)}{3\Gamma(5/4)},
$$
via Eqn.(\ref{eIVlxxviii}). Again following Refs.\onlinecite{controzzi} and \onlinecite{book},
we can rewrite $J_3(\omega,k)$, up to some overall constant, in the form
\begin{widetext}
\begin{eqnarray}\label{eAxxi}
J_3(\omega,k)&=&
\frac{v_F}{2g^2m^2}\exp\bigg(-2\int^\infty_0 \frac{dx}{x}
  \frac{G_v(x)}{\sinh(x)}\sinh^2(\frac{x}{4})\bigg)
F\biggl(1-\frac{K}{4},1-\frac{K}{4},1,
\frac{\omega^2-\tilde{k}^2}{2m^2}\biggr)\nn
&+&\frac{2^{K/4}v_F}{\pi m^2}\int_{-\infty}^\infty d\theta
\frac{1}{\left(\cosh\theta\right)^\frac{4-K}{2}}\big(\frac{\tanh(2\theta)^2}{2} +
\frac{\sinh(\theta)^2}{\cosh(2\theta)^2}\big)
F\biggl(1-\frac{K}{4},1-\frac{K}{4},1,
\frac{\omega^2-\tilde{k}^2}{4m^2c^2(\theta)}\biggr)\cr
&& \hskip 2.5in 
\times \exp\bigg[\int^\infty_0\frac{dx}{x}\frac{G_v(x)}{s(x)}(1-c(x)\cos(\frac{2\theta
  x}{\pi}))\bigg]. \nn
\end{eqnarray}
\end{widetext}
We plot $J_3(\omega,0)$ in Fig.\ref{Fig:j3} for the case $K=1$. 
The dominant feature is a narrow peak corresponding to the SO(6)
fermionic bound state. The peak is not sharp due to the admixture of
excitations from the gapless charge sector.

\subsection{Estimating the Amplitudes $A_{ij}$}
\label{app:ampl}

We now turn to estimating the various normalization constants,
i.e. the $A$'s appearing in Eqns \ref{eIIIxi}-\ref{eIIIxiii}. 
In the present context,
perhaps the best way of determining them would be to compare the 
field theory results to numerical computations following the lines of
B\cite{JGE}. However, such computations are presently not
available. A crude way of determining the order of magnitude of the
A's is to compare our results to the susceptibility calculated in a
Random Phase Approximation. The latter is given by 
\begin{eqnarray}\label{eAxxii}
\chi^{RPA}(\omega,{\bf q})&=&\chi_{11}(\omega,q_x)+
\chi_{22}(\omega,q_x)\cr
&&+2\cos(q_y)\chi_{12}(\omega,q_x)\ \cr\cr
\chi_{ab}(\omega,q_x)&=&\left(\frac{\chi^{(0)}(\omega,q_x)}
{1+2U\chi^{(0)}(\omega,q_x)}\right)_{ab}.
\end{eqnarray}
Here $\chi^{(0)}(\omega,q_x)$ is the susceptibility matrix of
non-interacting electrons ($U=0$) on a two-leg ladder.
In order to fix the constants $A_{ij}$, we now require
\begin{eqnarray}\label{eAxxiii}
\int_0^{\Lambda}\frac{d\omega}{\omega}{\rm Im}\chi^{RPA}(\omega,{\bf q}) 
=\int_0^\Lambda\frac{d\omega}{\omega}{\rm Im}\chi(\omega,{\bf q}),
\end{eqnarray}
where $\Lambda$ is an appropriately chosen energy scale. The rationale
behind Eqn.(\ref{eAxxiii}) is the following. We don't expect the RPA to
give an accurate description of the susceptibilities at low
energies as it for example fails to account for the dynamical
generation of spectral gaps. However, we expect RPA to give a
reasonable account of the static susceptibility (which corresponds to
taking $\Lambda\to\infty$) and hence also of the integrated spectral
weight at low frequencies. Field theory does not give a good account
of the static susceptibility, which is reflected in the fact that
one cannot take $\Lambda\to\infty$ in the field theory calculation as
the resulting integral would diverge. We therefore arbitrarily set
$\Lambda=10 m$.
A useful consistency check is that the results for
$A_{ij}$ are essentially unaffected by small changes in $\Lambda$.
We choose to consider Eqn.(\ref{eAxxiii}) rather than the integrated
spectral weights in order to suppress as much as possible the
neglected contributions that occur at higher energies in the field
theory calculation. On the field theory side we have
\begin{eqnarray}\label{eAxxiv}
&&-\frac{1}{\pi}{\rm Im}\chi(\omega,q,\frac{\pi}{2})
\approx\frac{3(A_{11}+A_{21})}{4\pi}J_1(\omega,q),\nn
&&-\frac{1}{\pi}{\rm Im}\chi(\omega,2K_{Fb},\frac{\pi}{2})
\approx\frac{3A_{12}}{4\pi}J_2(\omega,0),\nn
&&-\frac{1}{\pi}{\rm Im}\chi(\omega,2K_{Fab},\frac{\pi}{2})
\approx\frac{3A_{22}}{4\pi}J_2(\omega,0),\nn
&&-\frac{1}{\pi}{\rm Im}\chi(\omega,K_{Fb}-K_{Fab}+q,\frac{\pi}{2})
\approx\frac{3A_{31}}{4\pi}J_1(\omega,q),\nn
&&-\frac{1}{\pi}{\rm Im}\chi(\omega,K_{Fb}+K_{Fab},\frac{\pi}{2})
\approx\frac{3A_{32}}{4\pi}J_3(\omega,0).
\end{eqnarray}
The relevant frequency integrals are
\begin{eqnarray}\label{eAxxv}
\int_0^{10m}\frac{d\omega}{\omega}J_1(\omega,\frac{5m}{v_F})
&\approx & 1.04\frac{v_F}{m^2} ,\cr\cr
\int_0^{10m}\frac{d\omega}{\omega}J_2(\omega,\frac{5m}{v_F})
&\approx & 1.79\frac{v_F}{m^2} ,\cr\cr
\int_0^{10m}\frac{d\omega}{\omega}J_3(\omega,\frac{5m}{v_F})
&\approx & 0.23\frac{v_F}{m^2}.
\end{eqnarray}
In order to compare to compare to the RPA calculation we need to
estimate the gap $m$ as a function of $t$, $t_\perp$ and $U$.
We do this in two steps. Firstly, the spin gap at half-filling
$\Delta_s(1)$ has been determined numerically for small values of
$U/t$ in Refs [\onlinecite{noack,weihong}]. 
For $t_\perp =1.5 t$ and $U=t$ it was found that  
\be\label{eAxxvi}
\Delta_s(1)\approx 0.05 t\ .
\ee
In order to determine the kink gap $m$ we need to know how $\Delta_s$
evolves under doping. As numerical results are not available in the
literature for small $U$, we instead use the results of
Ref. (\onlinecite{so8}), where the doping dependence of the spin gap
was analyzed within the framework of a low-energy description of the
undoped ladder in terms of the SO(8) Gross-Neveu model. At
half-filling we have $m_{\rm SO(8)}=\Delta_s(1)$. Assuming that the
Fermi velocity is well approximated by the non-interacting value
$v_F\approx 1.32 t$ (for $t_\perp = 1.5t$), we find that the gap
$m$ for SO(6) kinks at four different doping levels is roughly equal
to 
\vskip .1in
\begin{center}
\begin{tabular}{|l|l|l|}
\hline
doping & $m_{SO(6)} ~ (m_{\rm 1/2 filling})$ & K \\
\hline
1\% & .47 & .94 \\
5\% & .29 & .915 \\
10\% & .21 & .93 \\
25\% & .13 & .945 \\
\hline
\end{tabular}
\end{center}
\vskip .1in
\noindent In this table we have additionally listed the corresponding
Luttinger parameter for the different dopings, also available from
Ref. (\onlinecite{so8}). Having fixed $v_F$ and $m$ in the low-energy
effective SO(6) theory in terms of $t$ and $U$ we now can use the
comparison between the field theory and RPA calculations of the
integrated susceptibilities to estimate the amplitudes $A_{ij}$.
We find 

\vskip .1in
\begin{center}
\begin{tabular}{|l|l|l|l|l|l|}
\hline
doping & $A_{11} + A_{21}$ & $A_{12}$ & $A_{22}$ & $A_{31}$ & $A_{32}$ \\
\hline
1\%  & 0.24 & 0.13  & 0.10  & 0.14 & 6.61 \\
5\%  & 0.25 & 0.045 & 0.070 & 0.14 & 4.70 \\
10\% & 0.26 & 0.12 & 0.091 & 0.22 & 3.24 \\
25\% & 0.61  & 0.091 & 0.79  & 0.24 & 4.94 \\
\hline
\end{tabular}
\center{\hbox{\hskip .3in A table of the field theoretic amplitudes, $A_{ij}$,}}
\center{\hbox{\hskip .3in computed for $U=t$ and $t_\perp=1.5t$.}}
\end{center}
\vskip -.1in
\noindent All values for the $A$'s are in units of 
$\frac{m^2}{t^2}$. 

We thus have all the ingredients at hand to calculate the 
spin-spin correlation functions of the doped ladders.

\section{Derivation of Form Factor Axioms for General Operators in SO(6) Gross-Neveu}
\label{app:axioms}
\def\nn{\nonumber\\}
\def\tt{\tilde{\theta}}
\def\r#1{(\ref{#1})}
\def\sgn{{\rm sgn}}
\def\eps{\epsilon}

In this appendix we derive the form-factor axioms for SO(6) Gross-Neveu for general semi-local operators.
We take the following approach.  We first consider the general form-factor axioms as
written by Smirnov in Ref. (\onlinecite{smir}) for the current and stress energy operators.  These
are written for a particle basis which is not C-symmetric.  In this basis the form-factor axioms are not easily
generalized.  To surmount this difficulty we transform to a particle basis which is C-symmetric.  
Here the form-factor axioms can be readily written down for all semi-local operators.  In particular
the semi-locality phases appearing in the axioms can be computed using operator product expansions.
We then recover the form-factor axioms for general semi-local operators in the original basis
by inverting the transformation.  As an important consistency check of both the transformation and
the operator product expansions, we recover the axioms for the currents and stress-energy tensor as 
originally stated.

\subsection{The N-Particle Form Factor Axioms for the Currents and Stress-energy Tensor in SO(6) Gross Neveu}

The N-particle form factor axioms for the currents and the stress-energy tensor in SO(6) Gross-Neveu
are given by Smirnov \cite{smir} as follows:
\begin{widetext}
\bea\label{eBi}
f^{\cal O}(\ldots,\theta_{j+1},\theta_{j},
\ldots)_{\ldots ,\eps'_{j+1},\eps'_{j},\ldots}
&=& S^{\eps_{j}\eps_{j+1}}_{\eps'_j\eps'_{j+1}}
(\theta_{j}-\theta_{j+1})
f^{\cal O}(\ldots,\theta_j,\theta_{j+1},
\ldots)_{\ldots ,\eps_j,\eps_{j+1},\ldots};\cr\cr
f^{\cal O}(\theta_1,\ldots,\theta_{n}+2\pi i)_{\eps_1,\ldots  ,\eps_n} &=&
(-1)^{\frac{1}{4}\sum_{j}l[\epsilon_j]}
f^{\cal O}(\theta_n,\theta_1,\ldots,\theta_{n-1})_{\eps_n,\eps_1,\ldots ,\eps_{n-1}};\cr\cr
i{\rm Res}\bigl|_{\theta_n=\theta_{n-1}+i\pi}\
f^{\cal O}(\theta_1,\ldots,\theta_n)_{\eps_1,\ldots,\eps_n}
&=& f^{\cal O}(\theta_1,\ldots,\theta_{n-2})_{\eps'_1,\ldots,\eps'_{n-2}}\
{C}_{\eps_n,\eps'_{n-1}}\cr\cr
&& \hskip -2in \times \Bigl[\delta^{\eps_1'}_{\eps_1}\cdots
  \delta^{\eps_{n-1}'}_{\eps_{n-1}}
-(-1)^{\frac{1}{4}\sum_{j=1}^{n-2}l[\epsilon_j]}
{S}^{\eps'_{n-1}\eps'_1}_{\gamma_1\eps_1}(\theta_{n-1}-\theta_1)
{S}^{\gamma_{1}\eps'_2}_{\gamma_2\eps_2}(\theta_{n-1}-\theta_2)\ \cdots
{S}^{\gamma_{n-3}\eps'_{n-2}}_{\eps_{n-1}
\eps_{n-2}}(\theta_{n-1}-\theta_{n-2})\Bigr].  
\eea
\end{widetext}
The first equation is the generalization of Eqn.(\ref{eIVxxii}) of scattering axiom 
to matrix elements with N-particles.  The second equation is the N-particle
generalization of Eqn.(\ref{eIVxxiii}), the periodicity axiom.  The phase that appears on the r.h.s. of the
periodicity axiom is defined through
\be
l[\epsilon]=
\begin{cases}\label{eBii}
1 & \text{for even parity kinks}\cr
2 & \text{for fermions}\cr
3 & \text{for odd parity kinks},\cr
\end{cases}
\ee
where the $l[\epsilon_j]$ are the rank of the representation the particle, $\epsilon_j$, falls in.
For the operators to which these axioms apply, the phase is always either $\pm 1$ as $\sum l[\epsilon_j]$ is
always a multiple of 4.
While the above axioms are completely general, we will 
concentrate solely on form factors involving particles which are either
kinks or anti-kinks.  
The form factors have simple poles at the points
$\theta_n=\theta_j+i\pi$.  The residues of these poles are the subject of the final axiom.
Here $C$ is the charge conjugation matrix and by necessity
$l[\epsilon_n]+l[\epsilon_{n-1}]=4$.
The annihilation pole condition relates form factors with different numbers of particles.
In the annihilation pole axiom a phase similar to that of the periodicity axiom appears.

From one viewpoint, the phases in the latter two axioms
can be justified by insisting the currents and the stress-energy tensor
satisfy both certain properties under charge conjugation and 
appropriate commutation relations.
From another, the phases can be seen as a result of the fact that the $C$ is not symmetric, i.e. $C^2=-1$, 
and the non-trivial semi-locality between the non-Cartan currents and the fundamental kink
fields.   Understanding the phases in this fashion provide the most ready path to generalize
the axioms to other fields.  But because the phases have two sources, 
this path needs to be indirect.
The first step is to transform the particle basis to a basis where the corresponding charge
conjugation matrix is symmetric.  In this basis the form-factor axioms can be written down
readily -- the only additional phases that appear are semi-locality factors and these
are well understood. To obtain the form-factor axioms in the original basis we then invert the transform.

\subsection{Transformation to a C-symmetric basis}

The path we then intend to follow is the one Ref. (\onlinecite{smir}) 
uses for SU(2).  There the transformation
to a C-symmetric basis is straightforward because of the simplified particle content.  Here, in
contrast, the transformation is more involved.  We define the transformation to a C-symmetric basis 
as follows.  A n-particle state behaves under the transformation, ${\cal T}$, via
\begin{eqnarray}\label{eBiii}
\tilde{A}_{\eps_1}\cdots\tilde{A}_{\eps_n}|0\rangle &=& {\cal T} A_{\eps_1} \cdots A_{\eps_n}|0\rangle\cr\cr
&=& \big(\prod_{i<j}P_{\eps_i\eps_j}) A_{\eps_1} \cdots A_{\eps_n}|0\rangle ,
\end{eqnarray}
where the $P_{\eps_i\eps_j}=\pm 1$ are given by
\begin{eqnarray}\label{eBiv}
P_{\bar{\alpha}\beta} &=& P_{\beta\bar{\alpha}} = P_{\bar{\alpha}\bar{\beta}}, ~~~ \alpha \neq \beta ;\cr\cr
P_{\bar{\beta}\beta} &=& -P_{\beta\bar{\beta}} = -1;\cr\cr
P_{\alpha\alpha} &=& -P_{\bar{\alpha}\bar{\alpha}} = 1;\cr\cr
P_{\alpha\beta} &=& -P_{\bar{\alpha}\bar{\beta}};
\end{eqnarray}
and
\begin{eqnarray*}
P_{\alpha\beta}P_{\beta\alpha} &=& -1 ~~~\text{if $\alpha\neq\beta$};\cr\cr
P_{\bar{\alpha}\bar{\beta}}P_{\bar{\beta}\bar{\alpha}} &=& -1 ~~~\text{if $\bar{\alpha}\neq\bar{\beta}$}.
\end{eqnarray*}
To meet this last set of conditions, we arbitrarily set
\begin{eqnarray*}
P_{+++,-+-} &=& P_{+++,+--}=P_{+++,--+}=P_{-+-,+--}\cr\cr
&=& P_{+--,--+}=P_{--+,-+-}=-1.
\end{eqnarray*}
Here $\alpha,\beta$ correspond to even chirality kinks while $\bar{\alpha},\bar{\beta}$ correspond
to odd chirality kinks.  This choice of the transformation is not
unique but is equivalent in effect to other possible choices.

In moving to this transformed basis, the S-matrix also undergoes transformation.  The transformed
$\tilde S$ must satisfy
\begin{equation}\label{eBv}
\tilde{A}_{\eps_1}\tilde{A}_{\eps_2} = \tilde{S}_{\eps_1\eps_2}^{\eps'_1\eps'_2} \tilde{A}_{\eps'_2} \tilde{A}_{\eps'_1}
\end{equation}
This implies the relation
\begin{equation}\label{eBvi}
\tilde{S}_{\eps_1\eps_2}^{\eps'_1\eps'_2} = S_{\eps_1\eps_2}^{\eps'_1\eps'_2} P_{\eps_1\eps_2}P_{\eps'_2\eps'_1}
\end{equation}
By construction $\tilde S$ is C-symmetric.  In particular it satisfies
\begin{equation}\label{eBvii}
\tilde{S}^{\alpha\beta}_{\gamma\delta}(i\pi - \theta) = 
\tilde{S}^{\bar{\gamma}\beta}_{\bar{\alpha}\delta}(\theta ) \tilde{C}_{\bar{\gamma}\gamma} \tilde{C}_{\alpha\bar{\alpha}},
\end{equation}
where $\tilde{C}= -i\gamma_2\gamma_4\gamma_6$, or equivalently, 
${\tilde C}_{\alpha\bar{\alpha}} = C_{\alpha\bar{\alpha}}P_{\alpha\bar{\alpha}}$, is now a symmetric matrix.

This transformation, as constructed, satisfies an important consistency condition.  Considering the
relation involving three particle states,
\begin{equation}\label{eBviii}
\tilde{A}_{\eps_1}\tilde{A}_{\eps_2}\tilde{A}_{\eps_3} = \tilde{S}_{\eps_1\eps_2}^{\eps'_1\eps'_2} \tilde{A}_{\eps'_2} \tilde{A}_{\eps'_1}\tilde{A}_{\eps_3},
\end{equation}
we see that to transform this relationship covariantly to the original basis, i.e. 
\begin{equation}\label{eBix}
A_{\eps_1}A_{\eps_2}A_{\eps_3} = S_{\eps_1\eps_2}^{\eps'_1\eps'_2} A_{\eps'_2} A_{\eps'_1}A_{\eps_3},
\end{equation}
we require that for any $\eps_3$ that
\begin{equation}\label{eBx}
P_{\eps_1\eps_3}P_{\eps_2\eps_3} = P_{\eps'_1\eps_3} P_{\eps'_2\eps_3},
\end{equation}
holds for any $\eps_1,\eps_2,\eps'_1,\eps'_2$ such that $S_{\eps_1\eps_2}^{\eps'_1\eps'_2}$ is non-zero.
The transformation as given above does indeed satisfy this condition.  This condition guarantees
both that $\tilde{S}$ satisfies the Yang-Baxter equation and as we will see, that we can recover
the axioms given in the original basis from the axioms stated in the C-symmetric basis.

\subsection{Form Factor Axioms in C-Symmetric Basis}

In the C-symmetric basis the axioms for arbitrary semi-local operators have the following form:
\begin{widetext}
\bea\label{eBxi}
\tilde{f}^{\cal O}(\ldots,\theta_{j+1},\theta_{j},
\ldots)_{\ldots ,\eps'_{j+1},\eps'_{j},\ldots}
&=& \tilde{S}^{\eps_{j}\eps_{j+1}}_{\eps'_j\eps'_{j+1}}
(\theta_{j}-\theta_{j+1})
\tilde{f}^{\cal O}(\ldots,\theta_j,\theta_{j+1},
\ldots)_{\ldots ,\eps_j,\eps_{j+1},\ldots};\cr\cr
\tilde{f}^{\cal O}(\theta_1,\ldots,\theta_{n}+2\pi i)_{\eps_1,\ldots  ,\eps_n} &=&
e^{2\pi i \omega({\cal O},\eps_{n})}
\tilde{f}^{\cal O}(\theta_n,\theta_1,\ldots,\theta_{n-1})_{\eps_n,\eps_1,\ldots ,\eps_{n-1}};\cr\cr
i{\rm Res}\bigl|_{\theta_n=\theta_{n-1}+i\pi}\
\tilde{f}^{\cal O}(\theta_1,\ldots,\theta_n)_{\eps_1,\ldots,\eps_n}
&=& \tilde{f}^{\cal O}(\theta_1,\ldots,\theta_{n-2})_{\eps'_1,\ldots,\eps'_{n-2}}\
\tilde{C}_{\eps_n,\eps'_{n-1}}\cr\cr
&& \hskip -2in \times \Bigl[\delta^{\eps_1'}_{\eps_1}\cdots
  \delta^{\eps_{n-1}'}_{\eps_{n-1}}
-e^{2\pi i \omega({\cal O},\eps_{n-1})}
\tilde{S}^{\eps'_{n-1}\eps'_1}_{\gamma_1\eps_1}(\theta_{n-1}-\theta_1)
\tilde{S}^{\gamma_{1}\eps'_2}_{\gamma_2\eps_2}(\theta_{n-1}-\theta_2)\ \cdots
\tilde{S}^{\gamma_{n-3}\eps'_{n-2}}_{\eps_{n-1}
\eps_{n-2}}(\theta_{n-1}-\theta_{n-2})\Bigr].  
\eea
\end{widetext}

The phase factor $e^{2\pi i \omega({\cal O},\eps)}$ arises from the 
semi-locality of the fields in the \sos Gross-Neveu model. 
Working in Euclidean space, fields can be said to be semi-local if
their product sees a phase change when the fields are taken around one
another in the plane via the analytic continuation $z \rightarrow
ze^{i2\pi}, \bar{z} \rightarrow \bar{z}e^{-i2\pi}$ 
(in conventions where right-moving fields depend only on $\bar{z}$). 
More specifically we have
\begin{equation}\label{eBxii}
\psi_{\eps} (ze^{i2\pi},\bar{z}e^{-i2\pi}){\cal O}(0) = e^{2\pi i \omega({\cal O},\eps)}.
\psi_{\eps} (z,\bar{z}){\cal O}(0)
\end{equation}
Here $\psi_{\eps}$ is the field that is associated with the particle
$A_{\eps}^\dagger (\theta)$.  
$\psi_{\eps}$ is chosen in such a fashion that it is both self-local
and bosonic.  For the case at hand we represent the kinks through the vertex operators
\begin{equation}\label{eBxiii}
A_{\sigma_1,\sigma_2,\sigma_3} \leftrightarrow e^{\frac{i}{4}(\sigma_1\phi_2+\sigma_2\phi_3+\sigma_3\phi_4)},
\end{equation}
where here $\sigma = \pm 1$ and $\phi_i$, $i=2,3,4$, are the bosons as discussed in Section 2.
The choice of field, $\psi_{\eps}$, influences the phase, $e^{2\pi i \omega({\cal O},\eps)}$, as different
choices will have different operator product expansions (OPEs) with $\cal O$.
Our choice passes an important check: the phase, $e^{2\pi i \omega({\cal O},\eps)}$,
so obtained from this assignment allows one to recover the form-factors
axioms as stated in the original basis.

We now write down the semi-locality phases for the currents and the stress energy tensor.
The stress energy tensor, $T^{\mu\nu}$, is both chargeless and local.  Its OPE with the kink fields
as described above uniformly leads to $e^{2\pi i \omega(T^{\mu\nu},\eps)} = 1$.
The phases for the currents can be written in a compact form.  Any current, $J_{\alpha\bar{\beta}}$ can (group
theoretically) be thought of as a product of an even and odd kink, $\alpha$ and $\bar{\beta}$.
If the $\eps$ corresponds to a kink $\gamma$ (of arbitrary chirality) 
the OPE then leads to the phase being given by 
\begin{equation}\label{eBxiv}
e^{2\pi i \omega(J_{\alpha\bar{\beta}},\gamma)} = 
-P_{\alpha\gamma}P_{\gamma\alpha}P_{\bar{\beta}\gamma}P_{\gamma\bar{\beta}}.
\end{equation}
This can be checked directly.
With the phase in this form, we recover the form factors in their original form more readily.

\subsection{Recovery of Form-Factor Axioms in Non-C-Symmetric Basis}

We now show that we can recover the form-factor axioms for the currents and stress-energy tensor
as given in the original basis (i.e. Eqn.(\ref{eBi})) by transforming the form-factor axioms in the C-symmetric
basis (i.e. Eqn.(\ref{eBxi})).  This shows both that the transformation and our choice
of fundamental fields to represent the kinks in the C-symmetric basis are consistent.

The demonstration that we recover the scattering axiom, the first equation in Eqn.(ref{eBi}), is trivial.
It follows immediately from the complementary fashion by which the S-matrix transforms relative
to the basis (i.e. Eqn.(\ref{eBvi})) and the fact that the condition in Eqn.(\ref{eBx}) holds.

The periodicity axiom provides the first non-trivial check of the transformation.  We will first
consider the topologically neutral operators, for example, the Cartan currents and the stress-energy
tensor.  In this case the semi-locality phase is $1$.  If we apply the transformation to the
second line of Eqn.(\ref{eBxi}), we see that to recover the periodicity axiom as stated in its original form,
we require that
\begin{equation}\label{eBxv}
\prod_{i=1}^{n-1}P_{\eps_n\eps_i}P_{\eps_i\eps_n} = (-1)^{\frac{1}{4}\sum_{j}l[\epsilon_j]}.
\end{equation}
To establish this we first note the any set of particles coupling to a topologically neutral operator
can be decomposed (not uniquely) into the following three subsets:
\begin{eqnarray}\label{eBxvi}
&&\text{1. $n_e$ complete sets of the four even kinks,}\cr
&&\text{$\{(+++),(+--),(-+-),(--+)\}$;}\cr\cr
&&\text{2. $n_o$ complete sets of the four odd kinks,}\cr
&&\text{$\{(---),(-++),(+-+),(++-)\}$;}\cr\cr
&&\text{3. $n_p$ pairs of particle-antiparticles $\{\alpha,\bar{\alpha}\}$.}
\end{eqnarray}
We thus decompose the n-states as follows, $n = 4n_e + 4n_o + 2n_p$.
It is straightforward to show that the $P's$ of the transformation satisfy the
following conditions
\begin{eqnarray}\label{eBxvii}
\prod_{\alpha \in {\rm even~kinks}} P_{\alpha\gamma}P_{\gamma\alpha} &=& -1;\cr\cr
\prod_{\bar{\alpha} \in {\rm odd~kinks}} P_{\bar{\alpha}\gamma}P_{\gamma\bar{\alpha}} &=& -1;\cr\cr
P_{\alpha\gamma}P_{\gamma\alpha}P_{\bar{\alpha}\gamma}P_{\gamma\bar{\alpha}} &=& -1 ,
\end{eqnarray}
where $\gamma$ is a kink of arbitrary chirality.
The two sides of Eqn.(\ref{eBxv}) can then be rewritten as
\begin{eqnarray}\label{eBxviii}
\prod_{i=1}^{n-1}P_{\eps_n\eps_{i}}P_{\eps_{i}\eps_n} &=& (-1)^{n_e+n_o+n_p};\cr\cr
(-1)^{\frac{1}{4}\sum_{j}l[\epsilon_j]} &=& (-1)^{n_e+3n_o+n_p}.
\end{eqnarray}
These two are equal and we thus see that we recover 
the periodicity axiom for topologically neutral operators.

The case of the non-topologically neutral operators, 
specifically the non-Cartan currents, is dealt with similarly.
Denote by $J_{\alpha\bar{\beta}}$ the particular current of concern.  Any n-particle state coupling
to it will consist of 
some topologically neutral combination of
$n_e$ complete sets of even kinks,
$n_o$ complete sets of odd kinks, and $n_p$ particle-antiparticle pairs plus
one of the following:
\begin{eqnarray}\label{eBxix}
&1.&\text{the kink pair $\bar{\alpha}$ and $\beta$;}\cr\cr
&2.&\text{four even chirality kinks of the form}\cr
&&\text{$\{\beta,\beta$ and two even kinks $\gamma\neq\gamma'$,
$\gamma,\gamma'\neq\alpha,\beta)\}$};\cr\cr
&3.&\text{four odd chirality kinks of the form}\cr
&&\text{$\{\bar{\alpha},\bar{\alpha}$ and two odd kinks, $\bar\gamma\neq\bar\gamma'$,
$\gamma,\gamma'\neq\alpha,\beta\}$};\nn
\end{eqnarray}
Each of the three combinations listed above carry the quantum numbers
of $J_{\alpha\bar{\beta}}$. In case 1, $n = 2 + 4n_e + 4n_o + 2n_p$.
To recover the periodicity axiom in its initial form we must satisfy
\begin{equation}\label{eBxx}
e^{2\pi i \omega(J_{\alpha\bar{\beta}},\eps_n)}
\prod_{i=1}^{n-1}P_{\eps_n\eps_i}P_{\eps_i\eps_n} = (-1)^{\frac{1}{4}\sum_{j}l[\epsilon_j]}.
\end{equation}
Using the expression for the semi-locality phase in Eqn.(\ref{eBxiv}) we see that
\begin{equation}\label{eBxxi}
e^{2\pi i \omega(J_{\alpha\bar{\beta}},\eps_n)}
\prod_{i=1}^{n-1}P_{\eps_n\eps_i}P_{\eps_i\eps_n} = -(-1)^{n_e+n_o+n_p+2}
\end{equation}
while 
\begin{equation}\label{eBxxii}
(-1)^{\frac{1}{4}\sum_{j}l[\epsilon_j]} = (-1)^{n_e+3n_o+n_p+1}.
\end{equation}
We thus see that in this case we recover Smirnov's axiom for the non-Cartan currents.  The other two cases
in (\ref{eBxix}) proceed identically.

The final axiom to consider is the annihilation pole axiom.  If we transform the final equation of
(\ref{eBxi}) we obtain 
\begin{widetext}
\begin{eqnarray}\label{eBxxiii}
i{\rm Res}\bigl|_{\theta_n=\theta_{n-1}+i\pi}\
{f}^{\cal O}(\theta_1,\ldots,\theta_n)_{\eps_1,\ldots,\eps_n}
&=& f^{\cal O}(\theta_1,\ldots,\theta_{n-2})_{\eps'_1,\ldots,\eps'_{n-2}}\
{C}_{\eps_n,\eps'_{n-1}}\prod^{n-2}_{i=1}P_{\eps_n\eps_i}P_{\eps_{n-1}\eps_i}\cr\cr
&& \hskip -2in \times \Bigl[\delta^{\eps_1'}_{\eps_1}\cdots
  \delta^{\eps_{n-1}'}_{\eps_{n-1}}
-e^{2\pi i \omega({\cal O},\eps_{n-1})}
\big(\prod_{i>j}^{n-2}P_{\eps_i\eps_j}P_{\eps'_i\eps'_j}\big)
P_{\gamma_1\eps_1}P_{\eps'_1\eps'_{n-1}}
\big(\prod_{i=2}^{n-3}P_{\gamma_i\eps_i}P_{\eps'_i\gamma_{i-1}}\big)
P_{\eps_{n-1}\eps_{n-2}}P_{\eps'_{n-2}\gamma_{n-3}}\cr\cr
&& \hskip -.5in \times
{S}^{\eps'_{n-1}\eps'_1}_{\gamma_1\eps_1}(\theta_{n-1}-\theta_1)
{S}^{\gamma_{1}\eps'_2}_{\gamma_2\eps_2}(\theta_{n-1}-\theta_2)\ \cdots
{S}^{\gamma_{n-3}\eps'_{n-2}}_{\eps_{n-1}\eps_{n-2}}(\theta_{n-1}-\theta_{n-2})\Bigr].  
\end{eqnarray}
\end{widetext}
We first demonstrate that
\begin{equation}\label{eBxxiv}
\prod^{n-2}_{i=1}P_{\eps_n\eps_i}P_{\eps_{n-1}\eps_i} = 1
\end{equation}
Given that $\eps_n$ and $\eps_{n-1}$ are particle-antiparticle, the $n-2$ states remaining must
be of the type given in Eqn.(\ref{eBxvi}) 
if the operator carries zero topological charge or of the form given in Eqn.(\ref{eBxix}) if the operator is a
non-Cartan current.  In either case this product over $\eps_i$ can be broken down into pairs of either
two even chirality kinks, \{$\beta, \gamma$\}, two odd chirality kinks, \{$\bar{\beta}, \bar{\gamma}$\}, 
or one even kink and one odd kink, \{$\beta,\bar{\gamma}$\} (where $\beta$ is not necessarily
the anti-particle of $\bar{\gamma}$).  
That the above product is 1 then reduces to the following equalities which can
easily be checked:
\begin{eqnarray}\label{eBxxv}
P_{\alpha\beta}P_{\bar{\alpha}\beta}P_{\alpha\gamma}P_{\bar{\alpha}\gamma} &=& 1;\cr\cr
P_{\alpha\bar{\beta}}P_{\bar{\alpha}\bar{\beta}}P_{\alpha\bar{\gamma}}P_{\bar{\alpha}\bar{\gamma}} &=& 1;\cr\cr
P_{\alpha\beta}P_{\bar{\alpha}\beta}P_{\alpha\bar{\gamma}}P_{\bar{\alpha}\bar{\gamma}} &=& 1;
\end{eqnarray}
where $\beta$ and $\gamma$ are arbitrary even kinks.  
Noting that $\eps'_{n-1}=\eps_{n-1}$, we now demonstrate
\begin{eqnarray}\label{eBxxvi}
&& \big(\prod_{i>j}^{n-2}P_{\eps_i\eps_j}P_{\eps'_i\eps'_j}\big) 
P_{\gamma_1\eps_1}P_{\eps'_1\eps'_{n-1}}\cr\cr
&& \times\big(\prod_{i=2}^{n-3}P_{\gamma_i\eps_i}P_{\eps'_i\gamma_{i-1}}\big)
P_{\eps_{n-1}\eps_{n-2}}P_{\eps'_{n-2}\gamma_{n-3}}\cr\cr
&& = \prod^{n-2}_{i=1}P_{\eps_{n-1}\eps_i}P_{\eps_i\eps_{n-1}}
\end{eqnarray}
If we can do so, the axiom reverts to the form in Eqn.(\ref{eBi}) if we can show
\begin{eqnarray}\label{eBxxvii}
e^{2\pi i \omega({\cal O},\eps_{n-1})}\prod^{n-2}_{i=1}P_{\eps_{n-1}\eps_i}P_{\eps_i\eps_{n-1}}
= (-1)^{\frac{1}{4}\sum_{i=1}^{n-2} l[\eps_j]}.\nn
\end{eqnarray}
But this equality is almost the same as Eqn.(\ref{eBxvii}) and Eqn.(\ref{eBxxii}) and
can be established on identical lines.

To show Eqn.(\ref{eBxxvi}) holds, we appeal to the behavior of the transformation under
scattering.  The phase on the l.h.s. of Eqn.(\ref{eBxxvi}) arises
from the scattering 
(note that we consider only the case $\epsilon_{n-1}'=\epsilon_{n-1}$)
\begin{eqnarray}\label{eBxxviii}
\tilde{A}_{\eps_{n-1}}\cdots\tilde{A}_{\eps_1} &=& 
\tilde{S}^{\eps'_{n-1}\eps'_1}_{\gamma_1\eps_1}(\theta_{n-1}-\theta_1)
\tilde{S}^{\gamma_{1}\eps'_2}_{\gamma_2\eps_2}(\theta_{n-1}-\theta_2)\ \cdots\cr\cr
&& \hskip -.8in \times \tilde{S}^{\gamma_{n-3}\eps'_{n-2}}_{\eps_{n-1}\eps_{n-2}}(\theta_{n-1}-\theta_{n-2})
\tilde{A}_{\eps'_{n-2}}\cdots\tilde{A}_{\eps'_1}\tilde{A}_{\eps'_{n-1}}.
\end{eqnarray}
Applying the transformation, we know this must become
\begin{eqnarray}\label{eBxxix}
{A}_{\eps_{n-1}}\cdots{A}_{\eps_1} &=& 
{S}^{\eps'_{n-1}\eps'_1}_{\gamma_1\eps_1}(\theta_{n-1}-\theta_1)
{S}^{\gamma_{1}\eps'_2}_{\gamma_2\eps_2}(\theta_{n-1}-\theta_2)\ \cdots\cr\cr
&& \hskip -.8in \times {S}^{\gamma_{n-3}\eps'_{n-2}}_{\eps_{n-1}\eps_{n-2}}(\theta_{n-1}-\theta_{n-2})
{A}_{\eps'_{n-2}}\cdots{A}_{\eps'_1}{A}_{\eps'_{n-1}}.
\end{eqnarray}
For this to happen the following consistency condition must hold
\begin{eqnarray}\label{eBxxx}
1 &=& \big(\prod_{i>j}^{n-2}P_{\eps_i\eps_j}P_{\eps'_i\eps'_j}\big) 
P_{\gamma_1\eps_1}P_{\eps'_1\eps'_{n-1}}\cr\cr
&& \times\big(\prod_{i=2}^{n-3}P_{\gamma_i\eps_i}P_{\eps'_i\gamma_{i-1}}\big)
P_{\eps_{n-1}\eps_{n-2}}P_{\eps'_{n-2}\gamma_{n-3}}\cr\cr
&& \times \prod_{i=1}^{n-2}P_{\eps_{n-1}\eps_i}P_{\eps'_i\eps_{n-1}}
\end{eqnarray}
This consistency condition is equivalent to what we need to establish in Eqn.(\ref{eBxxvi})
provided
\begin{equation}\label{eBxxxi}
\prod_{i=1}^{n-2}P_{\eps_i\eps_{n-1}} = \prod_{i=1}^{n-2}P_{\eps'_i\eps_{n-1}}
\end{equation}
If the $\{\eps_i\}_{i=1}^{n-2}$ are all kinks or anti-kinks then $\{\eps'_i\}_{i=1}^{n-2}$
is but a permutation and the above equality certain holds.  If there are kinks and anti-kinks,
then a kink-anti-kink pair $\alpha-\bar{\alpha}$ in $\{\eps_i\}_{i=1}^{n-2}$ may become
$\beta-\bar{\beta}$ in $\{\eps'_i\}_{i=1}^{n-2}$ but because 
\begin{equation}\label{eBxxxii}
P_{\alpha\eps_{n-1}}P_{\bar{\alpha}\eps_{n-1}} = P_{\beta\eps_{n-1}}P_{\bar{\beta}\eps_{n-1}},
\end{equation}
the above equality is unaffected.

\subsection{Periodicity Axiom for Semi-local Operators of Section IV F}

We are now able to write down the correct form for the periodicity axiom for the two-particle
form-factors of the Sections III B 4. and 5.  The periodicity axiom for the form factor
of the scalar operator transforming as the {\bf 10} (Section III B 4.) in the C-symmetric basis is
\begin{equation}\label{eBxxxiii}
\tilde{f}^{[abc]_A}_{\alpha\alpha}(\theta - 2\pi i) = \tilde{f}^{[abc]_A}_{\alpha\alpha}(-\theta )
\end{equation}
The semi-locality phase in this case is 1.  As this operator couples to two identical kinks, 
transforming to the non-C symmetric basis leaves the equality unchanged:
\begin{equation}\label{eBxxxiv}
{f}^{[abc]_A}_{\alpha\alpha}(\theta - 2\pi i) = {f}^{[abc]_A}_{\alpha\alpha}(-\theta )
\end{equation}

We now consider the periodicity axiom for the form factor of the scalar operator transforming as the 
{\bf 6}.  For specificity we consider the operator 
\begin{equation}\label{eBxxxv}
\psi_1 = e^{\frac{i}{2}(-\phi_2 + \theta_3 + \theta_4)}.
\end{equation}
This operator couples to two same chirality kinks.
In the C-symmetric basis that periodicity axiom reads
\begin{equation}\label{eBxxxvi}
\tilde{f}^{\psi_1}_{\alpha\beta}(\theta - 2\pi i) = 
\begin{cases}
-\tilde{f}^{\psi_1}_{\beta\alpha}(-\theta ) & \text{for $\alpha,\beta$ even}\cr
\tilde{f}^{\psi_1}_{\beta\alpha}(-\theta ) & \text{for $\alpha,\beta$ odd}\cr
\end{cases}
\end{equation}
We see that for this particular operator the even kinks coupling to the field have a non-trivial OPE
whereas the odd kinks coupling do not.
Transforming to the non-C symmetric basis introduces a sign changes as
$P_{\alpha\beta}P_{\beta\alpha}=-1$: 
\begin{equation}\label{eBxxxvii}
{f}^{\psi_1}_{\alpha\beta}(\theta - 2\pi i) = 
\begin{cases}
{f}^{\psi_1}_{\beta\alpha}(-\theta ) & \text{for $\alpha,\beta$ even}\cr
-{f}^{\psi_1}_{\beta\alpha}(-\theta ) & \text{for $\alpha,\beta$ odd}\cr
\end{cases}
\end{equation}
If we had instead considered an operator of the form, $\psi_1 =
e^{\frac{i}{2}(-\phi_2 - \theta_3 + \theta_4)}$ the role of even and
odd kinks would be swapped.  But regardless of which particular scalar
operator transforming as the {\bf 6} we consider, one type of kink
will carry an additional sign in the periodicity axiom relative to the other.

\section{Derivation of the Susceptibility of Non-Interacting Ladders}
\label{app:suscept}

The non-interacting susceptibility is given by
\begin{eqnarray}\label{eCi}
\chi^0_{11}(q_x,\omega ) &=& \chi^0_{22}(q_x,\omega )\cr\cr
&& \hskip -.8in =  -\frac{3}{4} \big(L_1(\omega,q_x)+L_2(\omega,q_x)+L_3(\omega,q_x)+L_4(\omega,q_x)\big);\cr\cr
\chi^0_{12}(q_x,\omega ) &=& \chi^0_{21}(q_x,\omega )\cr\cr
&& \hskip -.8in = -\frac{3}{4} \big(L_1(\omega,q_x)+L_2(\omega,q_x)-L_3(\omega,q_x)-L_4(\omega,q_x)\big),\cr
&&
\end{eqnarray}
where the $L_i$'s are defined as follows:
\begin{widetext}
\begin{eqnarray}\label{eCii}
L_1 (\omega , q_x) &=& \int d\tau dx e^{i\omega \tau - iq_x x} 
\langle T\bigg(\big(\sum_\sigma \sigma c^\dagger_{b\sigma}(x,\tau)c_{b\sigma}(x,\tau)\big)
\big(\sum_\sigma \sigma c^\dagger_{b\sigma}(0,0)c_{b\sigma}(0,0)\big)\bigg)
\rangle\big|_{\omega \rightarrow -i\omega + \epsilon}\cr\cr
&=&
-\frac{1}{2\pi} \frac{1}{\Gamma(\omega,q_x)}
\bigg(\tanh^{-1}{4t\sin(q_x/2)\cos(K_{Fb}+q_x/2) \over \Gamma(\omega,q_x)} 
- \tanh^{-1}{4t\sin(q_x/2)\cos(K_{Fb}-q_x/2) \over \Gamma(\omega,q_x)}\bigg);
\cr\cr\cr
L_2 (\omega , q_x) &=& \int d\tau dx e^{i\omega \tau - iq_x x} 
\langle T\bigg(\big(\sum_\sigma \sigma c^\dagger_{ab\sigma}(x,\tau)c_{ab\sigma}(x,\tau)\big)
\big(\sum_\sigma \sigma c^\dagger_{ab\sigma}(0,0)c_{ab\sigma}(0,0)\big)\bigg)
\rangle\big|_{\omega \rightarrow -i\omega + \epsilon}\cr\cr
&=&
-{1\over 2\pi} {1\over \Gamma(\omega,q_x)}
\bigg(\tanh^{-1}{4t\sin(q_x/2)\cos(K_{Fab}+q_x/2) \over \Gamma(\omega,q_x)} 
- \tanh^{-1}{4t\sin(q_x/2)\cos(K_{Fab}-q_x/2) \over \Gamma(\omega,q_x)}\bigg);
\cr\cr\cr
L_3(\omega , q_x) &=& \int d\tau dx e^{i\omega \tau - iq_x x} 
\langle T\bigg(\big(\sum_\sigma \sigma c^\dagger_{b\sigma}(x,\tau)c_{b\sigma}(x,\tau)\big)
\big(\sum_\sigma \sigma c^\dagger_{ab\sigma}(0,0)c_{ab\sigma}(0,0)\big)\bigg)
\rangle\big|_{\omega \rightarrow -i\omega + \epsilon}\cr\cr
&=& L_{31}(\omega, q_x) + L_{32}(\omega, q_x) + L_{33}(\omega, q_x) + L_{34}(\omega, q_x);\cr\cr
L_{31}(\omega, q_x) &=& 
{i\over 4\pi} \int dq_1 \theta (2t\cos (q_1) + t_\perp) 
{-i2t\cos (q_1) + i2t\cos (q_1-q) 
\over (-iw+\epsilon-i2t_\perp)^2 +4t^2(\cos(q_1) - \cos(q_1+q))^2}\cr\cr
&=& {1\over 2}L_1(\omega+2t_\perp,q_x)\cr\cr
L_{32}(\omega, q_x) &=& {i\over 4\pi} \int dq_1 \theta (2t\cos (q_1) - t_\perp) 
{-i2t\cos (q_1) + i2t\cos (q_1-q) 
\over (-iw+\epsilon-i2t_\perp)^2 +4t^2(\cos(q_1) - \cos(q_1+q))^2}\cr\cr
&=& {1\over 2}L_2(\omega+2t_\perp,q_x)\cr\cr
L_{33}(\omega, q_x) &=& 
{i\over 4\pi} \int dq_1 \theta (2t\cos (q_1) + t_\perp) 
{iw + i2t_\perp
\over (-iw+\epsilon-i2t_\perp)^2 +4t^2(\cos(q_1) - \cos(q_1+q))^2}\cr\cr
&=& - {i\over 4\pi} {1\over \Gamma(\omega+2t_\perp, q_x)}
\bigg\{\pi n_{K_{Fb}+|q_x|/2}+
\tan^{-1}\big({\Gamma(\omega+2t_\perp,q_x) \tan(K_{Fb}+|q_x|/2) \over (-i\omega-i2t_\perp+\epsilon)}\big)\cr\cr
&& \hskip .9in + \rm{sgn}(K_{Fb}-|q_x|/2)\bigg(\pi n_{\big|K_{Fb}-|q_x|/2\big|}+
\tan^{-1}\big({\Gamma(\omega+2t_\perp,q_x) \tan(|K_{Fb}-|q_x|/2|) \over (-i\omega-i2t_\perp+\epsilon)}
\big)\bigg)\bigg\}\cr\cr
L_{34}(\omega, q_x) &=& {i\over 4\pi} \int dq_1 \theta (2t\cos (q_1) - t_\perp) 
{-iw - i2t_\perp
\over (-iw+\epsilon -i2t_\perp)^2 +4t^2(\cos(q_1) - \cos(q_1+q))^2}\cr\cr
&=&  {i\over 4\pi} {1\over \Gamma(\omega+2t_\perp, q_x)}
\bigg\{\pi n_{K_{Fab}+|q_x|/2}+
\tan^{-1}\big({\Gamma(\omega+2t_\perp,q_x) \tan(K_{Fab}+|q_x|/2) \over (-i\omega-i2t_\perp+\epsilon)}\big)\cr\cr
&& \hskip .9in + \rm{sgn}(K_{Fab}-|q_x|/2)\bigg(\pi n_{\big|K_{Fab}-|q_x|/2\big|}+
\tan^{-1}\big({\Gamma(\omega+2t_\perp,q_x) \tan(|K_{Fab}-|q_x|/2|) \over (-i\omega-i2t_\perp+\epsilon)}
\big)\bigg)\bigg\};\cr\cr\cr
L_4(\omega , q_x) &=& \int d\tau dx e^{i\omega \tau - iq_x x} 
\langle T\bigg(\big(\sum_\sigma \sigma c^\dagger_{ab\sigma}(x,\tau)c_{ab\sigma}(x,\tau)\big)
\big(\sum_\sigma \sigma c^\dagger_{b\sigma}(0,0)c_{b\sigma}(0,0)\big)\bigg)
\rangle\big|_{\omega \rightarrow -i\omega + \epsilon}\cr\cr
 &=& L_3(\omega-4t_\perp,q_x,K_{Fb}\leftrightarrow K_{Fab}),
\end{eqnarray}
\end{widetext}
where $\Gamma (\omega, q_x) = (-(\omega+i\epsilon)^2+16t^2\sin^2(q_x/2))^{1/2}$
and $n_q$ is an integer minimizing $|n_q\pi -q|$.



\begin{thebibliography}{99}
\bibitem{note} 
We point out that $\psi^{[abc]_A}$ transforms under the
reducible representation ${\bf 10 \oplus 10}$.  However, we are only
interested in one of the ${\bf 10}$'s and so an appropriate projection
is understood when considering these fields.
\end{thebibliography}
\end{document}